\documentclass[english,aps,pra,superscriptaddress,notitlepage,longbibliography]{revtex4-1}
\usepackage{hyperref}
\usepackage[T1]{fontenc}
\usepackage[latin9]{inputenc}
\usepackage{verbatim}
\usepackage{float}
\usepackage{amsmath}
\usepackage{amssymb}
\usepackage{esint}

\makeatletter
\usepackage{babel}
\usepackage{amsfonts}
\usepackage{graphicx}
\usepackage{xcolor}
\definecolor{darkblue}{rgb}{0,0,0.5} 
\usepackage{transparent}

\makeatother

\usepackage[capitalize]{cleveref}

\begin{document}

\author{Emil Zeuthen}

\email{zeuthen@nbi.ku.dk}

\affiliation{Niels Bohr Institute, University of Copenhagen, DK-2100 Copenhagen,
Denmark}
\affiliation{Institute for Theoretical Physics and Institute for Gravitational
Physics (Albert Einstein Institute), Leibniz Universit\"{a}t Hannover,
Callinstra{\ss}e 38, 30167 Hannover, Germany}

\author{Albert Schliesser}

\affiliation{Niels Bohr Institute, University of Copenhagen, DK-2100 Copenhagen,
Denmark}

\author{Jacob M. Taylor}

\affiliation{Joint Quantum Institute, University of Maryland/National Institute of Standards and Technology, College Park, Maryland 20742, USA}
\affiliation{Joint Center for Quantum Information and Computer Science, University of Maryland, College Park, Maryland 20742, USA}

\author{Anders S. S{\o}rensen}

\affiliation{Niels Bohr Institute, University of Copenhagen, DK-2100 Copenhagen,
Denmark}

\title{Electro-optomechanical equivalent circuits for quantum transduction}
\begin{abstract}
Using the techniques of optomechanics, a high-$Q$ mechanical oscillator may serve as a link between
electromagnetic modes of vastly different frequencies. This approach has successfully been exploited for the frequency conversion of classical signals and has the potential of performing
quantum state transfer between superconducting circuitry and a traveling
optical signal. 
Such transducers are often operated in a linear regime, where the hybrid system can be described using linear response theory based on the Heisenberg-Langevin equations. While mathematically straightforward to solve, this approach yields little intuition about the dynamics of the hybrid system to aid the optimization of the transducer.
As an analysis and design tool for such electro-optomechanical transducers, we introduce an equivalent circuit formalism, where the entire transducer is represented by an electrical circuit.  
Thereby we integrate the transduction functionality of optomechanical (OM) systems into the toolbox of electrical engineering allowing the use of its well-established design techniques. This unifying impedance description can be applied both for static (DC) and harmonically varying (AC) drive fields, accommodates arbitrary linear circuits, and is not restricted to the resolved-sideband regime. Furthermore, by establishing the quantized input-output formalism for the equivalent circuit, we obtain the scattering matrix for linear transducers using circuit analysis, and thereby have a complete quantum mechanical characterization of the transducer.
Hence, this mapping of the entire transducer to the language of electrical engineering both sheds light on how the transducer performs and can at the same time be used to optimize its performance by aiding the design of a suitable electrical circuit.
\end{abstract}
\maketitle
\tableofcontents{}

\section{Introduction}

Quantum-coherent technology is envisioned to usher in a new era of information processing and communication. While quantum effects are already at play in semiconductor transistors and lasers, quantum-coherent effects are believed to take center stage in future quantum technology. To implement this, the entire infrastructure must be built on quantum-enabled components, and this has spurred a wide-ranging research effort into, e.g., quantum processors (gates), quantum memory, quantum transistors, and quantum error correction. Quantum computers, built from such components, are envisioned to be securely networked via optical fibers \cite{Gisin2002} on account of their long-distance transmission capabilities, thereby forming a quantum internet~\cite{Kimble2008}. Since several quantum computing architectures are working in the frequency domain of MHz or GHz--including superconducting qubits~\cite{Wendin2017,Chiorescu2003,Yamamoto2003,Wallraff2004,Lucero2012,Riste2013,Barends2016,Casparis2016,Otterbach2017,Wang2018}, quantum dots~\cite{Petta2005,Koppens2006,Shulman2012,Nichol2017,Stockklauser2017,Watson2017,Zajac2018}, electron spin ensembles~\cite{Wesenberg2009}, polar molecules~\cite{Andre2006}, 
and donor spin qubits in silicon~\cite{Dehollain2015,Tettamanzi2017}--this poses the demand for a quantum ``modem'' providing a hookup to the optical network. Such a device should provide quantum-coherent frequency conversion, also referred to as \textit{quantum transduction}, thereby bringing together the strengths of, e.g., on the one hand, superconducting circuitry in which quantum gates and state preparation can be performed efficiently and, on the other hand, the low-loss transmission and quantum-limited detection of optical signals. These advantageous features of the optical domain persist even at room temperature. Hence the successful transduction of low-frequency signals to the optical regime may also be exploited for the sensitive detection of weak, classical electrical signals.

Transduction typically involves a modulation mechanism. A strong and versatile candidate for such a mechanism can be found within optomechanics. This research field is concerned with the interaction between light fields and mechanical oscillators through the radiation pressure force~\cite{Aspelmeyer2013,CavityOptomechSpringer}. One can greatly enhance this force
in a strongly driven high-finesse cavity. This idea has been realized over a very wide mass spectrum ranging from nanoscale objects to the macroscopic mirrors employed in gravitational wave detection~\cite{Danilishin2012}.
With this approach, a new regime has been attained in the past decade in which the backaction of the light on the mechanical motion due to the radiation pressure force is significant~\cite{Arcizet2006,Schliesser2006,Gigan2006,Thompson2008,Wilson2009}.
Moreover, the availability of mechanical oscillators with quality factors in excess of $10^6$ combined with cryogenic cooling allows quantum operation, where the mapping of optical photons to and from the mechanical mode is performed faster than the thermal decoherence time of the mode (i.e., the regime of large quantum cooperativity).
This has enabled surprising levels of control with meso- and macroscopic mechanical oscillators allowing optical cooling of the mechanical motion to near its quantum-mechanical ground state~\cite{Chan2011,Verhagen2012,Underwood2015,Peterson2016,Nielsen2017}.

Since electromagnetic fields from all parts of the spectrum can exert a force on mechanical objects, the principles of optomechanics find their complete analogs in the context of radio-frequency and microwave electrical circuits. Electromechanical transduction has been exploited in the classical regime since the advent of the telephone in the late 19th century. Parallelling the rapid development in optomechanics, experiments in the micro- and nano-electromechanical (MEMS and NEMS) communities have also reached the quantum regime. Among the most notable achievements are cooling of a mechanical oscillator to the vicinity of its ground state~\cite{OConnell2010,Rocheleau2010,Teufel2011}, reversible state transfer of quantum-level signals~\cite{Palomaki2013,Reed2017} and entanglement~\cite{Palomaki2013b} between electromagnetic fields and a mechanical mode, mechanical microwave amplification near the quantum limit~\cite{Massel2011,Ockeloen2016,Toth2017}, and strong coupling between a mechanical oscillator and a super-conducting qubit~\cite{LaHaye2009,OConnell2010,Rouxinal2016,Chu2017}.

\begin{figure}
\centering
\includegraphics[width=0.6\textwidth]{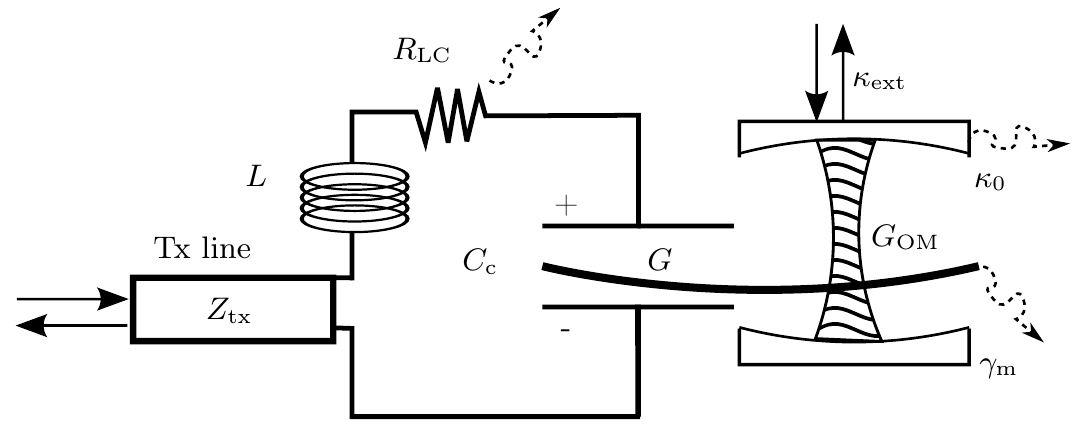}
\caption{Example electro-optomechanical transducer: A serial RLC circuit with intrinsic resistance $R_{\text{LC}}$ is loaded by a semi-infinite transmission line of characteristic impedance $Z_{\text{tx}}$. The circuit is capacitively coupled to a mechanical mode of intrinsic linewidth $\gamma_{\text{m}}$ with coupling strength $G$. Analogously, the optical mode has an intrinsic loss rate $\kappa_{0}$ and a readout rate $\kappa_{\text{ext}}$ and couples to the mechanical mode with strength $G_{\text{OM}}$.
\label{fig:Example-system}}
\end{figure}

Hence both optomechanics and electromechanics are mature research directions offering quantum-level operation. By combining the two, they would therefore offer a promising platform for quantum transduction between electrical and optical frequencies~\cite{Safavi-Naeini2011,Regal2011,Taylor2011,Tian2015,Ondrej2017} allowing
loss- and noiseless conversion of quantum states, but also
the generation of, e.g., two-mode squeezed hybrid states of light and microwaves for continuous-variable teleportation~\cite{Barzanjeh2012} and entanglement between distant superconducting qubits by transducer-mediated interaction with a common propagating optical field \cite{Ondrej2016}.
In fact, optomechanical interfaces suitable for integration with electrical circuits have already been devised~\cite{Jayich2008,Unterreithmeier2009,Schmid2014,Bagci2014,Andrews2014,Pitanti2015,Fink2016,Menke2017,Bochmann2013,Vainsencher2016,Balram2016,Zou2016,Takeda2017}. A prominent example of this is the Membrane-In-The-Middle optomechanical system in which a micromechanical membrane is placed inside a Fabry-P\'{e}rot cavity with fixed end mirrors~\cite{Jayich2008}. This geometry allows significant freedom to simultaneously optimize optical and mechanical properties. Moreover, the fact that the mechanical degree of freedom does not involve the optical components permits easy integration with electrical systems. In particular, it can be directly incorporated into floating-electrode geometries~\cite{Unterreithmeier2009}, where a mechanically compliant dielectric or conductor moves in the quasi-electrostatic field of fixed electrodes, thereby modulating their capacitance~\cite{Schmid2014}.
Using this approach, electro-optomechanical hybrid devices have already been built where a mechanical mode couples simultaneously to optical and electrical fields (see Fig.~\ref{fig:Example-system}), thereby realizing optical readout of radio-frequency signals in a room-temperature electrical circuit~\cite{Bagci2014,Takeda2017} as well as microwave-to-optical transduction in setups designed for cryogenic operation~\cite{Andrews2014,Pitanti2015,Fink2016,Menke2017}.
Alternatively, the electromechanical coupling can be implemented piezoelectrically~\cite{Bochmann2013,Vainsencher2016,Balram2016,Zou2016}.

Electro-optomechanics brings together the research fields of electrical engineering and quantum optics, both of which have a well-developed set of experimental and theoretical tools. To realize the full potential of marrying these two fields it is highly desirable to establish a common language for the resulting hybrid systems, thus making the knowledge and techniques from one system applicable to the other.

\subsection{Equivalent circuits}

\begin{figure}
\centering
\includegraphics[width=0.6\textwidth]{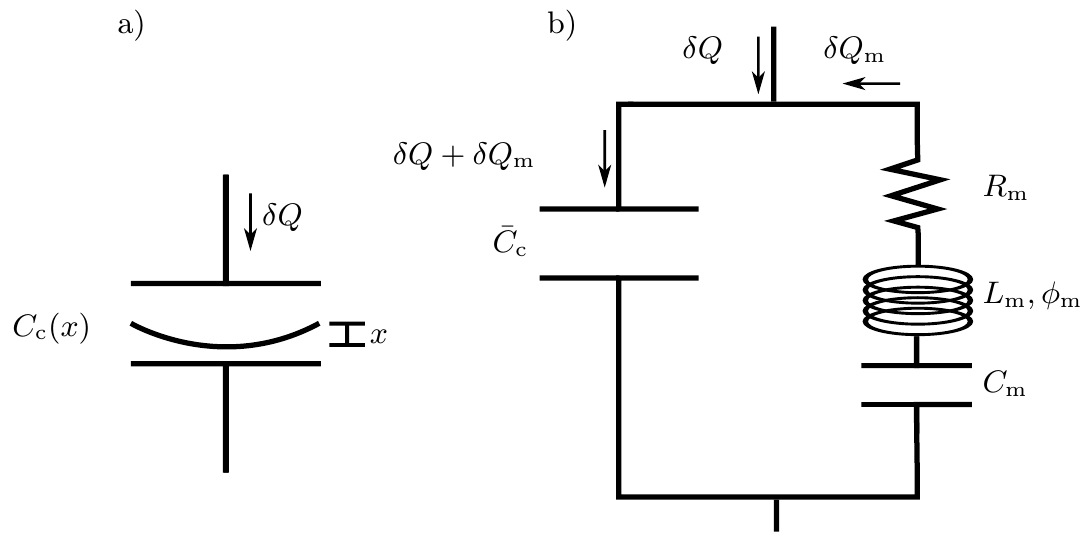}
\caption{DC-biased electromechanical interface based on capacitive coupling. a) Sketch of a mechanically modulated coupling capacitor, $C_{\text{c}}(x)$, where $x$ is a suitable mechanical position coordinate and $\delta Q$ is an added charge.
b) Equivalent circuit for the linearized dynamics of the system in a) in which the mechanical mode is represented by a serial RLC circuit in parallel to the \emph{unmodulated} coupling capacitor $\bar{C}_{\text{c}}$. The charge $\delta Q_{\mathrm{m}}$ and the magnetic flux $\phi_{\mathrm{m}}$ play the roles of position and momentum, respectively. The equivalent mechanical resistance $R_{\mathrm{m}}$, inductance $L_{\mathrm{m}}$, and capacitance $C_{\mathrm{m}}$ correspond to damping rate, mass, and (inverse) spring constant, respectively.}
\label{fig:Mech-equiv}
\end{figure}

In this article, we develop an equivalent circuit formalism for electro-optomechanical transducers that accommodates arbitrary linear electrical circuits. With this approach the entire coupled system of electrical modes, mechanical vibrations, and optical fields can be described in terms of a linear electrical circuit obeying Kirchhoff's laws. 
We consider the system shown in Fig.~\ref{fig:Mech-equiv}(a), where a membrane is inserted into a capacitor such that the capacitance $C_{\mathrm{c}}(x)$ depends on the excursion $x$ of the membrane from its equilibrium position. When the membrane moves it thereby modulates the capacitance, which leads to a coupling of the mechanical motion and the electric circuit. We show that in the linear regime this system can be exactly represented by the electrical circuit in Fig.~\ref{fig:Mech-equiv}(b). Here the mechanical oscillator is represented by a parallel arm containing an RLC circuit with resonance frequency and damping set by the corresponding mechanical parameters. The position of the membrane $x$ is then directly related to the charge $\delta Q_{\textrm{m}}$ sitting on the capacitor in the parallel arm, $x\propto \delta Q_{\textrm{m}}$.
The formal reason that such an equivalent-circuit formulation is possible, is that all the involved systems are conveniently described in the Heisenberg-Langevin input-output formalism for coupled oscillators~\cite{Hudson1984,Collett1985}. For an arbitrary linear circuit, the Heisenberg-Langevin equations are nothing but a quantized version of Kirchhoff's well-known circuit laws~\cite{Yurke1984}. As a consequence, all the involved degrees of freedom can be mapped to electrical analogs thereby allowing a common equivalent circuit description of the full electro-optomechanical system. Hence, one may adequately describe the system by Kirchhoff's laws with impedances determined by the electro-optomechanical couplings.
This approach becomes particularly advantageous if complicated electrical circuits are involved. For simple circuits a coupled oscillator model will suffice to describe the transducer~\cite{Wang2012,Tian2012,Andrews2014}, but for circuits involving multiple loops and branches the identification of the relevant harmonic oscillator modes is non-trivial, whereas Kirchhoff's laws provide a simple description~\cite{EZPhD}.

A similar representation of mechanical oscillation modes coupled to an electrical circuit by an equivalent circuit element is already an established tool in the MEMS community~\cite{Tilmans1996,Liwei1998, OConnell2014}, known as the Butterworth-van Dyke circuit, and has also been applied in the contexts of cavity-electromechanics~\cite{Brown2007} and piezo crystals~\cite{OConnell2010}. We take this idea further by making the following important extensions: Firstly, we derive a quite general, yet simple, equivalent circuit for AC-biased electromechanical (EM) systems as is typical of experiments in the microwave domain. To our knowledge such a generalization has not been presented before. Secondly, we derive the impedance of a mechanically coupled optical mode in a high-finesse optical resonator, allowing us to construct a full electro-optomechanical equivalent circuit. Moreover, we will discuss how to quantize the theory.
The main asset of our work, however, is that we merge all of these elements together in a single framework containing all elements of the transducer, including a full quantum description of the system. Below we will therefore reproduce several well-established results from the electrical engineering literature to make the description self-contained. The exact relation to these previous works will be pointed out in the relevant sections of the article.

\subsection{Structure of paper}

We will establish the electro-optomechanical equivalent circuit formalism
in the following steps: First we introduce a Hamiltonian description
for EM interfaces in Section \ref{sec:EM-interfaces};
based on this, we give an intuitive derivation of the mechanical equivalent
impedance for a DC-biased interface in Section \ref{sub:Intuitive-deriv-equiv}.
Next, we extend the EM equivalent circuit to the case
of an AC electrical drive in Section \ref{sec:EM-equiv-circ-AC}.
Then, in Section \ref{sec:Electrical-IO}, we supplement the impedance
formalism by introducing electrical input-output theory, which allows
us to consider the in- and outcoupling of signals and noise. Based on this, 
we describe how to quantize the theory in Section \ref{sec:Quantization} and demonstrate the formalism by calculating the squeezing spectrum of a simple electromechanical system using standard rules for combining impedances.
Having established all aspects of the EM equivalent
circuit, we then introduce the optical subsystem by analogy in Section
\ref{sec:Optical-impedance_full-equiv}, thereby arriving at the full
electro-optomechanical equivalent circuit.
As an example of the application of the formalism, we apply it
to a simple transducer in Section~\ref{sec:Example-application}.
In Section~\ref{sec:Adiab-elim} we derive a reduced equivalent circuit by adiabatic elimination of the electrical and optical modes. Finally, we conclude and give an outlook in Section~\ref{sec:Conclusion}.

\section{Electromechanical interfaces\label{sec:EM-interfaces}}

In this section we will introduce a Hamiltonian description for capacitive EM coupling and discuss the linearization of the interaction around the steady-state configuration induced by the drive field.
The recapitulation of these well-established methods will lay the foundation for the subsequent sections.

\subsection{Parametric electro-mechanical coupling}

We consider a single viscously damped mechanical mode with canonical position $x$ and momentum $p$, whose evolution is
governed by the Langevin equations (in the absence of EM coupling),
\begin{gather}
\dot{x}=p/m\nonumber \\
\dot{p}=-m\omega_{{\rm m},0}^{2}x-\gamma_{{\rm m},0}p+F(t),\label{eq:Mech_EOM}
\end{gather}
where dot signifies the time derivative, $m$ is the mass, $\omega_{\text{m,}0}$ is the resonance frequency, $\gamma_{\text{m,}0}$ is the damping rate, and $F$ is the associated stochastic force, whose fluctuation spectrum is~\cite{Gardiner}
\begin{equation}
\langle F^{*}(\omega) F(\omega') \rangle = 2 m \gamma_{\rm m,0} k_{\rm B} T_{\rm m} \delta (\omega - \omega'),\label{eq:mech-fluc-spectrum}
\end{equation}
where $k_{\rm B}$ is Boltzmann's constant and $T_{\rm m}$ is the ambient temperature of the mechanical component. The classical Eqs.~(\ref{eq:Mech_EOM},\ref{eq:mech-fluc-spectrum}) represent the high-temperature limit ($k_{\rm B}T_{\rm m} \gg \hbar \omega_{\rm m,0}$) of quantum Brownian motion, which is governed by Eq.~\eqref{eq:Mech_EOM} with $x$, $p$, and $F$ elevated to operators, and the fluctuation spectrum \eqref{eq:mech-fluc-spectrum} generalized to ($\omega,\omega'>0$)~\cite{Giovannetti2001}
\begin{eqnarray}
\langle \hat{F}^{\dagger}(\omega) \hat{F}(\omega') \rangle &=& 2 m \gamma_{\rm m,0} \hbar \omega n_{\rm m}(\omega) \delta (\omega - \omega'),\nonumber\\
\langle \hat{F}(\omega) \hat{F}^{\dagger}(\omega') \rangle &=& 2 m \gamma_{\rm m,0} \hbar \omega [ n_{\rm m}(\omega) + 1] \delta (\omega - \omega'),\label{eq:mech-fluc-spectrum-Q}
\end{eqnarray}
in terms of the Bose-Einstein distribution,
\begin{equation}
n_{i}(\omega)\equiv(e^{\hbar\omega/(k_{\text{B}}T_{i})}-1)^{-1}.\label{eq:Bose-Einstein}
\end{equation}
For later reference we note that the mechanical evolution (\ref{eq:Mech_EOM}) derives from the (classical) Hamiltonian
\begin{equation}
H_{\text{m},0}=\frac{p^{2}}{2m}+\frac{1}{2}m\omega_{\text{m},0}^{2}x^{2},\label{eq:H-m0}
\end{equation}
plus additional terms accounting for the coupling to the bath responsible for the damping and noise fluctuations.

In order to construct an EM transducer it is essential to have a coupling between  the mechanical and electronic degrees of freedom.  Motivated
by recent experiments \cite{Rocheleau2010,Massel2012,Palomaki2013,Bagci2014,Andrews2014},
we will consider a capacitive coupling where the mechanical oscillator modulates the capacitance $C_{\text{c}}$ of a capacitor in the circuit such that it acquires a dependence on the  mechanical position, $C_{\text{c}}(x)$, see Fig.~\ref{fig:Mech-equiv}a. The physical mechanism underlying this dependence
depends on the implementation, but examples include the Kelvin polarization
force from an inhomogeneous electric field on a dielectric mechanical
element \cite{Unterreithmeier2009,Schmid2010} and the quasi-electrostatic
interaction with a conductive mechanical element \cite{Rocheleau2010,Teufel2011,Massel2012,Zhou2013,Schmid2014}, but
the precise nature of the coupling is not important for this study (e.g., the formalism developed below for the linearized dynamics can also be extended to inductive~\cite{Crandall1968} or piezoelectric~\cite{OConnell2010,Bochmann2013} coupling).
Since the motion of the mechanical oscillator modulates the capacitance it also modulates the  resonance frequency of the circuit, thus giving rise to a dispersive interaction. As an alternative to this,  other types of coupling have been proposed  
including dissipative coupling \cite{Elste2009} and mechanical multimode
schemes \cite{Massel2012,Xuereb2012,Shkarin2014}. Here we restrict ourselves to the capacitive coupling, but the formalism is likely to be extendable to accommodate other kinds of parametric couplings as well as multiple mechanical modes.

If a charge $Q$ is added to the capacitor, the dependence of the capacitance on the position means that there is a force on the mechanical oscillator which will displace it to a new equilibrium value $\bar{x}$. Here we are interested in the small fluctuations $\delta x \equiv x-\bar{x}$ around this  equilibrium. We therefore expand the charging energy of the capacitor to obtain~\cite{Bagci2014}
\begin{equation}
H_{\text{C}}=\frac{Q^{2}}{2C_{\text{c}}(x)}\approx\frac{Q^{2}}{2\bar{C}_{\text{c}}}-\frac{1}{2}\frac{Q^{2}}{\bar{C}_{\text{c}}^{2}}\left.\frac{dC_{\text{c}}}{dx}\right|_{x=\bar{x}}\delta x+\frac{Q^{2}}{4\bar{C}_{\text{c}}^{2}}\left[\frac{2}{\bar{C}_{\text{c}}}\left.\left(\frac{dC_{\text{c}}}{dx}\right)^{2}\right|_{x=\bar{x}}-\left.\frac{d^{2}C_{\text{c}}}{dx^{2}}\right|_{x=\bar{x}}\right]\delta x^{2},\label{eq:H-charging}
\end{equation}
where $\bar{C}_{\text{c}}\equiv C_{\text{c}}(\bar{x})$
denotes the steady-state value of the coupling capacitance. Note, that the notion
of a position-dependent charging energy, presumed in Eq.~(\ref{eq:H-charging}),
is only meaningful in the quasi-electrostatic limit, where the charges on the capacitive element equilibrates much faster
than the timescale of the mechanical modulation $2\pi/\omega_{\text{m,}0}$.
Moreover, we remark that the mathematical problem of determining the equilibrium configuration of the biased electromechanical system involves the solution of non-linear equations and hence in general requires approximate or numerical methods (even more so for the electro-optomechanical systems to be considered later). However, we will not concern ourselves with this aspect in the present work and will henceforth presume the existence of a stable equilibrium configuration of the joint system in the presence of the given biasing/driving.

\subsection{Enhanced linearized interaction in presence of a drive field}

The EM interaction Hamiltonian~(\ref{eq:H-charging}) is non-linear
in nature and typically very weak. For transduction, however,  linear interaction
is typically sufficient and even desirable. A much stronger interaction can then be obtained
by applying a strong classical drive voltage which induces a
charge $\bar{Q}_{\text{c}}(t)$ of large amplitude on the capacitor.
Such biasing is a well-known technique in electronics, e.g., it is the operating principle behind condenser microphones, in which one plate of a capacitor is a diaphragm susceptible to sound waves.
We will consider two different situations: If the electrical signal is near resonance with the mechanical oscillator a DC bias is sufficient to couple the two, whereas if they are different an AC bias is required to bridge the difference in resonance frequencies.
As for the position variable above, we are interested in the small charge fluctuations $\delta Q$ around the mean and make the replacement 
$Q\rightarrow\bar{Q}_{\text{c}}(t)+\delta Q.$
We can then derive the effective Hamiltonian governing the interactions
among the fluctuation variables $\delta x,\delta Q$.
For simplicity, we assume a monochromatic electrical drive and, moreover,
that the circuit responds linearly, so as to induce a fluctuating
charge on the coupling capacitor
\begin{equation}
\bar{Q}_{\text{c}}(t)=\begin{cases}
\bar{Q}_{\text{c,}0}e^{i\omega_{\text{d}}t}+\bar{Q}_{\text{c,}0}^{*}e^{-i\omega_{\text{d}}t} & [\text{AC bias}]\\
\bar{Q}_{\text{c,}0} & [\text{DC bias}]
\end{cases}, \label{eq:Q-c-bar_def}
\end{equation}
where in the case of DC bias $\bar{Q}_{\text{c,}0}$ must be real. With this charge bias, Eq.~(\ref{eq:H-charging})
leads to a linear interaction among the fluctuation variables to lowest
order,
\begin{gather}
H_{\text{EM,int}}\approx-\frac{\bar{Q}_{\text{c}}(t)}{\bar{C}_{\text{c}}^{2}}\left.\frac{dC_{\text{c}}}{dx}\right|_{x=\bar{x}}\delta Q\delta x=\delta Q\delta x\times\begin{cases}
Ge^{i\omega_{\text{d}}t}+G^{*}e^{-i\omega_{\text{d}}t} & [\text{AC bias}]\\
G & [\text{DC bias}]
\end{cases}.\label{eq:H_int-lin_EM}
\end{gather}
Here we have introduced the drive-enhanced EM coupling parameter $G$
(SI units of V/m):
\begin{gather}
G\equiv-\frac{\bar{Q}_{\text{c,}0}}{\bar{C}_{\text{c}}^{2}}\left.\frac{dC_{\text{c}}}{dx}\right|_{x=\bar{x}}.\label{eq:G_def}
\end{gather}
Throughout this article we attempt to emphasize both the similarities and differences
between the DC- and AC-biased EM interfaces. Eq.~(\ref{eq:Q-c-bar_def}) implies that in the AC-biased case we define $|\bar{Q}_{\text{c,}0}|$ to be half of the charge amplitude whereas in the DC case it is the full amplitude; this choice allows for a simpler presentation below.
The EM coupling strength $G$ introduced in Eq.~(\ref{eq:G_def}) will play
a central role in the derivation of the equivalent picture later on
as it characterizes the strength of the interaction. A related, more familiar parameter in the optomechanics community is the linearized coupling rate $g_{\rm EM}$ between the two bosonic modes representing the circuit and mechanical resonances (as will be detailed below in Section~\ref{sec:Optical-impedance_full-equiv} for the equivalent OM case). An advantage
of $G$ over $g_{\rm EM}$ is that the former can be meaningfully defined without specifying
an electrical circuit resonance and therefore we will focus on $G$ in the derivation below.

\section{Intuitive derivation of electromechanical equivalent circuit for
DC bias\label{sub:Intuitive-deriv-equiv}}

We will now give a simple derivation of 
the equivalent circuit representation of a mechanical system
coupled to an electrical circuit. For simplicity we start with the
already well-established
case of DC-biased EM coupling~\cite{Liwei1998,OConnell2014}, postponing 
our extension to the
AC-biased scenario until Section~\ref{sec:EM-equiv-circ-AC}. In essence,
we are looking for a way to describe the linear response of the system
depicted in Fig.~\ref{fig:Mech-equiv}a with a circuit
diagram consisting of standard components (capacitors, inductors,
etc.). In the figure only the coupling capacitor is drawn, but it is assumed to be connected to an arbitrary linear circuit. Assume that we add a positive charge $\delta Q$ to the positive equilibrium charge
$\bar{Q}_{\text{c,}0}$ already present on the EM capacitor of capacitance $C_{\text{c}}(x=\bar{x})$. In this case the additional charge will introduce a force on the mechanical oscillator which pushes it towards a larger capacitance (so that $x=\bar{x}+\delta x$) in order to reduce the charging energy. As a consequence, the voltage fluctuation induced on the capacitor $\delta V(x)$ will be smaller than anticipated from the naive expectation  $\delta V(\bar{x})=\delta Q/C_{\text{c}}(\bar{x})$. Instead of modeling this as a capacitance which depends on $\delta x$ we instead introduce a fixed capacitance $\bar{C}_{\text{c}}\equiv C_{\text{c}}(\bar{x})$ and model the reduced voltage fluctuations as being due to a  part of the charge $-\delta Q_{\text{m}}$ not sitting on
the capacitor but instead being diverted to an equivalent \emph{mechanical} circuit branch
\emph{in parallel} to the coupling capacitor as shown in Fig.~\ref{fig:Mech-equiv}b.
Seen from the outside, the voltage fluctuation $\delta V$ on the capacitor will be exactly the same if we chose a suitable $\delta Q_{\mathrm{m}}$, and hence the two systems are equivalent.
Since the charge diverted to the parallel mechanical arm represents the mechanical motion, $\delta Q_{\rm m}\propto \delta x$, we expect it to obey similar equations of motion as the  viscously damped harmonic
oscillator~(\ref{eq:Mech_EOM}). Such an oscillator is mathematically equivalent to a serial RLC circuit and we therefore expect the mechanical arm
to be simply a serial RLC circuit,
\begin{equation}
Z_{\text{m}}(\omega)=-i\omega L_{\text{m}}+R_{\text{m}}+\frac{1}{-i\omega C_{\text{m}}},\label{eq:Z-m_EM-only}
\end{equation}
where $L_{\text{m}}$, $R_{\text{m}}$, and $C_{\text{m}}$ are mechanical equivalent
circuit parameters (see Fig.~\ref{fig:Mech-equiv}b). Below we confirm this ansatz for the mechanical impedance $Z_{\text{m}}(\omega)$ and derive explicit expressions for the individual components in terms of the  known physical parameters.

\subsection{Dynamical variables of the equivalent circuit\label{sub:Dyn-vars-equiv-circ}}

To derive the equivalent mechanical circuit, we consider the linearized Hamiltonian describing the mechanical system and the coupling capacitor,
\begin{equation}
H_{\text{C}} + H_{\text{m},0} \approx \frac{\delta Q^{2}}{2\bar{C}_{\text{c}}}+\frac{1}{2}m\omega_{\text{m},Q}^{2}\delta x^{2} + \frac{p^{2}}{2m}+G\delta Q\delta x,\label{eq:H-m,0_main-text}
\end{equation}
from Eqs.~(\ref{eq:H-m0},\ref{eq:H-charging}).
Here we have defined a modified mechanical frequency $\omega_{\text{m},Q}$ including the second order derivative in Eq.~(\ref{eq:H-charging}),
\begin{equation}
\omega_{\text{m,}Q}^{2}=\omega_{\text{m},0}^{2}+\frac{\bar{C}_{\text{c}}G^{2}}{m}-\frac{\langle\bar{Q}_{\text{c}}^{2}(t)\rangle}{2m\bar{C}_{\text{c}}^{2}}\left.\frac{d^{2}C_{\text{c}}}{dx^{2}}\right|_{x=\bar{x}}. \;\text{[DC bias]}\label{eq:omega-m,Q_phys}
\end{equation}
 The precise interpretation of this frequency will be discussed below. Note that we have introduced the time average $\langle \cdot \rangle$ of the square of the bias charge for consistency with the AC-biased case discussed below. For the DC-biased case this is simply given by $\langle\bar{Q}_{\text{c}}^{2}(t)\rangle=\bar{Q}_{\text{c},0}^2$.

The aim is now to see whether the system can be mapped to the equivalent circuit shown in Fig.~\ref{fig:Mech-equiv}b.
The Hamiltonian corresponding to this circuit is (ignoring the mechanical resistance $R_{\text{m}}$ at this stage)
\begin{eqnarray}
H' & = & \frac{(\delta Q+\delta Q_{\text{m}})^{2}}{2\bar{C}_{\text{c}}}+\frac{\delta Q_{\text{m}}^{2}}{2C_{\text{m}}}+\frac{\phi_{\text{m}}^{2}}{2L_{\text{m}}}\nonumber \\
 & = & \frac{\delta Q^{2}}{2\bar{C}_{\text{c}}}+\frac{1}{2}\left(\frac{1}{\bar{C}_{\text{c}}}+\frac{1}{C_{\text{m}}}\right)\delta Q_{\text{m}}^{2}+\frac{\phi_{\text{m}}^{2}}{2L_{\text{m}}}+\frac{\delta Q\delta Q_{\text{m}}}{\bar{C}_{\text{c}}},\label{eq:H-EM-equiv}
\end{eqnarray}
where $\delta Q_{\text{m}}$ and $\phi_{\text{m}}$ are charge and magnetic
flux variables of the virtual mechanical branch. The first two terms
of Eq.~(\ref{eq:H-EM-equiv}) are the charging energies
of the two capacitors, and the third term is the virtual magnetic
field energy of the equivalent mechanical inductor. The  fourth term, which is essential for the transducer, is a bilinear coupling 
between the mechanical oscillation and the charge on the coupling capacitor.

We can check the
ansatz~(\ref{eq:Z-m_EM-only}) and find expressions for the equivalent
circuit parameters in terms of EM quantities by comparing Eqs.~(\ref{eq:H-m,0_main-text}) and (\ref{eq:H-EM-equiv}) term by term (the terms in the two expressions are ordered in the same way). We first compare the last (coupling) terms. These become identical if we make  the identification 
\begin{equation}
\delta Q_{\text{m}}=\bar{C}_{\text{c}}G\delta x,\label{eq:Q-m_propto_deltaX}
\end{equation}
that is, the charge variable of the virtual mechanical arm is proportional
to the mechanical displacement.
Given the above correspondence, we expect a similar relationship among
the canonical conjugates, $\phi_{\text{m}}\propto p$. Taking the time derivative we indeed find
\begin{equation}
p=m\delta\dot{x}=\frac{m}{\bar{C}_{\text{c}}G}\delta\dot{Q}_{\text{m}}=\frac{m}{\bar{C}_{\text{c}}G}I_{\text{m}}=\frac{m}{\bar{C}_{\text{c}}GL_{\text{m}}}\phi_{\text{m}},\label{eq:p_propto_phi-m}
\end{equation}
using Eq.~(\ref{eq:Q-m_propto_deltaX}),  $\phi_{\text{m}}=L_{\text{m}}\delta\dot{Q}_{\text{m}}$, 
and $I_{\text{m}}\equiv\delta\dot{Q}_{\text{m}}$. Equating the inductive
energy term of Eq.~(\ref{eq:H-EM-equiv}) with the kinetic of Eq.~(\ref{eq:H-m,0_main-text}) and substituting using Eq.~(\ref{eq:p_propto_phi-m}),
we find an expression for $L_{\text{m}}$, 
\begin{equation}
\frac{p^{2}}{2m}=\frac{\phi_{\text{m}}^{2}}{2L_{\text{m}}}\Leftrightarrow L_{\text{m}}=\frac{m}{\bar{C}_{\text{c}}^{2}G^{2}}.\label{eq:L-m-DC_main-text}
\end{equation}
Substituting $L_{\text{m}}$ into Eq.~(\ref{eq:p_propto_phi-m})
 we then find 
\begin{equation}
\phi_{\text{m}}=\frac{1}{\bar{C}_{\text{c}}G}p.\label{eq:phi-m_propto_p}
\end{equation}
Taken together, Eqs.~(\ref{eq:Q-m_propto_deltaX},\ref{eq:phi-m_propto_p})
show that the dynamical variables of the equivalent circuit $\{\delta Q_{\text{m}},\phi_{\text{m}}\}$
are related to the original coordinates $\{\delta x,p\}$ by a simple
canonical scaling transformation.

Finally, we determine the equivalent mechanical resistance $R_{\text{m}}$,
which is most easily done by comparing equations of motion (where
damping can be incorporated straightforwardly). Equating the viscous
dissipation rate in Eq.~(\ref{eq:Mech_EOM}) with
$\dot{\phi}_{\text{m}}=-(R_{\text{m}}/L_{\text{m}})\phi_{\text{m}}+\ldots$
we get 
\begin{equation}
R_{\text{m}}=\gamma_{\text{m,0}}L_{\text{m}}=\frac{m\gamma_{\text{m,0}}}{\bar{C}_{\text{c}}^{2}G^{2}},\label{eq:R-m-DC_main-text}
\end{equation}
using the expression in Eq.~(\ref{eq:L-m-DC_main-text}) for $L_{\text{m}}$. Similarly, by comparing Eq.~(\ref{eq:Mech_EOM}) with $\dot{\phi}_{\text{m}}=2V_{\text{m}}+\ldots$, we find that the mechanical force $F$ maps to a voltage
\begin{equation}
2V_{\text{m}} \equiv \frac{F}{G\bar{C}_{c}},\label{eq:V-m_def}
\end{equation}
where the factor of two has been included to conform with the electrical input-output formalism to be presented in Section~\ref{sec:Electrical-IO}, cf.~Eq.~(\ref{eq:circuit_fluct-dissip-thm}). By considering the spectrum of the noise~(\ref{eq:mech-fluc-spectrum},\ref{eq:mech-fluc-spectrum-Q}) it is found that $V_{\text{m}}$, as given by Eq.~(\ref{eq:V-m_def}), is exactly the Johnson noise of a resistor with resistance $R_{\rm m}$.


\subsection{Effective mechanical resonance frequencies\label{sub:Eff-mech-res-freqs}}

There is a subtlety related to the effective resonance frequency of
the mechanical mode, which depends on how the timescale of the mechanical
mode $\delta Q_{\text{m}}$ compares to that of the electrical mode
$\delta Q$. Two different limits can be understood from Fig.~\ref{fig:Mech-equiv}b
and Eq.~(\ref{eq:H-EM-equiv}), namely fixed voltage versus fixed charge
dynamics. Fixed voltage across the terminals in Fig.~\ref{fig:Mech-equiv}b
corresponds to the situation where the voltage bias in the circuit
acts much faster than the mechanical modulation, i.e., supplying and
absorbing charge instantaneously so as to maintain a fixed voltage.
This will for instance be the case if the capacitor is connected to an ideal voltage source through a small resistance such that the corresponding $RC$ time is much smaller than the mechanical oscillation period.
The voltage across the mechanical arm will in this case be independent
of the capacitor arm; therefore we may read off the \emph{fixed
voltage} mechanical resonance frequency as the resonance frequency
of the mechanical branch of Fig.~\ref{fig:Mech-equiv}b,
\begin{equation}
\omega_{\text{m,}V}^{2}=\frac{1}{L_{\text{m}}C_{\text{m}}},\label{eq:omega-m,V_main-text}
\end{equation}
sometimes referred to as the mechanical series resonance \cite{OConnell2014}.
For a given applied voltage across
the coupling capacitor, the maximal mechanical response occurs at $\omega_{\text{m,}V}$.
If, on the other hand, the timescale of mechanical modulation is
much faster than that of $\delta Q$ (thus preventing the voltage
bias from reacting), we may effectively set $\delta Q\rightarrow0$. The \emph{fixed charge} mechanical resonance frequency will then be
the resonance frequency of the entire loop in Fig.~\ref{fig:Mech-equiv}b
in which case the capacitances $\bar{C}_{\text{c}},C_{\text{m}}$
are added in series, 
\begin{equation}
\omega_{\text{m,}Q}^{2}=\frac{1}{L_{\text{m}}}\left(\frac{1}{\bar{C}_{\text{c}}}+\frac{1}{C_{\text{m}}}\right).\;\text{[DC bias]}\label{eq:omega-m,Q_main-text}
\end{equation}
This is sometimes referred to as the mechanical parallel resonance \cite{OConnell2014}.
This is also the relation obtained by equating the second terms in Eqs.~(\ref{eq:H-m,0_main-text},\ref{eq:H-EM-equiv}).
For a given current running into the (physical) coupling capacitor
$C_{\text{c}}(x)$, the maximal mechanical response occurs at
$\omega_{\text{m,}Q}$. By comparing Eqs.~(\ref{eq:omega-m,V_main-text})
and (\ref{eq:omega-m,Q_main-text}) we see that the two limiting mechanical
frequencies are related by 
\begin{equation}
\omega_{\text{m,}Q}^{2}-\omega_{\text{m,}V}^{2}=\frac{1}{L_{\text{m}}\bar{C}_{\text{c}}}=\frac{\bar{C}_{\text{c}}G^{2}}{m},\;\text{[DC bias]}\label{eq:omega-m,Q-V_relation}
\end{equation}
from which we conclude that $\omega_{\text{m,}Q}\geq\omega_{\text{m,}V}$.
The mechanical oscillator thus has a different resonance frequency
depending on the circuit to which it is coupled (i.e., including circuit elements not shown in Fig.~\ref{fig:Mech-equiv}b). For instance, if the
bias voltage is applied via a low-pass filter with cut-off frequency
below the mechanical frequency (as in Ref.~\cite{Bagci2014}) this
entails fixed charge conditions.
From the expression for $\omega_{\text{m,}Q}^{2}$~(\ref{eq:omega-m,Q_phys}) 
and the relation in Eq.~(\ref{eq:omega-m,Q-V_relation}), we can find an expression for $\omega_{\text{m,}V}$ in terms of the known physical quantities,
\begin{equation}
\omega_{\text{m,}V}^{2}=\omega_{\text{m},0}^{2}-\frac{\langle\bar{Q}_{\text{c}}^{2}(t)\rangle}{2m\bar{C}_{\text{c}}^{2}}\left.\frac{d^{2}C_{\text{c}}}{dx^{2}}\right|_{x=\bar{x}}.\label{eq:omega-m,V_phys}
\end{equation}
Using Eqs.~(\ref{eq:L-m-DC_main-text},\ref{eq:omega-m,V_main-text}) we can then express the mechanical capacitance through quantities which can be calculated from first principles,
\begin{equation}
C_{\text{m}}=\frac{\bar{C}_{\text{c}}^{2}G^{2}}{\omega_{\text{m,}V}^{2}m}.\label{eq:C-m-DC_main-text}
\end{equation}

\section{Electromechanical equivalent circuit for AC bias\label{sec:EM-equiv-circ-AC}}

Above we have given an intuitive derivation of the equivalent circuit
in the case of a DC-biased capacitor. This allows us to describe how
electrical signals are converted into mechanical motion at the same
frequency. The typical purpose of a transducer is, however, to convert
signals from one frequency $\Omega$ to another $\omega_{\text{d}}\pm\Omega$
by harmonically driving the system with a frequency $\omega_{\text{d}}$.
In the following we shall develop an equivalent circuit formalism
to describe this situation in the regime where $\omega_{\rm d}$ is much larger than the mechanical resonance frequency.
This can, e.g., correspond to a mechanical oscillator in the MHz regime biased by an AC voltage in the GHz range. 

\begin{figure}[htbp]
\centering
\includegraphics[width=0.82\columnwidth]{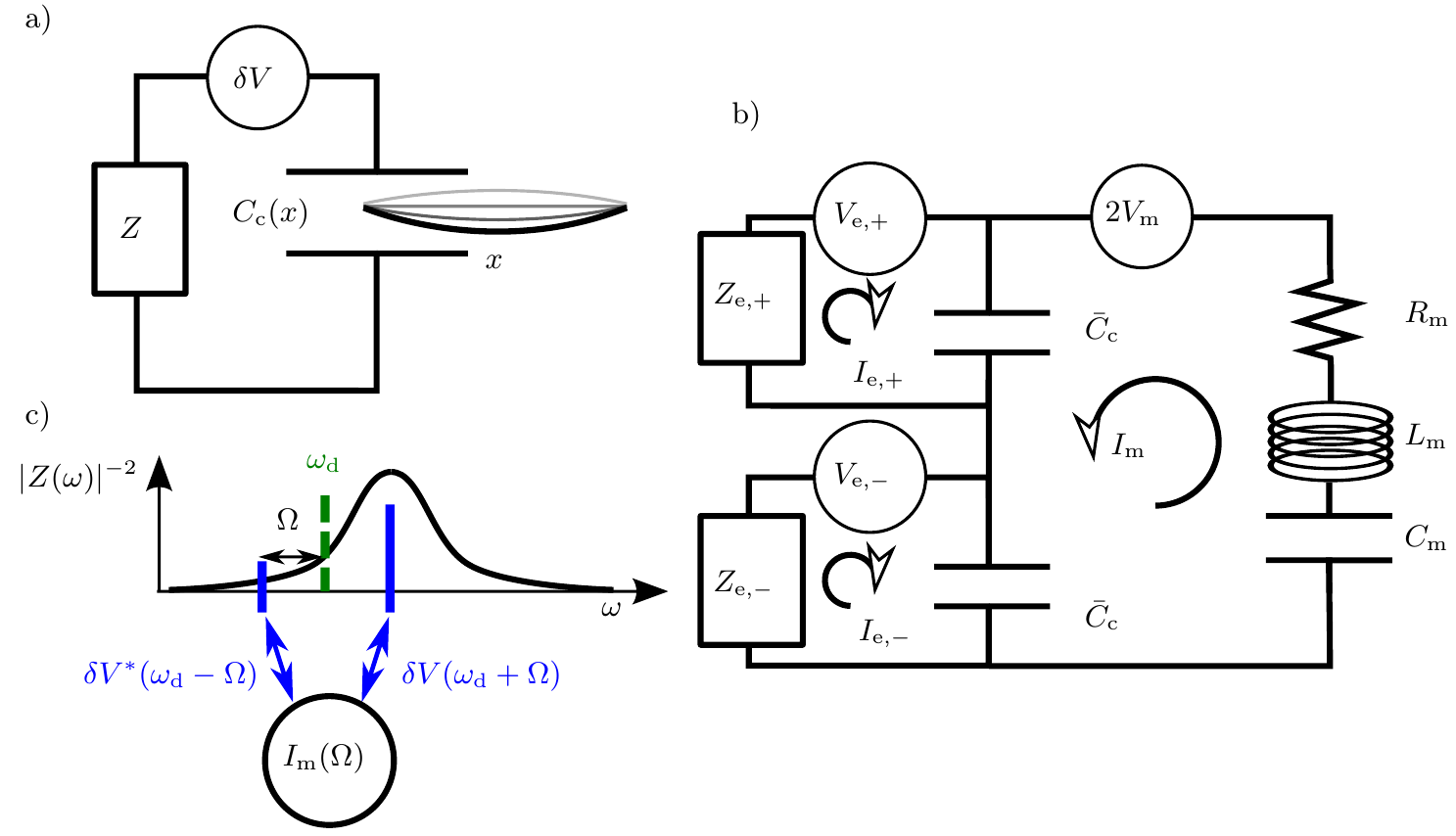}
\caption{AC-biased electromechanical interface. a) Th{é}venin equivalent circuit considered for the AC-biased system.
An ideal voltage source $\delta V(\omega)$ in series with an impedance
$Z(\omega)$ drives the coupled membrane capacitor system. b) EM equivalent
circuit. The mechanical motion is replaced by the central loop current
$I_{\text{m}}(\Omega)$. Through capacitors of capacitance $\bar{C}_{c}$,
the mechanical current is connected to loop currents $I_{\text{e,}+}$ and
$I_{\text{e,}-}$ representing the upper and lower sidebands of the electrical
current, which are driven by voltage sources $V_{\text{e,}+}(\Omega)=\delta V(\omega_{\text{d}}+\Omega)$
and $V_{\text{e,}-}(\Omega)=\delta V^{*}(\omega_{\text{d}}-\Omega)$. c) Sketch of the spectral mapping between the electrical and mechanical frequency scales. The electrical spectrum is folded around the carrier frequency $\omega_{\mathrm{d}}$ (with a gain profile determined by $Z(\omega)$) such that the upper and lower sidebands at $\omega_{\text{d}}\pm\Omega$ are mapped into a single mechanical frequency $\Omega$.
\label{Fig:AC-equivalent}}
\end{figure}

When the capacitor is biased by an alternating voltage the charge on the capacitor will take the form 
\begin{equation}
Q(t)=\bar{Q}_{\text{c}}(t)+\delta Q(t)=\bar{Q}_{\text{c},0}\left({\rm e}^{-i\omega_{\text{d}}t}+{\rm e}^{i\omega_{\text{d}}t}\right)+\delta Q(t),\label{eq:ACcharge}
\end{equation}
c.f.~Eq.~(\ref{eq:Q-c-bar_def}), where we for simplicity take the amplitude $\bar{Q}_{\text{c},0}$ to be real. Similar
to above, this amplitude should be found by self-consistently solving
for the equilibrium configuration of the electrical and mechanical
system, and $\delta Q(t)$ then represents the fluctuations around
this value.
It will be convenient to work in the Fourier domain following the standard linear response approach.
Exploiting that all charges, currents, and voltages are real valued, we introduce the
Fourier transform as an integral over positive frequencies so that,
e.g., the voltage fluctuations are denoted by 
\begin{equation}
\delta V(t)=\int_{0}^{\infty}\frac{d\omega}{\sqrt{2\pi}}\left[V(\omega){\rm e}^{-i\omega t}+V^{*}(\omega){\rm e}^{i\omega t}\right],
\label{eq:fourier_V}
\end{equation}
with similar expressions for the charge $\delta Q$ and current $\delta I$
fluctuations as well as the position $\delta x$ and momentum $p$
fluctuations. To proceed, it is convenient not to deal with the
specifics of the rest of the circuit and we therefore replace it with
its Th{é}venin equivalent diagram~\cite{dorf2010} as shown in Fig.~\ref{Fig:AC-equivalent}a, i.e., it is represented by an ideal voltage
source $\delta V$ and the input impedance $Z$.
If we now consider the contribution to Kirchhoff's voltage law coming from
the coupled EM system we have 
\begin{eqnarray}
\delta V(t) & = & ...+\frac{Q(t)}{C_{\text{c}}(x)}-\frac{\bar{Q}_{\text{c},0}\left({\rm e}^{-i\omega_{\text{d}}t}+{\rm e}^{i\omega_{\text{d}}t}\right)}{\bar{C}_{\text{c}}}\\
 & \approx & ...+\frac{\delta Q(t)}{\bar{C}_{\text{c}}}+G\delta x(t)\left({\rm e}^{-i\omega_{\text{d}}t}+{\rm e}^{i\omega_{\text{d}}t}\right),\label{eq:Vt}
\end{eqnarray}
where the ellipsis represents terms arising from the rest of the circuit.
In the last line we have expanded to lowest order in the fluctuations
and introduced the coupling constant $G$ as defined in Eq.~(\ref{eq:G_def}).

We will now assume that the mechanical component is the slowest frequency
scale in the problem so that we can neglect the mechanical response,
 $\delta x(\omega)\approx0$, at high frequencies $\omega>\omega_{\text{d}}$. 
This amounts to the assumption that the mechanical resonance frequency is small
compared to $\omega_{\text{\text{d}}}$.
With this assumption, the mechanical frequency component $\delta x(\Omega)\propto I_{\rm m}(\Omega)$ will couple to two frequency components of the electrical circuit located at the upper and lower sidebands of the drive, $\delta V(\omega_{\text{d}}+\Omega)$ and $\delta V^{*}(\omega_{\text{d}}-\Omega)$, as illustrated in Fig.~\ref{Fig:AC-equivalent}c.
We thus arrive at 
\begin{eqnarray}
\delta V(\omega_{\text{d}}+\Omega) & = & Z(\omega_{\text{d}}+\Omega)\delta I(\omega_{\text{d}}+\Omega)+\frac{\delta Q(\omega_{\text{d}}+\Omega)}{\bar{C}_{\text{c}}}-G \delta x(\Omega)\label{eq:KirchVtemp}\\
\delta V^{*}(\omega_{\text{d}}-\Omega) & = & Z^{*}(\omega_{\text{d}}-\Omega)\delta I^{*}(\omega_{\text{d}}-\Omega)+\frac{\delta Q^{*}(\omega_{\text{d}}-\Omega)}{\bar{C}_{\text{c}}}-G \delta x(\Omega).\label{eq:KirchVstartemp}
\end{eqnarray}
Here $\Omega>0$ denotes the small Fourier frequency of the excursion of the mechanical oscillator, which the electrical circuits sees as a frequency relative to the AC bias frequency $\omega_{\text d}$. Eqs.~(\ref{eq:KirchVtemp},\ref{eq:KirchVstartemp}) determine the circuit response above and below this center frequency, respectively, at the 'lab frame' frequency $\omega=\omega_{{\text d}} \pm \Omega$.
From this expression the principle of the transducer is apparent:
the oscillating bias field connects different frequency components of the
mechanical and electrical circuit. Eqs.~(\ref{eq:KirchVtemp},\ref{eq:KirchVstartemp})
also show that electrical frequency components are mapped into the
mechanical mode in a non-invertible, 2-to-1 manner. This characteristic
is the key to establishing a relatively simple equivalent circuit
for the AC case below. As we will see, \emph{we may consider
the upper and lower electrical sidebands, (\ref{eq:KirchVtemp}) and
(\ref{eq:KirchVstartemp}), to be distinct degrees of freedom coupling
to the mechanical mode.}

We now turn to the mechanical oscillator. By combining the interaction Hamiltonian~(\ref{eq:H_int-lin_EM}) with Eq.~(\ref{eq:Mech_EOM}) we may derive the equation of motion for the mechanical momentum
\begin{equation}
\dot{p}=-m\omega_{{\rm m},V}^{2}\delta x-\gamma_{{\rm m},0}p+F(t)-G{\left({\rm e}^{-i\omega_{\text{d}}t}+{\rm e}^{i\omega_{\text{d}}t}\right)}\delta Q-2G^{2}\bar{C}_{\text{c}}\delta x.\label{eq:p-dot_mech}
\end{equation}
In the above expression we have averaged
over oscillations occurring with a frequency $2\omega_{\text{d}}$ in accordance
with the assumption that the mechanical response only happens on a
slower timescale. Compared to the DC-biased case, where $\omega_{\text{m,}Q}$ and $\omega_{\text{m,}V}$ are related by Eq.~(\ref{eq:omega-m,Q-V_relation}), this averaging in the AC case gives a factor of two appearing on the right-hand side of Eq.~(\ref{eq:omega-m,Q-V_relation}) and hence in the last term of Eq.~(\ref{eq:p-dot_mech}).
For later convenience we have again introduced $\omega_{\text{m,}V}$ from
Eq.~(\ref{eq:omega-m,V_phys}) which we will again identify as the
resonance frequency at fixed voltage.

To find the equivalent circuit representation, we convert Eq.~(\ref{eq:p-dot_mech}) to Fourier
space and introduce the electrical analogs of the mechanical parameters
as defined in~\cref{eq:Q-m_propto_deltaX,eq:p_propto_phi-m,eq:L-m-DC_main-text,eq:R-m-DC_main-text,eq:V-m_def,eq:C-m-DC_main-text}. We thereby find
\begin{equation}
2V_{\text{m}}(\Omega)=-i\Omega L_{\text{m}}I_{\text{m}}(\Omega)+R_{\text{m}}I_{\text{m}}(\Omega)+\frac{\delta Q_{\text{m}}(\Omega)}{C_{\text{m}}}+\frac{2\delta Q_{\text{m}}(\Omega)+\delta Q(\omega_{\text{d}}+\Omega)+\delta Q^{*}(\omega_{\text{d}}-\Omega)}{\bar{C}_{\text{c}}}.\label{eq:MechKirch}
\end{equation}
This expression for the momentum (represented by $I_{\text{m}}$) resembles
Kirchhoff's voltage law, except for the mixing of different frequency components
and the appearance of the complex conjugate. To remove these differences,
we define new voltages and
charges for the upper and lower sidebands of the drive 
\begin{eqnarray}
V_{\text{e,}+}(\Omega)=\delta V(\omega_{\text{d}}+\Omega) & \qquad & V_{\text{e,}-}(\Omega)=\delta V^{*}(\omega_{\text{d}}-\Omega)\label{eq:V_U-L_defs}\\
Q_{\text{e,}+}(\Omega)=\delta Q(\omega_{\text{d}}+\Omega) & \qquad & Q_{\text{e,}-}(\Omega)=\delta Q^{*}(\omega_{\text{d}}-\Omega).\label{eq:Q_U-L_defs}
\end{eqnarray}
For the current we wish to retain the standard relation $I_{\text{e,}\pm}(\Omega)=-i\Omega Q_{\text{e,}\pm}(\Omega)$. We achieve this with the choice 
\begin{equation}
I_{\text{e,}+}(\Omega)=\frac{\Omega}{\omega_{\text{d}}+\Omega}I(\omega_{\text{d}}+\Omega)\qquad I_{\text{e,}-}(\Omega)=-\frac{\Omega}{\omega_{\text{d}}-\Omega}I^{*}(\omega_{\text{d}}-\Omega),\label{eq:I_U-L_defs}
\end{equation}
where the minus sign in the last expression is a consequence of the
complex conjugation. Correspondingly we have the upper and lower sideband impedances 
\begin{equation}
Z_{\text{e,}+}(\Omega)=\frac{\omega_{\text{d}}+\Omega}{\Omega}Z(\omega_{\text{d}}+\Omega)\qquad Z_{\text{e,}-}(\Omega)=-\frac{\omega_{\text{d}}-\Omega}{\Omega}Z^{*}(\omega_{\text{d}}-\Omega).\label{eq:Z_U-L_defs}
\end{equation}
Here the combination of the negative sign and the complex conjugation
in $Z_{\text{e,}-}$ means that reactances retain their sign whereas resistances
have their sign flipped. This reflects the instability associated with the coupling to
the lower sideband, which yields parametric amplification, manifesting the active character of the circuit (the bias field being the source of energy). Furthermore,
the prefactor in Eq.~(\ref{eq:Z_U-L_defs}) means that capacitors keep their usual expression
for the impedance, $1/(-i\Omega C)$, whereas inductances and resistances
are scaled up to reflect that it is harder to induce a given charge amplitude
at a higher frequency. Combining these definitions with the equations
of motion in Eqs.~(\ref{eq:KirchVtemp}), (\ref{eq:KirchVstartemp}),
and (\ref{eq:MechKirch}), we finally achieve 
\begin{eqnarray}
2V_{\text{m}}(\Omega) & = & \left[-i\Omega L_{\text{m}}+R_{\text{m}}+\frac{1}{-i\Omega C_{\text{m}}}\right]I_{\text{m}}(\Omega)+\frac{2I_{\text{m}}(\Omega)+I_{\text{e,}+}(\Omega)+I_{\text{e,}-}(\Omega)}{-i\Omega\bar{C}_{c}}\label{eq:KirchMech}\\
V_{\text{e,}\pm}(\Omega) & = & Z_{\text{e,}\pm}(\Omega)I_{\text{e,}\pm}(\Omega)+\frac{I_{\text{e,}\pm}(\Omega)+I_{\text{m}}(\Omega)}{-i\Omega\bar{C}_{\text{c}}}\label{eq:KirchV-U}
\end{eqnarray}
These equations of motion have a straightforward interpretation in
terms of the equivalent circuit diagram in Fig.~\ref{Fig:AC-equivalent}b. Here the mechanical system is represented by the loop current
$I_{\text{m}}$ in the central loop, whereas the outer loops represent the
upper and lower sidebands of the electrical system.
We remark that the divergent behavior of $L_{\text{m}}$, $R_{\text{m}}$, $V_{\text{m}}$, and $1/C_{\text{m}}$~(\ref{eq:L-m-DC_main-text},\ref{eq:R-m-DC_main-text},\ref{eq:V-m_def},\ref{eq:C-m-DC_main-text}) for vanishing EM coupling, $G\rightarrow 0$, serves to decouple the mechanical loop from the rest of the circuit as expected.

From the circuit, it is immediately apparent that $\omega_{\text{m},V}=1/\sqrt{L_{\text{m}}C_{\text{m}}}$
is the mechanical resonance frequency in the limit where the capacitor
is connected to an ideal voltage source $Z_{\text{e,}+}=Z_{\text{e,}-}=0$, so that
the outer arms are replaced by short circuits bypassing the capacitors of capacitance $\bar{C}_{\text{c}}$. On the other hand, for fixed charge $Z_{\text{e,}+},Z_{\text{e,}-}\rightarrow\infty$, the
mechanical resonance frequency $\omega_{\text{m},Q}$ is
shifted from $\omega_{\text{m},V}$ by twice the amount given in Eq.~(\ref{eq:omega-m,Q-V_relation})
since $\bar{C}_{\text{c}}$ appears for both sidebands.
This completes the derivation of the equivalent circuit. With the
results developed here the analysis of the coupled EM system can now be
reduced to finding voltages and current of linear circuits. This gives
a direct description of how voltage fluctuations are transduced
to the mechanical system and vice versa.

As an example, we consider a serial RLC with resonance frequency $\bar{\omega}_{\rm LC}=1/\sqrt{L \bar{C}_{\rm c}}\sim 1 \rm{GHz}$ capacitively coupled to the position of a mechanical element resonating in the regime $\omega_{\text{m,}Q}\sim 1 \rm{MHz}$ by means of a bias tone of frequency $\omega_{\rm d} \sim \bar{\omega}_{\rm LC} \gg \omega_{\text{m,}Q}$. In this case the Th\'{e}venin impedance of the electrical resonance $Z$ appearing in Fig.~\ref{Fig:AC-equivalent}a is $Z(\omega) = -i\omega L + R$, so that the equivalent impedances of the upper and lower electrical sidebands are, from Eqs.~(\ref{eq:Z_U-L_defs}),
\begin{gather}
Z_{{\rm e},+}(\Omega)=-i\frac{(\omega_{\text{d}}+\Omega)^{2}}{\Omega}L+\frac{\omega_{\text{d}}+\Omega}{\Omega}R\label{eq:Z-e-p_example}\\
Z_{{\rm e},-}(\Omega)=-i\frac{(\omega_{\text{d}}-\Omega)^{2}}{\Omega}L-\frac{\omega_{\text{d}}-\Omega}{\Omega}R.\label{eq:Z-e-m_example}
\end{gather}
As mentioned above, the negative resistance of the lower sideband gives rise to amplification effects reflecting the instability of the lower sideband.

\section{Electrical input-output formalism\label{sec:Electrical-IO}}

\begin{figure}
\centering
\includegraphics[width=0.5\textwidth]{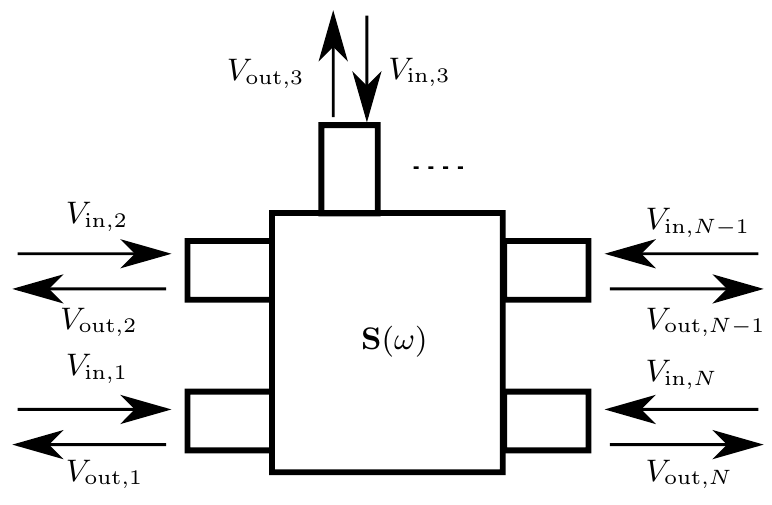}
\caption{N-port network. Each port is associated with an incoming and an outgoing traveling field, $V_{\text{in},i}$ and $V_{\text{out},i}$. The outgoing fields are linked to the incoming fields by the scattering matrix $\mathbf{S}(\omega)$.\label{fig:S-matrix}}
\end{figure}

In the preceding sections we have derived an equivalent impedance
description of EM systems. We will now extend this to describe how 
signals and noise enter and exit the system
via its various ports, which is essential to the analysis
of transducers. Our analysis generalizes the
well-established
network analysis employed in the characterization of passive, purely electrical radio-frequency and microwave circuits~\cite{Pozar} to accommodate the frequency conversion inherent to transduction.
Such electrical input-output formalism amounts to supplementing Kirchhoff's circuit laws with equations relating the input and output signals of (virtual) transmission lines to the currents in the circuit.
Solving these equations in the context of a linear N-port
circuit, see Fig.~\ref{fig:S-matrix}, the outgoing signals
can be related to the incoming ones by the classical scattering matrix
\begin{equation}
\vec{V}_{\text{out}}(\Omega)=\mathbf{S}(\Omega)\vec{V}_{\text{in}}(\Omega),\label{eq:circuit_scat-matrix}
\end{equation}
where 
$\vec{V}_{\text{in/out}}(\Omega)$ is a vector containing  the complex amplitudes of the incoming and outgoing traveling waves.

To derive a scattering relation of the form of Eq.~(\ref{eq:circuit_scat-matrix}) for our system, we shall use the equivalent circuit description derived above.
Physically, the input and output ports of the equivalent circuit are transmission lines, resistive elements, and mechanical dissipation. The mechanical force was converted into its Johnson voltage equivalent in Eq.~(\ref{eq:V-m_def}) and this suffices for DC-biased transducers. However, for the AC-biased EM circuit, frequency interconversion and spectral folding~[Fig.~\ref{Fig:AC-equivalent}c] must be accounted for. To accommodate this in the electrical input-output formalism, the vectors of inputs and outputs $\vec{V}_{\text{in/out}}(\Omega)$ are generalized to ($p\in\{\rm{in,out}\}$)
\begin{equation}
\vec{V}_{p}(\Omega) = (V_{p,\rm{m}}(\Omega),V_{p,1}(\omega_{\rm{d,}1}+\Omega),V^*_{p,1}(\omega_{\rm{d,}1}-\Omega),V_{p,2}(\omega_{\rm{d,}2}+\Omega),V^*_{p,2}(\omega_{\rm{d,}2}-\Omega),\ldots)^{T},\label{eq:V-inout_rot-frame}
\end{equation}
where $V_{\text{in,m}} \equiv V_{\text{m}}$ and we allow for the possibility of multiple drive fields~$\omega_{\rm{d,}i}$ (as will be relevant when adding, e.g., an optical subsystem). Hence, Eq.~(\ref{eq:circuit_scat-matrix}) together with Eq.~(\ref{eq:V-inout_rot-frame}) captures both the frequency interconversion between subsystems as well as the mixing of sideband signals.

\begin{figure}
\centering
\includegraphics[width=0.45\textwidth]{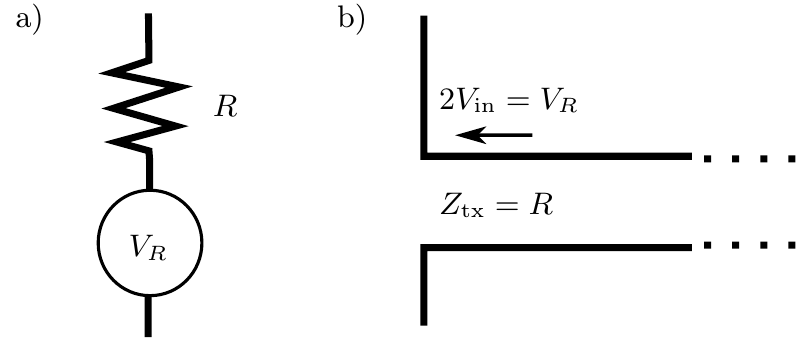}
\caption{Mapping between a) a resistor $R$ with Johnson
voltage noise $V_{R}$ and b) a semi-infinite lossless transmission
line with characteristic impedance $Z_{\text{tx}}=R$ and incoming
signal $2V_{\text{in}}=V_{R}$.\label{fig:Nyquist-trick}}
\end{figure}

To link the external input and output fields in $\vec{V}_{\text{in/out}}(\Omega)$ 
to the internal currents and voltages in
the impedance formalism, we recall
how the presence of a port in the circuit modifies Kirchhoff's equations.
To this end, we observe that the voltage $V_{i}$ across and the net (physical)
current amplitude $I_{i}$ into the $i$'th terminal can be expressed
in terms of the travelling wave amplitudes at the terminal as~\cite{Pozar}
\begin{eqnarray}
V_{i} & = & V_{\text{in},i}+V_{\text{out},i}\nonumber \\
I_{i} & = & \frac{1}{Z_{\text{tx},i}}\left[V_{\text{in},i}-V_{\text{out},i}\right].\label{eq:port-V-I-Vpm}
\end{eqnarray}
From Eqs.~(\ref{eq:port-V-I-Vpm}) we can derive the equivalent of
a fluctuation-dissipation relation for each port~\cite{Yurke1984},
\begin{equation}
V_{i}=-Z_{\text{tx,}i}I_{i}+2V_{\text{in},i}.\label{eq:circuit_fluct-dissip-thm}
\end{equation}
Eq.~(\ref{eq:circuit_fluct-dissip-thm}) is the key to extending Kirchhoff's
laws to an open system setting. Specifically, for Kirchhoff's voltage
law for any loop including one of the ports, it tells us that the
port introduces dissipation corresponding to the real-valued resistance
$R=Z_{\text{tx},i}$, as well as a source term $2V_{\text{in},i}$. Conversely,
as pointed out by Nyquist~\cite{Nyquist1928}, this implies that any resistive element
$R$ in the circuit can be mapped to an equivalent semi-infinite transmission
line of characteristic impedance $Z_{\text{tx},j}=R$, while the Johnson noise of the resistor is included by specifying a thermal mixed state for the corresponding source term
$V_{\text{in},j}$; this mapping is illustrated in Fig.~\ref{fig:Nyquist-trick}.
In this way, the resulting open circuit formalism puts noise
and signal inputs on equal footing.
From Eq.~(\ref{eq:port-V-I-Vpm}) we can also derive the input-output
relations for the ports,
\begin{equation}
V_{\text{out},i}=-Z_{\text{tx,}i}I_{i}+V_{\text{in},i},\label{eq:circuit_IO-rel}
\end{equation}
that allow us to determine the itinerant outgoing voltages.


With the above in place, we may, for an arbitrary N-port passive linear
circuit, use Kirchhoff's circuit laws supplemented with Eqs.~(\ref{eq:circuit_fluct-dissip-thm})
and~(\ref{eq:circuit_IO-rel}) to derive the scattering matrix $\mathbf{S}(\omega)$. 
In practice this can, e.g., be
done by applying standard impedance rules to the equivalent
circuit diagram under consideration as will be demonstrated below.
Once the scattering matrix $\mathbf{S}(\Omega)$ (\ref{eq:circuit_scat-matrix}) has
been obtained, we have a full characterization of the dynamics of the transducer. When the initial state of all the involved reservoirs is specified, the scattering matrix can therefore be used to evaluate the performance of the transducer for whichever  application one is interested in \cite{alpha}.

\begin{figure}
\centering
\includegraphics[width=0.8\textwidth]{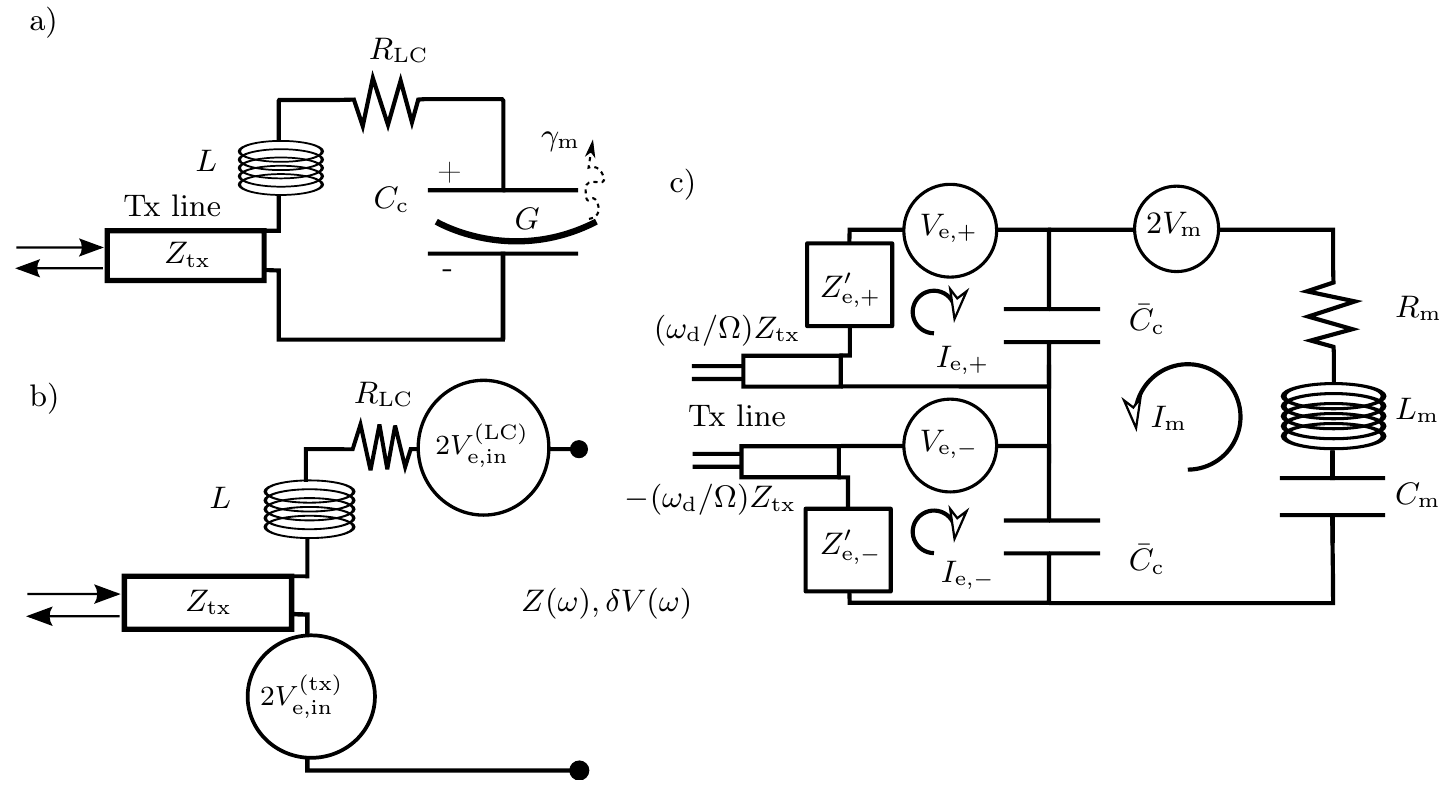}
\caption{AC-biased electromechanical example circuit. a) RLC circuit (ohmic resistance $R_{\text{LC}}$) coupled to a single transmission line of characteristic impedance $Z_{\text{tx}}$ and, at the same time, capacitively coupled to a mechanical mode (intrinsic damping rate $\gamma_{\text{m}}$) by means of an AC bias of frequency $\omega_{\text{d}}$ (not shown) resulting in an EM coupling strength $G$.
b) The electrical part of the example system in a) that determines the Th\'{e}venin equivalent quantities $Z(\omega)$ and $\delta V(\omega)$ (cf.~Fig.~\ref{Fig:AC-equivalent}a). Note that this subcircuit does not include the coupling capacitor. The voltage sources $V_{\text{e,in}}^{\text{(tx)}}$ and $V_{\text{e,in}}^{\text{(LC)}}$ represent the transmission line input and the Johnson noise of the resistor with resistance $R_{\text{LC}}$, respectively.
c) EM equivalent circuit for the system depicted in a). The transmission line impedance $Z_{\text{tx}}$ loads the upper/lower sideband loop by $\sim\pm (\omega_{\text{d}}/\Omega) Z_{\text{tx}}$ [Eqs.~(\ref{eq:Z-e-p_example},\ref{eq:Z-e-m_example}), for $\Omega \ll \omega_{\text{d}}$] (the inductor and the ohmic loss resistor are represented by $Z_{\text{e,}\pm}'$). In the quantum limit, the interaction of the electrical sideband fields $V_{\text{e,}\pm}$ with the mechanical system can lead to squeezing of the outgoing transmission line signal.
\label{fig:LC-overcoupled}}
\end{figure}

Before quantizing the theory and generalizing the formalism to accommodate optomechanical interfaces, we illustrate the usefulness of the equivalent circuit for determining the scattering matrix for an AC-biased EM interface. We treat the example of a serial RLC circuit with ohmic resistance $R_{\mathrm{LC}}$ coupled to a transmission line (of characteristic impedance $Z_{\text{tx}}$) and capacitively coupled to a mechanical mode (see Fig.~\ref{fig:LC-overcoupled}a).
Hence, the electrical subsystem is characterized by the Th\'{e}venin impedance~[Fig.~\ref{fig:LC-overcoupled}b] 
\begin{equation}
Z(\omega)=-i\omega L+R_{\text{LC}}+Z_{\text{tx}},\label{eq:Z-e_Example}
\end{equation}
where $L$ is the inductance in the electrical circuit and, for later
convenience, we define $\bar{\omega}_{\text{LC}}\equiv(L\bar{C}_{\text{c}})^{-1/2}$.
Since the current is the same through the circuit elements $R_{\text{LC}}$ and $Z_{\text{tx}}$, the fraction of the  power dissipated in the circuit that goes into the transmission line is
\begin{equation}
\eta_{\text{el}}\equiv Z_{\text{tx}}/(R_{\text{LC}}+Z_{\text{tx}}),\label{eq:eta-el_def}
\end{equation}
sometimes referred to as the electrical coupling efficiency.
Each of the decay channels induce a (Johnson) voltage source contributing
to the Th\'{e}venin voltage (see Fig.~\ref{fig:LC-overcoupled}b)
\begin{equation}
\delta V(\omega)=2V_{\text{e,}\text{in}}^{(\text{LC})}(\omega)+2V_{\text{e,}\text{in}}^{(\text{tx})}(\omega),\label{eq:Thevenin_Example}
\end{equation}
as follows from Eq.~(\ref{eq:circuit_fluct-dissip-thm}).
The electrical output fields are given by the input-output relations~(\ref{eq:circuit_IO-rel}).
For the serial RLC considered here~(\ref{eq:Z-e_Example}),
whose conventional circuit diagram only has a single loop current, the electrical transmission line output is given by
\begin{eqnarray}
V_{\text{e},\text{out}}^{\text{(tx)}}(\omega_{\text{d}}+\Omega) & = & -\frac{\omega_{\text{d}}+\Omega}{\Omega}Z_{\text{tx}}I_{\text{e},+}(\Omega)+V_{\text{e},\text{in}}^{\text{(tx)}}(\omega_{\text{d}}+\Omega)\label{eq:tx-line-U_IO-rel_Example}\\
V_{\text{e},\text{out}}^{\text{(tx)}*}(\omega_{\text{d}}-\Omega) & = & \frac{\omega_{\text{d}}-\Omega}{\Omega}Z_{\text{tx}}I_{\text{e},-}(\Omega)+V_{\text{e},\text{in}}^{\text{(tx)}*}(\omega_{\text{d}}-\Omega),\label{eq:tx-line-L_IO-rel_Example}
\end{eqnarray}
expressed in terms of the upper and lower sideband electrical loop currents
$I_{\text{e},\pm}(\Omega)$ by means of Eqs.~(\ref{eq:I_U-L_defs}).

To use Eqs.~(\ref{eq:tx-line-U_IO-rel_Example},\ref{eq:tx-line-L_IO-rel_Example}) to establish the scattering matrix \eqref{eq:circuit_scat-matrix}, we need to determine the electrical loop currents $I_{\text{e},\pm}$. Rather than solving Kirchhoff's equations (\ref{eq:KirchMech},\ref{eq:KirchV-U}) algebraically, the transfer coefficients can be obtained by applying simple impedance rules to the corresponding equivalent circuit, Fig.~\ref{fig:LC-overcoupled}b.
To this end, it is convenient to introduce the effective impedance of the central (mechanical) loop,
\begin{equation}
Z_{\text{m,eff}}(\Omega)=Z_{\text{m}}(\Omega)+\Delta Z(\Omega),\label{eq:Z-m-eff_def-EM}
\end{equation}
where the bare mechanical impedance $Z_{\text{m}}$ \eqref{eq:Z-m_EM-only} is loaded by the electrical sideband loops,
\begin{equation}
\Delta Z(\Omega)=\sum_{s=\pm}\left[-i\Omega \bar{C}_{\text{c}}+1/Z_{\text{e},s}(\Omega)\right]^{-1}=\sum_{s=\pm} \frac{1+\mathcal{Q}_{\text{e,}s}(\Omega)}{-i \Omega \bar{C}_{\text{c}}},
\label{eq:DeltaZ-EM_def}
\end{equation}
where in the final expression we have expressed the load from sideband $s$ as a sum of the impedance of its coupling capacitor and a contribution proportional to its susceptibility function (in the rotating frame of the AC bias),
\begin{equation}
\mathcal{Q}_{\text{e,}\pm}(\Omega) \equiv - \frac{1/(-i \Omega \bar{C}_{\text{c}})}{Z_{\text{e,}\pm}(\Omega)+1/(-i \Omega \bar{C}_{\text{c}})}.\label{eq:Q-e_def}
\end{equation}
The contributions $s=\pm$ to Eq.~\eqref{eq:DeltaZ-EM_def} are simply the Th\'{e}venin impedances of the respective sideband loops \emph{including} the coupling capacitor.
For the present example of an AC-biased serial RLC circuit, the electrical susceptibilities for the two sidebands have the standard form for the response of a resonant circuit,
\begin{equation}
\mathcal{Q}_{\text{e,}\pm}(\Omega)=-\frac{\bar{\omega}_{\text{LC}}^{2}}{\bar{\omega}_{\text{LC}}^{2}-(\omega_{{\rm d}}\pm\Omega)^{2}\mp i(\omega_{{\rm d}}\pm\Omega)R_{\text{tot}}/L},
\end{equation}
where $R_{\text{tot}} \equiv R_{\text{LC}} + Z_{\text{tx}}$.

Using the rule of voltage division across serially connected elements, the electrical voltage sources representing the incoming transmission line signal $V_{\text{e,}\pm}$ are seen to induce a voltage $-V_{\text{e,}\pm}(\Omega)(-i\Omega \bar{C}_{\text{c}})^{-1}/[Z_{\text{e,}\pm}(\Omega)+(-i\Omega \bar{C}_{\text{c}})^{-1}]$
across their respective (virtual) coupling capacitors (of capacitance $\bar{C}_{\text{c}}$), adding to the voltage $2V_{\text{m}}(\Omega)$ in the mechanical loop. The mechanical loop current response $I_{\text{m}}$ to this net voltage follows by multiplication by the loaded mechanical admittance $1/Z_{\text{m,eff}}$ \eqref{eq:Z-m-eff_def-EM}. Having determined $I_{\text{m}}$, it can be considered a current source from the point of view of the electrical sideband loops, and in terms of this the electrical sideband currents are simply
$I_{\text{e,}\pm}(\Omega)=[V_{\text{e,}\pm}(\Omega)-I_{\text{m}}(-i\Omega \bar{C}_{\text{c}})^{-1}]/[Z_{\text{e,}\pm}(\Omega)+(-i\Omega \bar{C}_{\text{c}})^{-1}]$.
Expressing these observations in terms of the functions $\mathcal{Q}_{\text{e,}\pm}(\Omega)$ \eqref{eq:Q-e_def} and putting them together,
the sideband loop currents are found to be,
\begin{equation}
I_{\text{e,}\pm}(\Omega)=\mathcal{Q}_{\text{e,}\pm}(\Omega) \left[\frac{\mathcal{Q}_{\text{e,}\pm}(\Omega)}{Z_{\text{m,eff}}(\Omega)}+i\Omega \bar{C}_{\text{c}}\right]V_{\text{e,}\pm}(\Omega)+\frac{\mathcal{Q}_{\text{e,}\pm}(\Omega)\mathcal{Q}_{\text{e,}\mp}(\Omega)}{Z_{\text{m,eff}}(\Omega)}V_{\text{e,}\mp}(\Omega)+\frac{\mathcal{Q}_{\text{e,}\pm}(\Omega)}{Z_{\text{m,eff}}(\Omega)}2V_{\text{m}}(\Omega).\label{eq:Scat_current_eo2}
\end{equation}
Using Eqs.~(\ref{eq:V_U-L_defs},\ref{eq:Thevenin_Example}-\ref{eq:tx-line-L_IO-rel_Example}) we arrive at the scattering relation in terms of the itinerant fields (considering the overcoupled limit, $R_{\text{LC}} \rightarrow 0$, for simplicity),
\begin{equation}
\left(\begin{array}{c}
V_{\text{e,out}}^{\text{(tx)}}(\omega_{\text{d}}+\Omega)\\
V_{\text{e,out}}^{\text{(tx)}*}(\omega_{\text{d}}-\Omega)
\end{array}\right)=\left(\begin{array}{cc}
D(\Omega) & O(\Omega)\\
O^{*}(-\Omega) & D^{*}(-\Omega)
\end{array}\right)\left(\begin{array}{c}
V_{\text{e,in}}^{\text{(tx)}}(\omega_{\text{d}}+\Omega)\\
V_{\text{e,in}}^{\text{(tx)}*}(\omega_{\text{d}}-\Omega)
\end{array}\right)+\left(\begin{array}{c}
M(\Omega)\\
M^{*}(-\Omega)
\end{array}\right)V_{\text{m}}(\Omega),\label{eq:scatrel-squeezing-classical}
\end{equation}
where the coefficients are given by
\begin{gather}
D(\Omega)\equiv 1-2i(\omega_{\text{d}}+\Omega) \bar{C}_{\text{c}}Z_{\text{tx}}\mathcal{Q}_{\text{e,}+}(\Omega)-2\frac{\omega_{\text{d}}+\Omega}{\Omega}Z_{\text{tx}}\frac{\mathcal{Q}_{\text{e,}+}^2(\Omega)}{Z_{\text{m,eff}}(\Omega)},\\
O(\Omega) \equiv -2\frac{\omega_{\text{d}}+\Omega}{\Omega}Z_{\text{tx}}\frac{\mathcal{Q}_{\text{e,}+}(\Omega)\mathcal{Q}_{\text{e,}-}(\Omega)}{Z_{\text{m,eff}}(\Omega)},\label{eq:transfer-O}\\
M(\Omega) \equiv -2\frac{\omega_{\text{d}}+\Omega}{\Omega}Z_{\text{tx}}\frac{\mathcal{Q}_{\text{e,}+}(\Omega)}{Z_{\text{m,eff}}(\Omega)},\label{eq:transfer-M}
\end{gather}
having made use of the properties $\mathcal{Q}^{*}_{\text{e,}-}(\Omega)=\mathcal{Q}_{\text{e,}+}(-\Omega)$ [Eqs.~(\ref{eq:Z_U-L_defs},\ref{eq:Q-e_def})] and $Z^{*}_{\text{m,eff}}(\Omega)=Z_{\text{m,eff}}(-\Omega)$.
The scattering relation~\eqref{eq:scatrel-squeezing-classical} fully characterizes the linearized interaction of the system and thus contains, e.g., all of the physical effects familiar from the analogous setup in linearized optomechanics: e.g., dynamical back-action on the mechanical mode (``optical spring effect'') \cite{Aspelmeyer2013}, Optomechanically Induced Transparency \cite{Weis2010}, and classical noise squashing \cite{Safavi-Naeini2013b}.

\section{Quantization of the equivalent circuit\label{sec:Quantization}}

We now turn to the quantization of our circuit theory. Because our system is described by a bilinear Hamiltonian, the Heisenberg-Langevin equations are algebraically equivalent to their classical counterpart. Hence, the scattering matrix is identical in the quantum and classical cases. We therefore do not need to consider the internal degrees of freedom of the circuit (i.e., its quasi-localized normal modes) and it suffices to expand the itinerant voltage amplitudes~(\ref{eq:V-inout_rot-frame}), including noise sources [with the representation in Fig.~\ref{fig:Nyquist-trick}], using quantized ingoing and outgoing fields following the standard procedure~\cite{QNreview}. Writing this in the frequency domain, we have
\begin{equation}
\hat{V}_{p,i}(t)=\int_{0}^{\infty}\frac{d\omega}{\sqrt{2\pi}}\sqrt{\frac{\hbar\omega Z_{\text{tx,}i}}{2}}\left[\hat{b}_{p,i}(\omega)e^{-i\omega t}+\text{H.c.}\right],\label{eq:V-pm-t-hat}
\end{equation}
where the annihilation operators $\hat{b}_{p,i}(\omega)$, $p\in\{\text{in},\text{out}\}$
obey the commutation relations
\begin{equation}
[\hat{b}_{p,i}(\omega),\hat{b}_{p,j}^{\dagger}(\omega')]=\delta(\omega-\omega')\delta_{i,j},\label{eq:bosonic-fields_commutator}
\end{equation}
with all other commutators being zero. Eqs.~(\ref{eq:V-pm-t-hat},\ref{eq:bosonic-fields_commutator})
specify the correct ohmic noise operator that enters the quantum version
of Eq.~(\ref{eq:circuit_fluct-dissip-thm}), i.e., within the Markov approximation of a memoryless reservoir~\cite{Gardiner}. This expansion has the same form as the Fourier transform introduced in Eq.~(\ref{eq:fourier_V}), and hence we can immediately identify 
the corresponding voltage operators which replace their classical counterparts,
\begin{align}
V_{p,i}(\omega)&\rightarrow\hat{V}_{p,i}(\omega)=\sqrt{\frac{\hbar \omega Z_{\text{tx,}i}}{2}}\hat{b}_{p,i}(\omega)
\nonumber\\
V_{p,i}^*(\omega)&\rightarrow\hat{V}_{p,i}^\dagger(\omega)=\sqrt{\frac{\hbar\omega Z_{\text{tx,}i}}{2}}
\hat{b}_{p,i}^{\dagger}(\omega),
\label{eq:V-pm-omega-hat}
\end{align}
where again $p \in \{\text{in,out}\}$. From these expressions  we can then find the corresponding upper and lower sideband operators entering the equivalent circuit using Eq.~(\ref{eq:V_U-L_defs}).


To characterize the noise we will assume that all reservoirs are in their thermal state as specified by the expectation values ($\omega,\omega'>0$)
\begin{eqnarray}
\langle\hat{V}_{\text{in},i}^{\dagger}(\omega)\hat{V}_{\text{in},j}(\omega')\rangle & = & \frac{\hbar\omega Z_{\text{tx,}i}}{2}n_{i}(\omega)\delta(\omega-\omega')\delta_{i,j}\nonumber \\
\langle\hat{V}_{\text{in},i}(\omega)\hat{V}_{\text{in},j}^{\dagger}(\omega')\rangle & = & \frac{\hbar\omega Z_{\text{tx,}i}}{2}[n_{i}(\omega)+1]\delta(\omega-\omega')\delta_{i,j}\label{eq:VV-thermal_exp-vals}
\end{eqnarray}
and $\langle\hat{V}_{\text{in},i}(\omega)\hat{V}_{\text{in},j}(\omega')\rangle=0=\langle\hat{V}_{\text{in},i}^{\dagger}(\omega)\hat{V}_{\text{in},j}^{\dagger}(\omega')\rangle$, where the thermal flux per unit bandwidth is given by the Bose-Einstein distribution \eqref{eq:Bose-Einstein}. Note that Eqs.~\eqref{eq:VV-thermal_exp-vals} are completely equivalent to those pertaining to quantum Brownian motion \eqref{eq:mech-fluc-spectrum-Q} considered previously.
Due to the linearly rising term $\propto \omega$ in the second line of Eqs.~(\ref{eq:VV-thermal_exp-vals}), a cutoff is required to avoid divergences when integrating over all spectral components~\cite{Gardiner}; in this article, however, we shall only consider finite frequency ranges.
The thermal expectation values of the mechanical Johnson voltage $V_{\text{m}}(\Omega)$~(\ref{eq:V-m_def}) take the same form as Eqs.~\eqref{eq:VV-thermal_exp-vals}
with the replacements $Z_{\text{tx,}i}\rightarrow R_{\text{m}},T_{i}\rightarrow T_{\text{m}}$~\cite{Giovannetti2001}. With these replacements the scattering matrix derived from the equivalent circuit yields a complete characterization of the performance of the transducer also in the quantum regime.

We demonstrate the quantization procedure by continuing the above example of an AC-biased EM interface [Fig.~\ref{fig:LC-overcoupled}]. The scattering matrix~\eqref{eq:scatrel-squeezing-classical} for the voltage amplitudes is unchanged under quantization. However, in the quantum domain it is more customary to work with the bosonic operators~\eqref{eq:bosonic-fields_commutator}, whose scattering matrix differs from that of the voltage amplitudes due to the different zero-point amplitudes~\eqref{eq:V-pm-omega-hat} of the various itinerant fields. Hence, from Eqs.~(\ref{eq:scatrel-squeezing-classical},\ref{eq:V-pm-omega-hat}) we have
\begin{equation}
\left(\begin{array}{c}
\hat{b}_{\text{out}}^{\text{(tx)}}(\omega_{\text{d}}+\Omega)\\
\hat{b}_{\text{out}}^{\text{(tx)}\dagger}(\omega_{\text{d}}-\Omega)
\end{array}\right)=\left(\begin{array}{cc}
D(\Omega) & O'(\Omega)\\
O'^{*}(-\Omega) & D^{*}(-\Omega)
\end{array}\right)\left(\begin{array}{c}
\hat{b}_{\text{in}}^{\text{(tx)}}(\omega_{\text{d}}+\Omega)\\
\hat{b}_{\text{in}}^{\text{(tx)}\dagger}(\omega_{\text{d}}-\Omega)
\end{array}\right)+\left(\begin{array}{c}
M'(\Omega)\\
M'^{*}(-\Omega)
\end{array}\right)\hat{c}_{\text{in}}(\Omega),\label{eq:scatrel-squeezing}
\end{equation}
where $\hat{c}_{\text{in}}(\Omega)$ is the mechanical noise input field operator analogous to $\hat{b}_{\text{in}}^{\text{(tx)}}$ [using the representation of Fig.~\ref{fig:Nyquist-trick}], and we have defined the modified transfer functions [cf.~Eqs.~(\ref{eq:transfer-O},\ref{eq:transfer-M})]
\begin{gather}
O'(\Omega) \equiv \sqrt{\frac{\omega_{\text{d}}-\Omega}{\omega_{\text{d}}+\Omega}} O(\Omega),\\
M'(\Omega) \equiv \sqrt{\frac{|\Omega|}{\omega_{\text{d}}+\Omega}\frac{R_{\text{m}}}{Z_{\text{tx}}}} M(\Omega).
\end{gather}

\begin{figure}
\begin{centering}
\includegraphics[width=0.6\columnwidth]{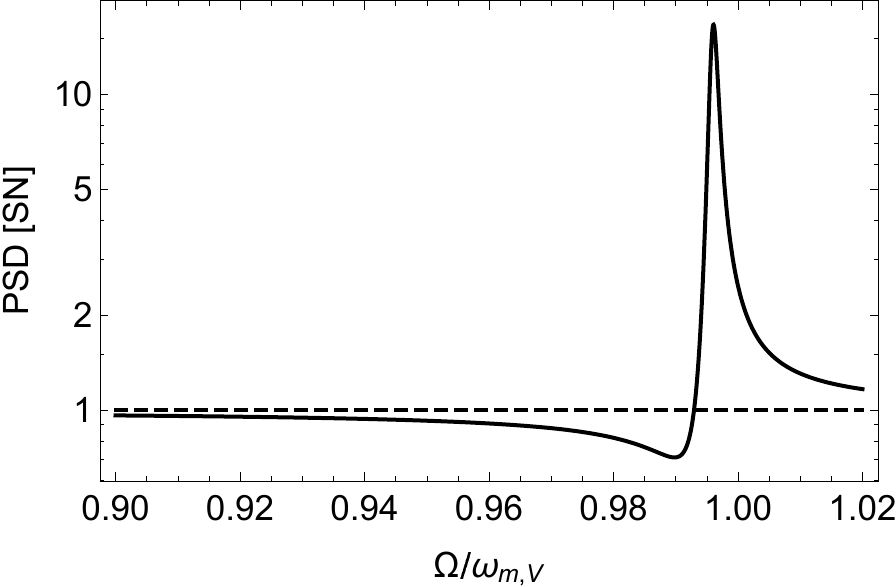}
\end{centering}
\caption{Squeezing of a traveling electro-magnetic field produced by a cold microwave LC resonator ($n_{\text{tx}}\approx 0$, $\bar{\omega}_{\text{LC}}=2\pi \times 1 \text{GHz}$) coupled to a mechanical mode ($\omega_{\text{m,}V}=2\pi \times 1 \text{MHz}$) at an ambient temperature $T_\text{m}=4 \mathrm{K}$. Plotted is the homodyne spectrum $S(\Omega)$ of $\hat{X}_{\theta}(\Omega)$ for a local oscillator phase $\theta\approx 0.18 \pi$, which maximizes the squeezing obtainable for a single spectral component (at $\Omega \approx 0.99 \omega_{\text{m,}V}$ for these parameters). The shot noise limit at 1 (in the normalization used here) is indicated by the dashed line. Other system parameters used are $Z_{\text{tx}}=50 \mathrm{\Omega},Q_{\text{LC}}=\bar{\omega}_{\text{LC}}L/Z_{\text{tx}}=100,\omega_{\text{d}}=\bar{\omega}_{\text{LC}}-2\pi\times 5 \text{MHz},\gamma_{\text{m}}=2\pi\times 0.1 \text{Hz},G/\sqrt{m}=3.2\times 10^{11}\text{V}/(
\text{m}\sqrt{
\text{kg}})$.}\label{fig:EM-squeezing}
\end{figure}

As an example, the quantized theory allows us to discuss the ponderomotive squeezing of the outgoing electromagnetic field~\cite{Safavi-Naeini2013,Purdy2013,Nielsen2017} that can result for our example circuit {[}Fig.~\ref{fig:LC-overcoupled}{]}. This can be understood as two-mode squeezing~\cite{Lvovsky2014} of the upper and lower sidebands. Hence, using Eq.~\eqref{eq:scatrel-squeezing} along with Eqs.~(\ref{eq:Bose-Einstein},\ref{eq:VV-thermal_exp-vals}), we evaluate the spectrum of the  quadrature~\cite{Mancini1994,Fabre1994},
\begin{equation}
\hat{X}_{\theta}(\Omega) \equiv [e^{-i \theta}\hat{b}_{\text{out}}^{\text{(tx)}}(\omega_{\text{d}}+\Omega)+e^{i \theta}\hat{b}_{\text{out}}^{\text{(tx)}\dagger}(\omega_{\text{d}}-\Omega)]/2+\text{H.c.},
\end{equation}
corresponding to the measured observable of a homodyne measurement of the field $\hat{b}_{\text{out}}^{\text{(tx)}}(\omega)$ using a local oscillator with frequency $\omega_{\text{d}}$ and phase $\theta$. The power spectral density $S(\Omega)$ for the zero-mean field $\hat{X}_{\theta}(\Omega)$ is
\begin{equation}
S(\Omega)\delta (\Omega - \Omega') \equiv \langle \hat{X}_{\theta}(\Omega) \hat{X}_{\theta}(\Omega') \rangle,\label{eq:S-def}
\end{equation}
and can be directly determined from the results above.
As shown in Fig.~\ref{fig:EM-squeezing}, the resulting outgoing field exhibits squeezing, $S<1$, for suitable parameters.

\section{Optical impedance and full electro-optomechanical equivalent circuit\label{sec:Optical-impedance_full-equiv}}

In Section~\ref{sec:EM-equiv-circ-AC}, we derived the equivalent circuit
for an AC-biased EM interface involving an arbitrary
linear electrical circuit. We now extend the theory to accommodate optomechanical coupling.
We consider a single optical cavity mode whose frequency $\omega_{\text{cav}}(\hat{x})$
is modulated by the same mechanical position $\hat{x}$ entering the
EM coupling. To this end, we exploit that the OM interface is largely equivalent to the AC-biased EM interface. Following the standard procedure \cite{Aspelmeyer2013},
we take as our starting point the quantum Hamiltonians for the mechanical and optical modes; the former is given by the quantum analog of Eq.~(\ref{eq:H-m0}) whereas the optical single-mode Hamiltonian is
\begin{equation}
H_{\text{opt}}=\hbar\omega_{\text{cav}}(\hat{x})\hat{a}^{\dagger}\hat{a}.\label{eq:H-opt}
\end{equation}
Applying a coherent optical laser drive of frequency $\omega_{\text{l}}$
to the optical cavity (represented by a Hamiltonian $H_{\text{l}}$),
we expand the total Hamiltonian of the optomechanical system $H=H_{\text{opt}}+H_{\text{m,}0}+H_{\text{l}}+\ldots$
around the ensuing steady-state configuration $(\bar{x},\alpha)$ (where the ellipsis represents the EM coupling and terms responsible for coupling to the environment
of the hybrid system). The linearized dynamics of the displaced variables,
$\hat{x}=\bar{x}+\delta\hat{x}$ and $\hat{a}=\alpha+\delta\hat{a}$,
is then described by the Hamiltonian 
\begin{gather}
H_{\text{OM}}=\hbar\bar{\omega}_{\text{cav}}\delta\hat{a}^{\dagger}\delta\hat{a}+\left[\frac{\delta\hat{p}^{2}}{2m}+\frac{1}{2}m\omega_{\text{m}}^{2}\delta\hat{x}^{2}\right]+H_{\text{OM,int}}\label{eq:H-OM-lin}\\
H_{\text{OM,int}}\equiv \sqrt{\hbar} G_{\text{OM}}\delta\hat{x}(e^{i\omega_{\text{l}}t}e^{-i\theta}\delta\hat{a}+e^{-i\omega_{\text{l}}t}e^{i\theta}\delta\hat{a}^{\dagger})/\sqrt{2},\label{eq:H-OM-RWA}
\end{gather}
where the first and second terms of Eq.~(\ref{eq:H-OM-lin}) are the ``free-evolution''
Hamiltonians of the displaced optical and mechanical modes, whereas
the third term, $H_{\text{OM,int}}$, is the drive-enhanced linear
coupling between them. In the above equations, we have introduced the steady-state cavity resonance $\bar{\omega}_{\text{cav}}$, the optically shifted mechanical frequency $\omega_{\text{m}}$, the optomechanical coupling strength $G_{\text{OM}}$ (units of
$\sqrt{\text{energy}\times\text{frequency}}/\text{length}$), and the phase of the intracavity drive field $\theta$,
\begin{equation}
\omega_{\text{m}}^{2}\equiv\omega_{\text{m},0}^{2}+\frac{\hbar|\alpha|^{2}}{m}\left.\frac{d^{2}\omega_{\text{cav}}}{dx^{2}}\right|_{x=\bar{x}},\;G_{\text{OM}}\equiv\sqrt{2\hbar}\left.\frac{d\omega_{\text{cav}}}{dx}\right|_{x=\bar{x}}|\alpha|,\;\theta\equiv\text{Arg}[\alpha].\label{eq:optical-shift_G-OM}
\end{equation}
The coupling strength $G_{\text{OM}}$ can be related to the more familiar coupling rate $g_{\text{OM}}$ that appears in the conventional form of the coupling Hamiltonian~(\ref{eq:H-OM-RWA}),
\begin{equation}
H_{\text{OM,int}}=\hbar g_{\text{OM}} (\delta\hat{c}+\delta\hat{c}^{\dagger})(e^{i\omega_{\text{l}}t}e^{-i\theta}\delta\hat{a}+e^{-i\omega_{\text{l}}t}e^{i\theta}\delta\hat{a}^{\dagger}),\label{eq:H-OM-RWA2}
\end{equation}
where $\delta \hat{c}$ is the mechanical bosonic annihilation operator, $[\delta \hat{c},\delta \hat{c}^{\dagger}]=1$. Using $\delta \hat{x}=x_{\text{ZPF}}(\delta\hat{c}+\delta\hat{c}^{\dagger})$ and $x_{\text{ZPF}}\equiv \sqrt{\hbar/2 m \omega_{\text{m}}}$ and comparing Eqs.~(\ref{eq:H-OM-RWA}) and (\ref{eq:H-OM-RWA2}), the two optomechanical coupling parameters are seen to be related by
\begin{equation}
g_{\text{OM}}=G_{\text{OM}}/\sqrt{4 m \omega_{\text{m}}}.
\label{eq:g-G-OM}
\end{equation}

The typical approach from here is to consider the linearized optomechanical interaction, \eqref{eq:H-OM-RWA} or \eqref{eq:H-OM-RWA2}, in a rotating frame with respect to $\hat{H}_{0}=\hbar\omega_{\text{l}}\delta \hat{a}^{\dagger}\delta \hat{a}$, thereby removing its explicit time-dependence. Here, however, we take a different approach
in order to achieve equations of motion
equivalent to those governing the EM coupling considered
in Section~\ref{sec:EM-equiv-circ-AC}. To this end, we note that within the rotating
wave approximation (RWA), we may approximate the OM interaction Hamiltonian~(\ref{eq:H-OM-RWA}) by an expression equivalent in form to that of the EM interaction~\eqref{eq:H_int-lin_EM},
\begin{equation}
H_{\text{OM,int}}\approx \sqrt{\hbar} G_{\text{OM}}\delta\hat{x}(e^{i\omega_{\text{l}}t}+e^{-i\omega_{\text{l}}t})\hat{X},\label{eq:H-OM-int_inverseRWA}
\end{equation}
where we have introduced the dimensionless light quadratures
\begin{equation}
\hat{X}\equiv(e^{-i\theta}\delta\hat{a}+e^{i\theta}\delta\hat{a}^{\dagger})/\sqrt{2},\;\hat{P}\equiv(e^{-i\theta}\delta\hat{a}-e^{i\theta}\delta\hat{a}^{\dagger})/(\sqrt{2}i),\label{eq:X-P-optical_defs}
\end{equation}
obeying $[\hat{X},\hat{P}]=i$. Note that here we have gone in the opposite direction of what is typically done in the RWA, where expressions similar to Eq.~(\ref{eq:H-OM-int_inverseRWA}) are replaced by Eq.~(\ref{eq:H-OM-RWA}). As opposed to electromechanics, where this may not be the case, the RWA 
is typically a very good approximation in optomechanics since the
dynamics on the mechanical timescale $2\pi/\omega_{\text{m}}$ are much
slower than that of the optical drive $2\pi/\omega_{\text{l}}$. As a consequence, derivations of the coupling Hamiltonian~\eqref{eq:H-OM-RWA} typically assume the RWA from the outset. 
Which of the two forms \eqref{eq:H-OM-RWA} or \eqref{eq:H-OM-int_inverseRWA} is the more correct model is thus not clear, 
and there is a priori no reason to prefer one form over the other. By choosing the unconventional form in Eq.~(\ref{eq:H-OM-int_inverseRWA}), however, the Hamiltonian  linearly couples $\delta\hat{x}$
to $\hat{X}$ with a strength $G_{\text{OM}}$ in a manner similar
to the linearized EM interaction Hamiltonian~(\ref{eq:H_int-lin_EM})
considered above. With the form in Eq.~(\ref{eq:H-OM-int_inverseRWA}), we can thus obtain the equivalent circuit in an analogous way.


We now consider how the
optical mode couples to its environment via its loss and drive ports.
This can conveniently be treated using the input-output
formalism from quantum optics~\cite{Collett1985}, which is analogous to the input-output theory for circuits considered in Section~\ref{sec:Electrical-IO}. In quantum optics, this formalism is traditionally only discussed within the RWA, and the microscopic details of the optical bath coupling is in general not known, leaving it an open question how the bath couples to the quadrature
variables $(\hat{X},\hat{P})$~\cite{Gardiner}. Assuming a linear coupling to the bath
modes, the precise microscopic model is, however, unimportant within the RWA.
Thus, in a spirit similar to Eq.~(\ref{eq:H-OM-int_inverseRWA}), this
permits us to assume that the optical bath modes couple to the quadrature
$\hat{X}$, resulting in the usual viscous damping and noise terms
in the equation of motion of the conjugate quadrature $\hat{P}$ (within the Markov approximation)
\begin{gather}
\dot{\hat{X}}=\bar{\omega}_{\text{cav}}\hat{P}\nonumber \\
\dot{\hat{P}}=-\bar{\omega}_{\text{cav}}\hat{X}-\kappa\hat{P}+\sqrt{2\kappa}\hat{P}_{\text{in}}+\ldots,\label{eq:optical-EOM}
\end{gather}
where, as previously, a dot above an operator indicates the time derivative, $\kappa$ is the decay rate of the optical mode, and the operator $\hat{P}_{\text{in}}$ represents the noise and/or signal input leaking into the mode (the ellipsis represents OM coupling terms). The input operator $\hat{P}_{\text{in}}$ and its output counterpart $\hat{P}_{\text{out}}$ can be expanded on a set of itinerant bosonic modes, in analogy to Eq.~(\ref{eq:V-pm-t-hat}), as
\begin{equation}
\hat{P}_{p}(t)=\int_{0}^{\infty}\frac{d\omega}{\sqrt{2\pi}}\sqrt{\frac{\omega}{\bar{\omega}_{\text{cav}}}}\left[\hat{a}_{p}(\omega)e^{-i\omega t}+\hat{a}_{p}^{\dagger}(\omega)e^{i\omega t}\right],\label{eq:P-in-x_def}
\end{equation}
where $p \in \{\text{in,out}\}$ and we have introduced bosonic field
operators obeying $[\hat{a}_{p}(\omega),\hat{a}_{p}^{\dagger}(\omega')]=\delta(\omega-\omega')$
and $[\hat{a}_{p}(\omega),\hat{a}_{p}(\omega')]=0$.
The normalization of Eq.~(\ref{eq:P-in-x_def}) was chosen so as to achieve a simple dimensionless form of Eqs.~(\ref{eq:optical-EOM}) (resulting in the appearance of the cavity resonance frequency $\bar{\omega}_{\text{cav}}$).
If $\hat{P}_{\text{in}}$ is in a thermal state then Eq.~(\ref{eq:P-in-x_def}) represents an ohmic bath with the following expectation value in Fourier space ($\omega,\omega'>0$)
\begin{eqnarray}
\langle\hat{P}_{\text{in}}^{\dagger}(\omega)\hat{P}_{\text{in}}(\omega')\rangle & = & \frac{\omega}{\bar{\omega}_{\text{cav}}}n_{\text{opt}}(\omega)\delta(\omega-\omega')\nonumber\\
\langle\hat{P}_{\text{in}}(\omega)\hat{P}_{\text{in}}^{\dagger}(\omega')\rangle & = & \frac{\omega}{\bar{\omega}_{\text{cav}}}[n_{\text{opt}}(\omega)+1]\delta(\omega-\omega'),\label{eq:PP-exp-val}
\end{eqnarray}
where $n_{\text{opt}}$ is given by Eq.~(\ref{eq:Bose-Einstein}) in terms of the temperature of the optical system $T_{\text{opt}}$. For all practical purposes the magnitude of optical frequencies is such that $|\hbar\omega/k_{\text{B}}T_{\text{opt}}| \gg 1$, which entails that to very good approximation we may take the optical noise to be vacuum, $n_{\text{opt}}(\omega) \approx 0$.
With the conventions implicit in~\cref{eq:X-P-optical_defs,eq:optical-EOM,eq:P-in-x_def},
the optical input-output relation for a single-sided cavity reads (for $\omega>0$),  
\begin{equation}
\hat{a}_{\text{out}}(\omega)=i\sqrt{\frac{\omega}{\bar{\omega}_{\text{cav}}}}\sqrt{\kappa}e^{-i\theta}\hat{a}(\omega)+\hat{a}_{\text{in}}(\omega),\label{eq:Ext_IO-rel-a}
\end{equation}
where the intracavity drive phase $\theta$ was introduced in Eq.~\eqref{eq:optical-shift_G-OM}.

\begin{figure}
\centering
\includegraphics[width=0.6\textwidth]{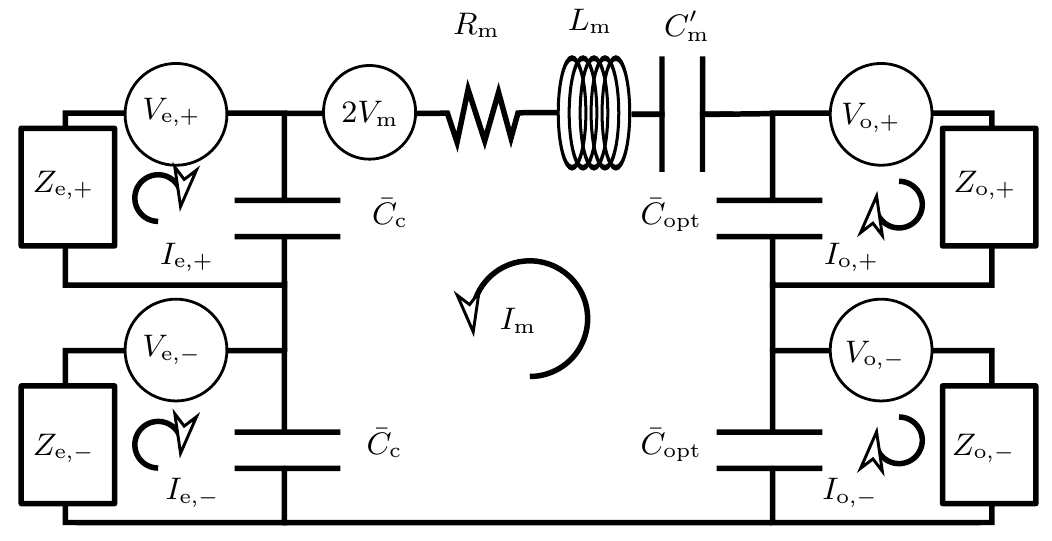}
\caption{Electro-optomechanical equivalent circuit for a mechanical mode acting
as an intermediary between an arbitrary linear electrical circuit
and a single optical mode. Each of the electrical or optical sidebands
are represented by a loop current $I_{\text{e,}\pm}$ or $I_{\text{o},\pm}$
in the diagram and are coupled capacitively to the mechanical loop
current $I_{\text{m}}$ via $\bar{C}_{\text{c}}$ or $\bar{C}_{\text{opt}}$.
The effective voltage sources $V_{\text{e,}\pm},V_{\text{o},\pm},V_{\text{m}}$
represent electrical, optical, and mechanical noise or signal inputs.
Using standard circuit rules to determine the current in an external
loop, we may determine the output at the corresponding sideband.\label{fig:EOM-equiv-circuit}}
\end{figure}

Having achieved OM equations of motion
similar to the EM equations, we may straightforwardly retrace the steps of Section
\ref{sec:EM-equiv-circ-AC} to derive an OM equivalent circuit.
Rather than considering this on its own, we proceed immediately
to the transduction scenario with simultaneous electro- and optomechanical
couplings. In this case the displaced variables $\delta Q$, $\delta x$, and $\delta a$
are defined with respect to the equilibrium configuration of the electro-optomechanical hybrid
system subjected to simultaneous electrical and optical driving (we
again neglect higher harmonics of the system response). Kirchhoff's law for the mechanical loop (\ref{eq:KirchMech}) then 
generalizes to
\begin{equation}
2V_{\text{m}}(\Omega) = \left[-i\Omega L_{\text{m}}+R_{\text{m}}+\frac{1}{-i\Omega C_{\text{m}}'}\right]I_{\text{m}}(\Omega)
+\frac{2I_{\text{m}}(\Omega)+I_{\text{e,}+}(\Omega)+I_{\text{e,}-}(\Omega)}{-i\Omega \bar{C}_{\text{c}}}+\frac{2I_{\text{m}}(\Omega)+I_{\text{o,}+}(\Omega)+I_{\text{o,}-}(\Omega)}{-i\Omega \bar{C}_{\text{opt}}},\label{eq:Mech-loop_EO}
\end{equation}
whereas the effective electrical equations (\ref{eq:KirchV-U})
are unaltered, but supplemented by the optical counterparts
\begin{eqnarray}
V_{\text{o},\pm}(\Omega) & = & Z_{\text{o},\pm}(\Omega)I_{\text{o},\pm}(\Omega)+\frac{I_{\text{o},\pm}(\Omega)+I_{\text{m}}(\Omega)}{-i\Omega \bar{C}_{\text{opt}}}\label{eq:V-o-U_KVL}
\end{eqnarray}
Here we define optical upper/lower sideband quantities $V_{\text{o},\pm}(\Omega),Q_{\text{o},\pm}(\Omega),I_{\text{o},\pm}(\Omega)$
and $Z_{\text{o},\pm}(\Omega)$ analogously to the electrical quantities in
Eqs.~\eqref{eq:V_U-L_defs}--\eqref{eq:Z_U-L_defs}
with the replacements $\omega_{\text{d}}\rightarrow\omega_{\text{l}},Z\rightarrow Z_{\text{opt}},\delta V\rightarrow 2 V_{\text{o,}\text{in}},\delta Q\rightarrow\delta Q_{\text{o}}$,
and according to the definitions (here and henceforth replacing $\hat{X}$ by its classical counterpart $X$ without loss of generality, see discussion in Section~\ref{sec:Quantization})
\begin{equation}
\delta Q_{\text{o}}\equiv\sqrt{\hbar}\bar{\omega}_{\text{cav}}\frac{\bar{C}_{\text{c}}G}{G_{\text{OM}}}X,\;I_{\text{o}}\equiv\delta\dot{Q}_{\text{o}}\label{eq:Q-o_X}
\end{equation}
\begin{equation}
Z_{\text{opt}}(\omega)=-i\omega L_{\text{opt}}+R_{\text{opt}}\label{eq:Z-opt_def}
\end{equation}
\begin{equation}
L_{\text{opt}}\equiv\frac{G_{\text{OM}}^{2}}{\bar{C}_{\text{c}}^{2}G^{2}\bar{\omega}_{\text{cav}}^{3}},\quad \bar{C}_{\text{opt}}\equiv\bar{\omega}_{\text{cav}}\frac{\bar{C}_{\text{c}}^{2}G^{2}}{G_{\text{OM}}^{2}},\quad R_{\text{opt}}\equiv\kappa L_{\text{opt}}\label{eq:R-opt_def}
\end{equation}
and
\begin{equation}
1/C_{\text{m}}'\equiv1/C_{\text{m}}-2/\bar{C}_{\text{opt}}.\label{eq:C-m-prime}
\end{equation}
Furthermore, while the mechanical capacitance $C_{\text{m}}$ is still defined by Eq.~(\ref{eq:C-m-DC_main-text}), the mechanical frequency $\omega_{\text{m,}V}$ entering this definition now
contains static shifts from both the EM and OM interactions, (\ref{eq:omega-m,V_phys}) and (\ref{eq:optical-shift_G-OM}), combining to
\begin{equation}
\omega_{\rm{m,}V}=\omega_{\text{m},0}^{2}-\frac{\langle\bar{Q}_{\text{c}}^{2}(t)\rangle}{2m\bar{C}_{\text{c}}^{2}}\left.\frac{d^{2}C_{\text{c}}}{dx^{2}}\right|_{x=\bar{x}} +\frac{\hbar|\alpha|^{2}}{m}\left.\frac{d^{2}\omega_{\text{cav}}}{dx^{2}}\right|_{x=\bar{x}}.
\end{equation}
Note that the need to define the modified $C_{\text{m}}'$ (\ref{eq:C-m-prime}) appearing in Eq.~(\ref{eq:Mech-loop_EO}), which was not required for the electrical coupling, can be traced to the difference in how we define the coupling constant (whether we take derivatives of the resonance frequency or the capacitance).
Converting Eq.~(\ref{eq:P-in-x_def}) to electrical units we obtain the
equivalent optical input and output voltage fields ($p \in \{\text{in,out}\}$),
\begin{equation}
\hat{V}_{\text{o,}p}(t)\equiv\sqrt{\frac{\hbar\bar{\omega}_{\text{cav}}R_{\text{opt}}}{2}}\hat{P}_{p}(t)=\int_{0}^{\infty}\frac{d\omega}{\sqrt{2\pi}}\sqrt{\frac{\hbar\omega R_{\text{opt}}}{2}}\left[\hat{a}_{p}(\omega)e^{-i\omega t}+\hat{a}_{p}^{\dagger}(\omega)e^{i\omega t}\right]\label{eq:V-o-p_def},
\end{equation}
which is completely analogous to Eq.~(\ref{eq:V-pm-t-hat}).
We likewise convert the optical input-output relation (\ref{eq:Ext_IO-rel-a}) into electrical units using Eqs.~(\ref{eq:X-P-optical_defs},\ref{eq:Q-o_X},\ref{eq:V-o-p_def}),
\begin{equation}
V_{\text{o},\text{out}}(\omega)=-R_{\text{opt}}I_{\text{o}}(\omega)+V_{\text{o},\text{in}}(\omega),\label{eq:Optical_IO-rel_equiv_Example}
\end{equation}
completely analogous to its electrical counterpart (\ref{eq:circuit_IO-rel}) [Eq.~\eqref{eq:Optical_IO-rel_equiv_Example} is stated in terms of classical variables without loss of generality, see Section \ref{sec:Quantization}].
This full set of equations can be represented by the combined electro-optomechanical equivalent circuit diagram shown in Fig.~\ref{fig:EOM-equiv-circuit}, which generalizes Fig.~\ref{Fig:AC-equivalent}.

From the equivalence between the optical mode and the serial RLC, the equations derived for the latter in the last paragraph of Section~\ref{sec:Electrical-IO} carry over straightforwardly to the analogous optical system of a single cavity mode with two decay channels, $\kappa=\kappa_{\rm int}+\kappa_{\text{ext}}$,
where $\kappa_{\rm int}$ is the intrinsic loss rate and $\kappa_{\text{ext}}$
is the coupling rate to the optical signal port. Hence, the optical impedance is
\begin{equation}
Z_{\text{opt}}(\omega)=-i\omega L_{\text{opt}}+R_{\rm int}+Z_{\text{ext}},\label{eq:Z-o_Example}
\end{equation}
where we use Eq.~(\ref{eq:R-opt_def}) to define $R_{\rm int}\equiv\kappa_{\rm int}L_{\text{opt}},Z_{\text{ext}}\equiv\kappa_{\text{ext}}L_{\text{opt}}$.
Similar to Eqs.~(\ref{eq:Z-e_Example},\ref{eq:eta-el_def}), we have the optical coupling efficiency
\begin{equation}
\eta_{\text{opt}}\equiv Z_{\text{ext}}/(R_{\rm int}+Z_{\text{ext}})=\kappa_{\text{ext}}/(\kappa_{\rm int}+\kappa_{\text{ext}}).
\end{equation}
Accordingly, the equivalent optical input and output voltages $V_{\text{o,}\text{in/out}}$ each split into two independent contributions of the same form (\ref{eq:V-o-p_def}) obtained by the replacements $R_{\text{opt}}\rightarrow R_{\rm int},Z_{\text{ext}}$, respectively, and introducing appropriate bosonic operators.
Analogously, the optical readout is found by reexpressing Eq.~(\ref{eq:Optical_IO-rel_equiv_Example}) in terms of
the upper and lower sideband optical currents $I_{\text{o,}\pm}(\Omega)$ ($\Omega>0$)
\begin{eqnarray}
V_{\text{o},\text{out}}^{\text{(ext)}}(\omega_{\text{l}}+\Omega) & = & -\frac{\omega_{\text{l}}+\Omega}{\Omega}Z_{\text{ext}}I_{\text{o},+}(\Omega)+V_{\text{o},\text{in}}^{\text{(ext)}}(\omega_{\text{l}}+\Omega)\label{eq:V-o-U_current_Example}\\
V_{\text{o},\text{out}}^{\text{(ext)}}(\omega_{\text{l}}-\Omega) & = & \frac{\omega_{\text{l}}-\Omega}{\Omega}Z_{\text{ext}}I_{\text{o},-}^{*}(\Omega)+V_{\text{o},\text{in}}^{\text{(ext)}}(\omega_{\text{l}}-\Omega).\label{eq:V-o-L_current_Example}
\end{eqnarray}

Considering the electro-optomechanical circuit diagram~[Fig.~\ref{fig:EOM-equiv-circuit}], we are led to define the mechanical impedance in the presence of OM coupling (cf.~Eq.~(\ref{eq:Z-m_EM-only})),
\begin{equation}
Z_{\text{m}}'(\Omega)\equiv-i\Omega L_{\text{m}}+R_{\text{m}}+\frac{1}{-i\Omega C_{\text{m}}'},\label{eq:Z-mp_def}
\end{equation}
and observe that the central loop in Fig.~\ref{fig:EOM-equiv-circuit} has an effective impedance (cf. Eq.~(\ref{eq:Z-m-eff_def-EM}))
\begin{equation}
Z_{\text{m,eff}}(\Omega)=Z_{\text{m}}'(\Omega)+\Delta Z(\Omega),\label{eq:Z-m-eff_def0}
\end{equation}
where the latter term represents the four parallel connections.
Applying standard impedance combination rules we find (generalizing Eq.~(\ref{eq:DeltaZ-EM_def})),
\begin{equation}
\Delta Z(\Omega)=\sum_{l}\left[-i\Omega C_{l}+1/Z_{l}(\Omega)\right]^{-1}
=\frac{1}{-i\Omega}\sum_{s=\pm}\left(\frac{1+\mathcal{Q}_{\text{e},s}(\Omega)}{\bar{C}_{\text{c}}}+\frac{1+\mathcal{Q}_{\text{o},s}(\Omega)}{\bar{C}_{\text{opt}}}\right),\label{eq:Z-m-eff_def}
\end{equation}
where the optical susceptibility functions are defined in analogy to their electrical counterparts \eqref{eq:Q-e_def}, 
\begin{equation}
\mathcal{Q}_{\text{o,}\pm}(\Omega)\equiv-\frac{1}{1-i\Omega \bar{C}_{\text{opt}}Z_{\text{o,}\pm}(\Omega)}.\label{eq:Q-func_def}
\end{equation}
The resulting effective impedance $Z_{\text{m,eff}}$~(\ref{eq:Z-m-eff_def0})
captures the dynamical back-action modifications to the mechanical response induced by each of the sideband couplings.
The effective mechanical resonance
frequency $\Omega_{\text{m}}$, including all static and dynamical
shifts from the electrical and optical interactions, can be found from Eq.~(\ref{eq:Z-m-eff_def}) as the solution to the equation (assuming we are below the threshold for normal-mode splitting \cite{Dobrindt2008})
\begin{equation}
\text{Im}[Z_{\text{m,eff}}(\Omega_{\text{m}})]=0,\label{eq:Omega-m_def}
\end{equation}
leading us to define the effective mechanical resistance
\begin{equation}
R_{\text{m,eff}}\equiv Z_{\text{m,eff}}(\Omega_{\text{m}}),\label{eq:R-m_eff}
\end{equation}
which yields the effective transducer bandwidth $\gamma_{\text{m,eff}}\equiv R_{\text{m,eff}}/L_{\text{m}}$
in the weak-coupling regime to be discussed in Section~\ref{sec:Adiab-elim}.

\section{Example of application\label{sec:Example-application}}
To demonstrate the full electro-optomechanical circuit formalism we now consider a specific example of a transducer. We choose the simplest possible scenario of a single mechanical mode
serving as the intermediary between a serial RLC circuit via its
capacitance $C_{\text{c}}(x)$ and a single mode
of an optical cavity via a parametric dispersive coupling, see Fig.~\ref{fig:Example-system}.
The electromechanical part of this hybrid system is thus identical to the system considered in Fig.~\ref{fig:LC-overcoupled}. 
This particular example is an instance of a three-oscillator cascade. This problem can also be conveniently solved using coupled mode theories~\cite{Wang2012,Tian2012,Andrews2014}, and it is in principle not necessary to employ our equivalent circuit formalism. Nevertheless we use this example because it is the simplest and it demonstrates how the familiar solution arises in our formalism. We emphasize, however,  that the strength of our formalism is that it allows for the analysis of arbitrary linear circuits, where, e.g., a mode description is less apparent. A detailed analysis of such more complex circuits is beyond the scope of this article and will be considered elsewhere~\cite{EZPhD}.

Referring to the electro-optomechanical equivalent circuit~[Fig.~\ref{fig:EOM-equiv-circuit}], 
we note that the current responses of the various sideband loops $I_{l}(\Omega)$ to the input fields can be determined by straightforward generalization of Eq.~(\ref{eq:Scat_current_eo2}). Combining these with the input-output relations (\ref{eq:tx-line-U_IO-rel_Example},\ref{eq:tx-line-L_IO-rel_Example},\ref{eq:V-o-U_current_Example},\ref{eq:V-o-L_current_Example})
yields the scattering matrix between the sidebands of the electrical and optical transmission lines (as well as the noise sources).
Let us for specificity evaluate the optical output at the upper sideband~(\ref{eq:V-o-U_current_Example})
at the effective mechanical peak frequency $\Omega_{\text{m}}$ defined
in Eq.~(\ref{eq:Omega-m_def}), assuming red-detuned driving for both the optical and electrical subsystems
$\bar{\omega}_{\text{LC}}-\omega_{\text{d}}=\Omega_{\text{m}}=\bar{\omega}_{\text{cav}}-\omega_{\text{l}}$. The current response of interest $I_{\text{o,}+}$ depends on the susceptibility functions (\ref{eq:Q-e_def},\ref{eq:Q-func_def}), which at the mechanical peak take the values
\begin{gather}
\mathcal{Q}_{\text{o},+}(\Omega_{\text{m}})=-iQ_{\text{cav}},\;\mathcal{Q}_{\text{o},-}(\Omega_{\text{m}})\approx\frac{iQ_{\text{cav}}}{1-4iQ_{\text{cav}}\Omega_{\text{m}}/\bar{\omega}_{\text{cav}}},\label{eq:Q-o_eval_Example}\\
\mathcal{Q}_{\text{e},+}(\Omega_{\text{m}})=-iQ_{\text{LC}},\;\mathcal{Q}_{\text{e},-}(\Omega_{\text{m}})\approx\frac{iQ_{\text{LC}}}{1-4iQ_{\text{LC}}\Omega_{\text{m}}/\bar{\omega}_{\text{LC}}},\label{eq:Q-e_eval_Example}
\end{gather}
where $Q_{\rm cav}\equiv \bar{\omega}_{\rm cav}/\kappa$ and $Q_{\rm LC}\equiv\bar{\omega}_{\rm LC}L/(R_{\rm LC}+Z_{\rm tx})$ are the loaded quality factors of the optical and electrical resonances, respectively; the approximations are valid in the limit of high quality factors $Q_{\text{cav}},Q_{\text{LC}}\gg1$ and small mechanical frequency $\Omega_{\text{m}}\ll\bar{\omega}_{\text{cav}},\bar{\omega}_{\text{LC}}$.
These quantities signify the signal enhancement of the various sidebands. Let us, for simplicity, assume the optomechanically resolved-sideband regime, $\kappa/(4\Omega_{\text{m}}) \ll 1$. From Eq.~(\ref{eq:Q-o_eval_Example}) we see that this implies that $|\mathcal{Q}_{\text{o},-}(\Omega_{\text{m}})/\mathcal{Q}_{\text{o},+}(\Omega_{\text{m}})| \ll 1$ and hence we may disregard the loop $(\text{o,}-)$ altogether (although for applications where quantum noise is important, one has to carefully consider to what extent this limit is fulfilled~\cite{alpha}).
In this scenario, we arrive at a scattering relation of the form
\begin{multline}
\hat{a}_{\text{out}}^{\text{(ext)}}(\bar{\omega}_{\text{cav}})=S_{\text{tx,}+}\hat{b}_{\text{in}}^{\text{(tx)}}(\bar{\omega}_{\text{LC}})+S_{\text{LC,}+}\hat{b}_{\text{in}}^{\text{(LC)}}(\bar{\omega}_{\text{LC}})+S_{\text{tx,}-}\hat{b}_{\text{in}}^{\text{(tx)}\dagger}(\bar{\omega}_{\text{LC}}-2\Omega_{\text{m}})\\+S_{\text{LC,}-}\hat{b}_{\text{in}}^{\text{(LC)}\dagger}(\bar{\omega}_{\text{LC}}-2\Omega_{\text{m}})
+S_{\text{m}}\hat{c}_{\text{in}}(\Omega_{\text{m}})+S_{\text{ext,}+}\hat{a}_{\text{in}}^{\text{(ext)}}(\bar{\omega}_{\text{cav}})+S_{\rm{int},+}\hat{a}_{\text{in}}^{(\rm int)}(\bar{\omega}_{\text{cav}}),\label{eq:scat-rel_superpos}
\end{multline}
where $\hat{a}_{\text{in}}^{(X)},\hat{b}_{\text{in}}^{(X)},\hat{c}_{\text{in}}^{(X)}$ are optical, electrical, and mechanical input operators, respectively.
Using the dimensionless voltage mapping factor $\zeta$ from the mechanical
loop to the upper optical sideband,
\begin{equation}
\zeta\equiv-2\frac{\bar{\omega}_{\text{cav}}}{\Omega_{\text{m}}}Z_{\text{ext}}\frac{-iQ_{\text{cav}}}{R_{\text{m,eff}}}= i \eta_{\text{opt}}\frac{4 g_{\text{OM}}^2}{\bar{\omega}_{\text{cav}}\gamma_{\text{m,eff}}},\label{eq:T-def_Example}
\end{equation}
expressed in terms of $g_{\text{OM}}=G_{\text{OM}}/\sqrt{4 m \Omega_{\text{m}}}$ and $\gamma_{\text{m,eff}}$, defined below \cref{eq:R-m_eff}, the scattering matrix elements are
\begin{equation}
S_{\text{m}}=\zeta\sqrt{\frac{\Omega_{\text{m}}R_{\text{m}}}{\bar{\omega}_{\text{cav}}Z_{\text{ext}}}}\label{eq:S-m}
\end{equation}
\begin{eqnarray}
S_{\text{tx,}+}=\zeta\sqrt{\frac{\bar{\omega}_{\text{LC}}Z_{\text{tx}}}{\bar{\omega}_{\text{cav}}Z_{\text{ext}}}}\mathcal{Q}_{\text{e},+}(\Omega_{\text{m}}) & \; & S_{\text{tx,}-}=\zeta\sqrt{\frac{(\bar{\omega}_{\text{LC}}-2\Omega_{\text{m}})Z_{\text{tx}}}{\bar{\omega}_{\text{cav}}Z_{\text{ext}}}}\mathcal{Q}_{\text{e},-}(\Omega_{\text{m}})\label{eq:S-tx}\\
S_{\text{LC,}+}=\zeta\sqrt{\frac{\bar{\omega}_{\text{LC}}R_{\text{LC}}}{\bar{\omega}_{\text{cav}}Z_{\text{ext}}}}\mathcal{Q}_{\text{e},+}(\Omega_{\text{m}}) & \; & S_{\text{LC,}-}=\zeta\sqrt{\frac{(\bar{\omega}_{\text{LC}}-2\Omega_{\text{m}})R_{\text{LC}}}{\bar{\omega}_{\text{cav}}Z_{\text{ext}}}}\mathcal{Q}_{\text{e},-}(\Omega_{\text{m}})\label{eq:S-LC}\\
S_{\text{ext,}+}=1-2\eta_{\text{opt}} - iQ_{\text{cav}}\zeta & \quad & S_{\text{\rm{int},}+}=\sqrt{\eta_{\text{opt}}^{-1}-1}\left[-2\eta_{\text{opt}} - iQ_{\text{cav}}\zeta\right]\label{eq:S-opt},
\end{eqnarray}
where $\mathcal{Q}_{\text{e},\pm}(\Omega_{\text{m}})$
are given in Eqs.~(\ref{eq:Q-e_eval_Example}). All of the scattering
elements contain a frequency and impedance conversion factor of the
form $\sqrt{\omega_{\text{in}}R_{\text{in}}/\omega_{\text{out}}R_{\text{out}}}$;
this is merely the ratio of conversion factors between voltage and the itinerant bosonic modes. The electrical and mechanical scattering coefficients~\eqref{eq:S-m}--\eqref{eq:S-LC} only consist of single terms because for these sources only a single path exists to the optical readout port 'ext'. In contrast, the effective reflection coefficient for the 'ext' port $S_{\text{ext,}+}$~(\ref{eq:S-opt}) results from interference of various scattering paths.

A useful characterization of the transducer can be given in terms of its signal transfer efficiency $\eta$, its added noise flux per unit bandwidth $N$ (referenced to the input)~\cite{alpha}, and its bandwidth. These are readily extracted from the scattering matrix. As we are considering signal conversion from the upper optical sideband to the upper electrical sideband, the (peak) signal transfer efficiency is
\begin{equation}
\eta = |S_{\text{tx,}+}|^2 = 4 \frac{\eta_{\text{el}} \eta_{\text{opt}}}{R_{\text{m,eff}}^2}\frac{Q_{\text{LC}}}{\Omega_{\text{m}}\bar{C}_\text{c}} \frac{Q_{\text{cav}}}{\Omega_{\text{m}}\bar{C}_\text{opt}}.\label{eq:eta-peak_example}
\end{equation}
The added flux of noise quanta per unit bandwidth $N$ contaminating the output at the upper optical sideband, referenced to the input, is $N(\omega)\delta(\omega-\omega') = \left.\eta^{-1}\langle \hat{a}_{\text{out}}^{\text{(ext)}\dagger}(\omega) \hat{a}_{\text{out}}^{\text{(ext)}}(\omega')\rangle\right|_{\text{no signal}}$, where the input at the signal port is disregarded since it is considered as signal and not noise. For our scattering relation we find from Eqs.~\eqref{eq:scat-rel_superpos}--\eqref{eq:S-opt} that at the transducer peak the added noise is
\begin{multline}
N =  \frac{1}{\eta_{\text{el}}}\frac{R_{\text{m}}\Omega_{\text{m}}\bar{C}_{\text{c}}}{Q_{\text{LC}}} n_{\text{m}}(\Omega_{\text{m}}) + \left(\frac{1}{\eta_{\text{el}}}-1\right) n_{\text{LC}}(\bar{\omega}_{\text{LC}})\\
+ \left(1+\left[\frac{4 \Omega_{\mathrm m} L}{R_{\text{LC}}+Z_{\text{tx}}}\right]^{2}\right)^{-1} \left[\left(\frac{1}{\eta_{\text{el}}}-1\right) (n_{\text{LC}}(\bar{\omega}_{\text{LC}} - 2 \Omega_{\text{m}}) + 1) + (n_{\text{tx}}(\bar{\omega}_{\text{LC}} - 2 \Omega_{\text{m}}) + 1)\right].
\label{eq:N-example}
\end{multline}
In Eq.~(\ref{eq:N-example}) we see that in this case the electrical added noise is only affected by $\eta_{\text{el}}$ for the upper sideband (term proportional to $n_{\text{LC}}(\bar{\omega}_{\text{LC}}$), whereas for the lower sideband (second line), the noise is further suppressed by the degree to which the sideband is off-resonant from the RLC resonator. Also this term has extra vacuum noise contributions (unity terms added to $n_{\text{LC}}$ and $n_{\text{tx}}$) since it stems from the lower sideband (i.e., from $\hat{b}_{\text{in}}^{(X)\dagger}$). Turning to the mechanical contribution $\propto n_{\text{m}}$, the suppression factor is $1/(\eta_{\text{el}}\mathcal{C}_{\text{EM}})$, where we have introduced the EM cooperativity,
\begin{equation}
\mathcal{C}_{\text{EM}} \equiv Q_{\mathrm{LC}}/(R_{\mathrm m}\Omega_{\mathrm m}\bar{C}_{\mathrm c})=4 g_{\text{EM}}^{2}/(\gamma_{\mathrm{LC}} \gamma_{\text{m,}0}),\label{eq:C-EM}
\end{equation}
in terms of LC linewidth $\gamma_{\mathrm{LC}} \equiv R_{\mathrm{LC}}/L$ and the EM coupling rate $g_{\text{EM}} = G/\sqrt{4mL\Omega_{\text{m}}\bar{\omega}_{\text{LC}}}$ between annihilation operators (analogous to its OM counterpart~(\ref{eq:g-G-OM})).
This reduction in the mechanical noise is a consequence of the EM signal rate overwhelming the intrinsic mechanical decay rate by the factor $\eta_{\text{el}}\mathcal{C}_{\text{EM}}$.
Since we have assumed the OM resolved-sideband regime and are driving optically red-detuned, the contribution to $N$ from the optical vacuum inputs vanish.
Finally, as pointed out in connection with Eq.~\eqref{eq:R-m_eff}, the effective transducer bandwidth in the regime of adiabatic coupling to the optical and electrical subsystems can be determined as
\begin{equation}
\gamma_{\text{m,eff}}=\frac{R_{\text{m,eff}}}{L_{\text{m}}}=\frac{1}{L_{\text{m}}} \left(R_{\text{m}}+\frac{Q_{\text{cav}}}{\Omega_{\text{m}}\bar{C}_{\text{opt}}}+\frac{Q_{\text{LC}}}{\Omega_{\text{m}}\bar{C}_{\text{c}}}\left[1-\frac{1}{1+(4Q_{\text{LC}}\Omega_{\text{m}}/\bar{\omega}_{\text{LC}})^2}\right] \right),\label{eq:gamma-m,eff}
\end{equation}
where we have used Eqs.~(\ref{eq:Z-m-eff_def0},\ref{eq:Z-m-eff_def},\ref{eq:Q-o_eval_Example},\ref{eq:Q-e_eval_Example}) as well as the assumption of being in the optomechanically resolved-sideband regime, $\kappa/(4\Omega_{\text{m}}) \ll 1$. We remark that Eq.~\eqref{eq:gamma-m,eff} can be reexpressed in terms of the EM cooperativity $\mathcal{C}_{\text{EM}}$~\eqref{eq:C-EM} and its OM analog $\mathcal{C}_{\text{OM}}\equiv 4 g_{\text{OM}}^{2}/(\kappa \gamma_{\text{m,}0})$ so that, e.g., in the fully optically and electrically resolved-sideband limit we can use Eq.~\eqref{eq:R-m-DC_main-text} to get $\gamma_{\text{m,eff}}=\gamma_{\text{m,0}}(1+\mathcal{C}_{\text{OM}}+\mathcal{C}_{\text{EM}})$. 

\section{Adiabatic elimination of electrical and optical modes}\label{sec:Adiab-elim}

\begin{figure}
\centering
\includegraphics[width=0.6\columnwidth]{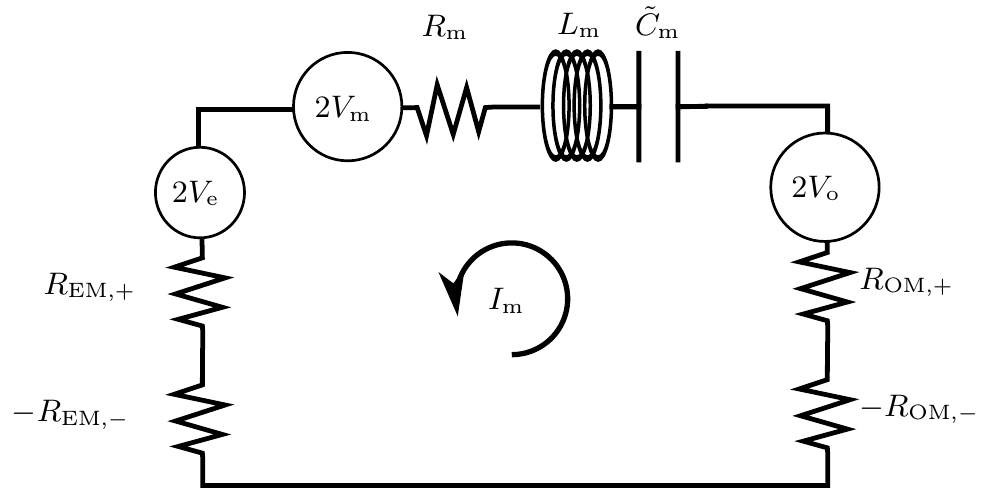}
\caption{Reduced electro-optomechanical equivalent circuit in which electrical
and optical modes have been adiabatically eliminated. It consists
of an effective mechanical loop of shifted resonance frequency loaded
by a resistive element for each sideband coupling. The resistances
are positive for the upper sidebands, $R_{\text{EM/OM},+}$, and negative
for the lower ones, $-R_{\text{EM/OM},-}$, leading to amplification
effects. Each sideband coupling drives the effective mechanical loop
with a Th\'{e}venin voltage representing the noise and signal ports of
that subsystem. The readout of the system corresponds to the signal
dissipated in the various ports.\label{fig:Reduced-equiv-circ}}

\end{figure}

In the preceding sections, we have demonstrated how the electro-optomechanical equivalent circuit
in Fig.~\ref{fig:EOM-equiv-circuit} can be used to deduce the elements
of the scattering matrix. However, if the effective mechanical linewidth is narrow compared to the relevant linewidths of the electrical and optical subsystems, we may
derive an even simpler, reduced equivalent circuit (Fig.~\ref{fig:Reduced-equiv-circ})
by adiabatically eliminating the electrical and optical modes in Fig.~\ref{fig:EOM-equiv-circuit}. Such elimination amounts, in the context of complex impedances, to neglecting the weak frequency dependence of the electrical and optical loads on the mechanical mode as given by Eq.~(\ref{eq:Z-m-eff_def}).

We start by expressing the effective mechanical resonance frequency,
\begin{equation}
\Omega_{\text{m}}\equiv \frac{1}{L_{\text{m}}\tilde{C}_{\text{m}}},\label{eq:Omega-m-pert}
\end{equation}
i.e., including dynamical shifts from the coupling to electrical and optical subsystems, in terms of an effective mechanical capacitance $\tilde{C}_{\rm m}$. The frequency $\Omega_{\text{m}}$ was defined in Eq.~(\ref{eq:Omega-m_def}), but here we will evaluate it perturbatively. To zeroth order in $\mathcal{Q}_{i,\pm}$, we have $\Omega_{\text{m}}=\omega_{\text{m,}Q}$ from Eqs.~(\ref{eq:Z-mp_def}-\ref{eq:Z-m-eff_def},\ref{eq:Omega-m_def}); we therefore approximate the shift from $\mathcal{Q}_{i,\pm}(\Omega)$ by evaluating it at this frequency, yielding the effective mechanical capacitance
\begin{equation}
\frac{1}{\tilde{C}_{\rm m}} = \frac{1}{C_{\text{m}}} + \frac{2}{\bar{C}_{\text{c}}} + \frac{1}{\bar{C}_{\text{c}}}\left(\text{Re}[\mathcal{Q}_{\rm{e},+}(\omega_{\text{m,}Q})+\mathcal{Q}_{\rm{e},-}(\omega_{\text{m,}Q})]\right)
+\frac{1}{\bar{C}_{\text{opt}}}\left(\text{Re}[\mathcal{Q}_{\text{o},+}(\omega_{\text{m,}Q})+\mathcal{Q}_{\text{o},-}(\omega_{\text{m,}Q})]\right),\label{eq:C-m-tilde}
\end{equation}
where we have replaced $C_{\text{m}}'$ with the original $C_{\text{m}}$ using Eq.~(\ref{eq:C-m-prime}). Note that the sum of the two first terms of Eq.~(\ref{eq:C-m-tilde}) corresponds to the frequency $\omega_{\text{m,}Q}$ (as per Eq.~(\ref{eq:omega-m,Q_main-text}) with an extra factor of 2 in front of $1/\bar{C}_{\text{c}}$ in the AC case as stated below Eq.~(\ref{eq:p-dot_mech})).
In the following, we will derive the effective resistive loads imposed on the mechanical mode.

\subsection{Elimination of optical mode}
The real part of the OM load from the respective sidebands give rise to effective resistive elements $\pm R_{\text{EM},\pm}$. Evaluating the relevant parts of Eq.~(\ref{eq:Z-m-eff_def}) at the mechanical frequency $\omega_{\text{m,}Q}$, we find
\begin{equation}
R_{\text{OM},\pm}\equiv\pm\left.\text{Re}\left[\frac{\mathcal{Q}_{\text{o},\pm}(\Omega)}{-i\Omega\bar{C}_{\text{opt}}}\right]\right|_{\Omega=\omega_{\text{m,}Q}}.\label{eq:R-OM_def}
\end{equation}
This can be put on a more specific form since we are assuming a single
cavity mode throughout this work, thus fixing the form of the optical impedance $Z_{\text{opt}}$~(\ref{eq:Z-opt_def}). Considering optical frequencies
$\omega$ close to the cavity resonance, $|\omega-\bar{\omega}_{\text{cav}}|\ll\bar{\omega}_{\text{cav}}$,
the optical susceptibility $\mathcal{Q}_{\text{o,}\pm}(\Omega)$ is well-approximated by a Lorentzian. Ignoring
corrections of order $\omega_{\text{m,}Q}/\omega_{\text{l}}$, this
allows us to reexpress Eq.~(\ref{eq:R-OM_def}) as
\begin{gather}
R_{\text{OM},\pm}=L_{\text{m}}\gamma_{\text{OM,}\pm}\\
\gamma_{\text{OM,}\pm} \equiv \gamma_{\text{m,}0}\mathcal{C}_{\text{OM}}\mathcal{L}_{\pm}^{2},\quad \mathcal{C}_{\text{OM}}\equiv\frac{4g_{\text{OM}}^{2}}{\gamma_{\text{m,}0}\kappa},\label{eq:gamma-m_C-OM_main-text}
\end{gather}
where $\gamma_{\text{OM,}\pm}$ are the OM anti-Stokes and Stokes rates expressed in terms of the OM cooperativity $\mathcal{C}_{\text{OM}}$ and the Lorentzian sideband strengths $\mathcal{L}_{\pm}$
at the upper/lower OM sidebands, 
\begin{gather}
\mathcal{L}(\omega)\equiv\frac{\kappa/2}{-i(\omega-\bar{\omega}_{\text{cav}})+\kappa/2},\label{eq:Lorentzian-distrib_main-text}\\
\mathcal{L}_{\pm}\equiv|\mathcal{L}(\omega_{\text{l}}\pm\omega_{\text{m,}Q})|,\quad\theta_{\pm}\equiv\text{Arg}[\mathcal{L}(\omega_{\text{l}}\pm\omega_{\text{m,}Q})],\label{eq:Lorentz-pm}
\end{gather}
also introducing the sideband phases $\theta_{\pm}$.
The effective optical Th\'{e}venin voltage $V_{\text{o}}$, which has
contributions from $V_{\text{o,}\pm}$, can be determined by voltage
division (see discussion above Eq.~\eqref{eq:Scat_current_eo2}),
\begin{gather}
V_{\text{o}}(\Omega)\approx-iQ_{\text{cav}}[e^{i\theta_{+}}\mathcal{L}_{+}V_{\text{o,}+}(\Omega)-e^{-i\theta_{-}}\mathcal{L}_{-}V_{\text{o,}-}(\Omega)],\label{eq:V-o_def}
\end{gather}
where we have ignored the frequency dependence of $\mathcal{L}(\omega)$
over the bandwidth of interest as in Eq.~\eqref{eq:gamma-m_C-OM_main-text}. For the typical example of an optical
mode with two decay channels considered in Section \ref{sec:Example-application},
$V_{\text{o}}(\Omega)$ in Eq.~(\ref{eq:V-o_def}) can be stated explicitly
in terms of the bosonic operators of the itinerant optical fields
 (again ignoring corrections of order $\omega_{\text{m,}Q}/\omega_{\text{l}}$),
\begin{equation}
\hat{V}_{\text{o}}(\Omega)\approx-i\frac{2g_{\text{OM}}}{\sqrt{\kappa}}\frac{\sqrt{\hbar m\omega_{\text{m,}Q}/2}}{\bar{C}_{\text{c}}G}[e^{i\theta_{+}}\mathcal{L}_{+}\hat{a}_{\text{in}}(\omega_{\text{l}}+\Omega)-e^{-i\theta_{-}}\mathcal{L}_{-}\hat{a}_{\text{in}}^{\dagger}(\omega_{\text{l}}-\Omega)],\label{eq:V-o-a}
\end{equation}
where $\hat{a}_{\text{in}}(\omega)=\sqrt{\eta_{\text{opt}}}\hat{a}_{\text{in}}^{\text{(ext)}}(\omega)+\sqrt{1-\eta_{\text{opt}}}\hat{a}_{\text{in}}^{\text{(int)}}(\omega)$
is a linear combination of the two optical input fields. The effective
OM input-output relation can be found by combining Eq.~(\ref{eq:Optical_IO-rel_equiv_Example})
with Eq.~(\ref{eq:V-o-U_KVL}) and the optical counterparts
of Eq.~(\ref{eq:I_U-L_defs}).
For the readout port 'ext' we find the outgoing field
\begin{eqnarray}
a_{\text{out}}^{(\text{ext})}(\omega_{\text{l}}+\Omega) & = & ie^{i\theta_{+}}\sqrt{\eta_{\text{opt}}}\sqrt{\frac{2R_{\text{OM},+}}{\hbar\omega_{\text{m,}Q}}}\frac{\omega_{\text{m,}Q}}{\Omega}I_{\text{m}}(\Omega)+a_{\text{in}}^{(\text{eff})}(\omega_{\text{l}}+\Omega)\label{eq:Eff-OM-IO-rel-U}\\
a_{\text{out}}^{(\text{ext})}(\omega_{\text{l}}-\Omega) & = & -ie^{i\theta_{-}}\sqrt{\eta_{\text{opt}}}\sqrt{\frac{2R_{\text{OM},-}}{\hbar\omega_{\text{m,}Q}}}\frac{\omega_{\text{m,}Q}}{\Omega}I_{\text{m}}^{*}(\Omega)+a_{\text{in}}^{(\text{eff})}(\omega_{\text{l}}-\Omega),\label{eq:Eff-OM-IO-rel-L}
\end{eqnarray}
where we have defined the effective optical noise operator ($\Omega>0$)
\begin{equation}
\hat{a}_{\text{in}}^{(\text{eff})}(\omega_{\text{l}}\pm\Omega)\equiv\left[1-2\eta_{\text{opt}}\mathcal{L}_{\pm}e^{i\theta_{\pm}}\right]\hat{a}_{\text{in}}^{(\text{ext})}(\omega_{\text{l}}\pm\Omega)-2\sqrt{\eta_{\text{opt}}(1-\eta_{\text{opt}})}\mathcal{L}_{\pm}e^{i\theta_{\pm}}\hat{a}_{\text{in}}^{(\text{int})}(\omega_{\text{l}}\pm\Omega).\label{eq:a-in-eff_def}
\end{equation}

\subsection{Elimination of electrical modes}

We define the effective EM resistances from Eq.~(\ref{eq:Z-m-eff_def}) analogously to Eq.~(\ref{eq:R-OM_def}),
\begin{equation}
R_{\text{EM},\pm}\equiv\left.\pm\text{Re}\left[\frac{\mathcal{Q}_{\text{e},\pm}(\Omega)}{-i\Omega\bar{C}_{\text{c}}}\right]\right|_{\Omega=\omega_{\text{m,}Q}}.\label{eq:R-EM_def}
\end{equation}
From this expression we find (neglecting corrections of order $\omega_{\text{m,}Q}/\omega_{\text{d}}$)
\begin{eqnarray}
R_{\text{EM},\pm} & \approx & \frac{\omega_{\text{d}}}{\omega_{\text{m,}Q}}\left. \text{Re} \left[\left(-i\omega \bar{C}_{\text{c}}+1/Z(\omega)\right) ^{-1} \right]\right|_{\omega=\omega_{\text{d}}\pm\omega_{\text{m,}Q}},\label{eq:R-EM-U_result}
\end{eqnarray}
where $Z(\omega)$ is the arbitrary impedance illustrated in Fig.~\ref{Fig:AC-equivalent}a, here entering in a parallel combination with the coupling capacitor impedance.
The Th\'{e}venin voltage $V_{\text{e}}$ of the reduced circuit (Fig.~\ref{fig:Reduced-equiv-circ}) is easily derived in terms of the Th\'{e}venin voltages $V_{\text{e,}\pm}$ of Fig.~\ref{fig:EOM-equiv-circuit} (paralleling Eq.~\eqref{eq:V-o_def}),
\begin{equation}
2 V_{\text{e}}(\Omega) \equiv \mathcal{Q}_{\text{e,}+}(\omega_{\text{m,}Q})V_{\text{e,}+}(\Omega) + \mathcal{Q}_{\text{e,}-}(\omega_{\text{m,}Q})V_{\text{e,}-}(\Omega).\label{eq:Ve_adiab}
\end{equation}
For purposes of practical calculation, however, a simpler strategy is to calculate $V_{\text{e}}$ directly without the intermediate step of determining $V_{\text{e,}\pm}$. To this end, note that the contributions of the individual loops $l$ to the effective electro-optical load $\Delta Z$ \eqref{eq:Z-m-eff_def} are simply the Th\'{e}venin impedances of those loops \emph{including} their respective coupling capacitors. Specifically, denoting the corresponding (lab frame) Th\'{e}venin voltage for the electrical circuit $\delta V'(\omega)$, the electrical Th\'{e}venin voltage of the reduced equivalent circuit is
\begin{equation}
2 V_{\text{e}}(\Omega) = \delta V'(\omega_{\text{d}}+\Omega) + \delta V'^{*}(\omega_{\text{d}}-\Omega).\label{eq:Ve-from-Vprime}
\end{equation}
The effective EM input-output relations can then be determined as in the OM case once the electrical circuit has been specified.

Recapitulating the effective description that we have now obtained (see Fig.~\ref{fig:Reduced-equiv-circ}):
The electrical and optical modes enter as effective loads attached
to the mechanical loop. The combined electrical and optical loading
of the mechanical current loop $\Delta Z(\Omega)$ was derived in Eq.~(\ref{eq:Z-m-eff_def}). The real part of $\Delta Z(\Omega)$ adds resistance while the imaginary part shifts the mechanical resonance frequency. The effective resistance of the mechanical
loop thus has a contribution from each sideband of each coupled subsystem, positive from the upper sidebands
and negative for the lower ones
\begin{equation}
R_{\text{m,eff}}=R_{\text{m}}+R_{\text{EM,}+}-R_{\text{EM,}-}+R_{\text{OM,}+}-R_{\text{OM,}-}.\label{eq:Rm-eff_Adiab}
\end{equation}
The resistances $R_{\text{EM/OM,}\pm}$ are the electrical equivalents of the EM/OM anti-Stokes and Stokes rates for the scattering of mechanical phonons into the respective sidebands as electrical/optical photons,
\begin{equation}
\gamma_{\text{EM/OM},\pm} \equiv R_{\text{EM/OM},\pm}/L_{\text{m}}.\label{eq:Stokes_defs}
\end{equation}
Moreover, each of the eliminated electrical and optical subsystems contribute a voltage source, $V_{\text{e}}(\Omega)$ and $V_{\text{o}}(\Omega)$, to the effective mechanical loop.

As a particular example we again consider a mechanical mode coupled to a serial RLC circuit and a single optical mode [Fig.~\ref{fig:Example-system}] and derive the reduced equivalent circuit. For specificity, we take both the electrical and optical drives to be red-detuned by $\omega_{\text{m,}Q}$.
First we determine the effective mechanical resonance frequency $\Omega_{\mathrm{m}}$ (\ref{eq:Omega-m-pert}) within the Lorentzian approximation (\ref{eq:Q-o_eval_Example},\ref{eq:Q-e_eval_Example}) as the solution to the equation
\begin{equation}
\Omega_{\rm{m}}^{2} = \omega_{\text{m,}Q}^2 - \frac{\bar{C}_{\text{c}}G^2}{m} \frac{4Q_{\text{LC}}^2\Omega_{\text{m}}/\bar{\omega}_{\text{LC}}}{1+(4Q_{\text{LC}}\Omega_{\text{m}}/\bar{\omega}_{\text{LC}})^2} - \frac{G_{\text{OM}}^2}{m\bar{\omega}_{\text{cav}}}\frac{4Q_{\text{cav}}^2\Omega_{\text{m}}/\bar{\omega}_{\text{cav}}}{1+(4Q_{\text{cav}}\Omega_{\text{m}}/\bar{\omega}_{\text{cav}})^2}.\label{eq:Example_Omega-m}
\end{equation}
A perturbative solution to Eq.~\eqref{eq:Example_Omega-m} can be obtained by substituting $\Omega_{\text{m}}\rightarrow \omega_{\text{m,}Q}$ on its right-hand side, which amounts to applying the result \eqref{eq:C-m-tilde}.
This determines the frequency at which the mechanical response is maximal (in the limit of high effective mechanical quality factor $Q_{\text{m,eff}}\equiv \Omega_{\text{m}}L_{\text{m}}/R_{\text{m,eff}}\gg 1$) and hence typically the desirable (center) frequency for signal input (in the rotating frame of the drive fields).
Next, we determine the resistances $R_{\text{OM},\pm},R_{\text{EM},\pm}$ from Eqs.~(\ref{eq:R-OM_def},\ref{eq:R-EM_def}) again making use of the sideband strengths~(\ref{eq:Q-o_eval_Example},\ref{eq:Q-e_eval_Example})
\begin{eqnarray}
R_{\text{OM},+}=\frac{Q_{\text{cav}}}{\omega_{\text{m,}Q}\bar{C}_{\text{opt}}}, &\quad & R_{\text{OM},-}=\frac{Q_{\text{cav}}}{\omega_{\text{m,}Q}\bar{C}_{\text{opt}}}\frac{1}{1+(4Q_{\text{cav}}\omega_{\text{m,}Q}/\bar{\omega}_{\text{cav}})^2}\label{eq:R-OM-pm_example}\\
R_{\text{EM},+}=\frac{\bar{\omega}_{\text{LC}}}{\omega_{\text{m,}Q}}\frac{L/\bar{C}_{\text{c}}}{R_{\text{LC}}+Z_{\text{tx}}},& &R_{\text{EM},-}=\frac{\bar{\omega}_{\text{LC}}}{\omega_{\text{m,}Q}}\frac{L/\bar{C}_{\text{c}}}{R_{\text{LC}}+Z_{\text{tx}}}\frac{1}{1+(4Q_{\text{LC}}\omega_{\text{m,}Q}/\bar{\omega}_{\text{LC}})^2}.\label{eq:R-EM-pm_example}
\end{eqnarray}
The optical and electrical Th\'{e}venin voltages of the reduced equivalent circuit are given respectively by  Eq.~(\ref{eq:V-o-a}) and, using Eqs.~(\ref{eq:Ve_adiab},\ref{eq:Q-e_eval_Example},\ref{eq:V_U-L_defs}), we find
\begin{equation}
2 V_{\text{e}}(\Omega) \approx -iQ_{\text{LC}}\delta V(\omega_{\rm d}+\Omega) + \frac{iQ_{\text{LC}}}{1-4iQ_{\text{LC}}\omega_{\text{m,}Q}/\bar{\omega}_{\text{LC}}} \delta V^{*}(\omega_{\rm d}-\Omega).\label{eq:Ve_adiab_example}
\end{equation}
In view of Eqs.~(\ref{eq:R-OM-pm_example},\ref{eq:R-EM-pm_example}) and~(\ref{eq:eta-peak_example}), the peak transfer efficiency between the upper sidebands can in our example case be compactly expressed as
\begin{equation}
\eta = \eta_{\text{el}}\eta_{\text{opt}}\frac{4 R_{\text{OM,}+} R_{\text{EM,}+}}{R_{\text{m,eff}}^2},
\end{equation}
where the effective mechanical resistance $R_{\text{m,eff}}$ is given by Eq.~(\ref{eq:Rm-eff_Adiab}).
The reduced equivalent circuit [Fig.~\ref{fig:Reduced-equiv-circ}] with parameters given by Eqs.~(\ref{eq:Omega-m-pert},\ref{eq:V-o-a},\ref{eq:Example_Omega-m}-\ref{eq:Ve_adiab_example}) provides a simplified description of the example hybrid system [Fig.~\ref{fig:Example-system}] which is accurate for frequency components $\Omega$ in a band around $\Omega_{\text{m}}$ much narrower than the frequency scale over which the effective electro-optical load $\Delta Z(\Omega)$ \eqref{eq:Z-m-eff_def} varies, i.e., $|\Omega - \Omega_{\text{m}}| \ll \text{min}\{(R_{\text{LC}}+Z_{\text{tx}})/L,\kappa\}$ for the present example. The exact description is recovered by retaining the $\Omega$-dependence of the electrical and optical susceptibility functions $\mathcal{Q}_{i,\pm}(\Omega)$ and (consequently) that of the electro-optical load $\Delta Z(\Omega)$.

\section{Conclusion and outlook}\label{sec:Conclusion}
Electro-optomechanical hybrid systems are promising candidates for providing low-noise transduction between the electrical and optical domains with applications both in the classical and quantum regime. To fully realize the technological potential of such hybrid systems, a common framework unifying their disparate components is needed. 
In this article we have developed an equivalent circuit formalism for electro-optomechanical transduction which is an exact mapping of the linearized dynamics of the hybrid system. This allows the scattering matrix of the transducer to be determined by standard circuit analysis. Importantly, the formalism accommodates AC-biased interfaces and incorporate amplification effects due to finite sideband resolution. This makes the formalism particularly useful for describing transduction between different frequencies. The  equivalent circuit description is, however, also applicable when considering scenarios involving only EM or OM components and may help provide a more intuitive understanding of these systems.

Importantly, the framework developed here provides a common language that bridges the gap between the electrical engineering and quantum optics communities. In particular, it allows the engineering community to readily explore the prospects of integrating OM functionalities into electrical systems using the vast set of tools available for circuit design. In this way, one can exploit the low-noise sensing capabilities and optical communication compatibility
of optomechanics~\cite{Bagci2014} in real-world applications such as nuclear magnetic resonance (NMR) detection~\cite{Takeda2017} and radio-astronomy.

Our formalism captures the quantum mechanical aspects of transduction in a simple manner and one only has to account for the quantum noise in the input and output.
This means that describing the quantum mechanical properties of the device is not more complicated than describing the classical dynamics of the system. The formalism is therefore highly suited for assessing the potential of electro-optomechanical transducers in future quantum communication based on
optical networks~\cite{Stannigel2010}.

The transduction is treated as a scattering scenario, where asymptotic input fields are mapped to output fields. In essence, this means that we consider the source of the signal to be at the other end of an infinitely long transmission line such that it experiences no back action from the transducer. For some transduction applications, such as the conversion of a signal from a superconducting qubit to an optical photon, it may be an advantage to consider a more integrated structure where the qubit is an integral part of the electrical circuit. In such situation, one can envision combining the impedance description of OM systems presented here with the quantum theory of nonlinear circuits, e.g., superconducting flux~\cite{Burkard2004} or charge~\cite{Burkard2005} qubits, or one may use a black-box quantization approach for weakly anharmonic circuits~\cite{Nigg2012} to obtain an efficient description of the joint system.

\begin{acknowledgments}
We acknowledge helpful discussions with A.~Simonsen, K.~Lehnert, J.~Aumentado, R.~Simmonds, and K.~Usami. We thank J.~Foley for reading the manuscript.
EZ thanks the JQI for hosting him. JMT thanks the NBI for their hospitality during his many stays.

The research leading to these results was funded by The European Union Seventh Framework Programme through SIQS (grant no.~600645), ERC Grants QIOS (grant no.~306576), and by the European Union's Horizon 2020 research and innovation programme (ERC project Q-CEOM, grant agreement no.~638765 and project HOT, grant agreement no.~732894), a starting grant from the Danish Council for Independent Research (grant number 4002-00060) as well as the Physics Frontier Center at the JQI, DARPA DSO. EZ acknowledges funding from the Carlsberg Foundation.

\end{acknowledgments}

\bibliography{beta}

\begin{thebibliography}{104}%
\makeatletter
\providecommand \@ifxundefined [1]{%
 \@ifx{#1\undefined}
}%
\providecommand \@ifnum [1]{%
 \ifnum #1\expandafter \@firstoftwo
 \else \expandafter \@secondoftwo
 \fi
}%
\providecommand \@ifx [1]{%
 \ifx #1\expandafter \@firstoftwo
 \else \expandafter \@secondoftwo
 \fi
}%
\providecommand \natexlab [1]{#1}%
\providecommand \enquote  [1]{``#1''}%
\providecommand \bibnamefont  [1]{#1}%
\providecommand \bibfnamefont [1]{#1}%
\providecommand \citenamefont [1]{#1}%
\providecommand \href@noop [0]{\@secondoftwo}%
\providecommand \href [0]{\begingroup \@sanitize@url \@href}%
\providecommand \@href[1]{\@@startlink{#1}\@@href}%
\providecommand \@@href[1]{\endgroup#1\@@endlink}%
\providecommand \@sanitize@url [0]{\catcode `\\12\catcode `\$12\catcode
  `\&12\catcode `\#12\catcode `\^12\catcode `\_12\catcode `\%12\relax}%
\providecommand \@@startlink[1]{}%
\providecommand \@@endlink[0]{}%
\providecommand \url  [0]{\begingroup\@sanitize@url \@url }%
\providecommand \@url [1]{\endgroup\@href {#1}{\urlprefix }}%
\providecommand \urlprefix  [0]{URL }%
\providecommand \Eprint [0]{\href }%
\providecommand \doibase [0]{http://dx.doi.org/}%
\providecommand \selectlanguage [0]{\@gobble}%
\providecommand \bibinfo  [0]{\@secondoftwo}%
\providecommand \bibfield  [0]{\@secondoftwo}%
\providecommand \translation [1]{[#1]}%
\providecommand \BibitemOpen [0]{}%
\providecommand \bibitemStop [0]{}%
\providecommand \bibitemNoStop [0]{.\EOS\space}%
\providecommand \EOS [0]{\spacefactor3000\relax}%
\providecommand \BibitemShut  [1]{\csname bibitem#1\endcsname}%
\let\auto@bib@innerbib\@empty
\bibitem [{\citenamefont {Gisin}\ \emph {et~al.}(2002)\citenamefont {Gisin},
  \citenamefont {Ribordy}, \citenamefont {Tittel},\ and\ \citenamefont
  {Zbinden}}]{Gisin2002}%
  \BibitemOpen
  \bibfield  {author} {\bibinfo {author} {\bibfnamefont {Nicolas}\ \bibnamefont
  {Gisin}}, \bibinfo {author} {\bibfnamefont {Gr\'egoire}\ \bibnamefont
  {Ribordy}}, \bibinfo {author} {\bibfnamefont {Wolfgang}\ \bibnamefont
  {Tittel}}, \ and\ \bibinfo {author} {\bibfnamefont {Hugo}\ \bibnamefont
  {Zbinden}},\ }\bibfield  {title} {\enquote {\bibinfo {title} {Quantum
  cryptography},}\ }\href {\doibase 10.1103/RevModPhys.74.145} {\bibfield
  {journal} {\bibinfo  {journal} {Rev. Mod. Phys.}\ }\textbf {\bibinfo {volume}
  {74}},\ \bibinfo {pages} {145--195} (\bibinfo {year} {2002})}\BibitemShut
  {NoStop}%
\bibitem [{\citenamefont {Kimble}(2008)}]{Kimble2008}%
  \BibitemOpen
  \bibfield  {author} {\bibinfo {author} {\bibfnamefont {H.~J.}\ \bibnamefont
  {Kimble}},\ }\bibfield  {title} {\enquote {\bibinfo {title} {The quantum
  internet},}\ }\href {http://dx.doi.org/10.1038/nature07127} {\bibfield
  {journal} {\bibinfo  {journal} {Nature}\ }\textbf {\bibinfo {volume} {453}},\
  \bibinfo {pages} {1023--1030} (\bibinfo {year} {2008})}\BibitemShut {NoStop}%
\bibitem [{\citenamefont {Wendin}(2017)}]{Wendin2017}%
  \BibitemOpen
  \bibfield  {author} {\bibinfo {author} {\bibfnamefont {G}~\bibnamefont
  {Wendin}},\ }\bibfield  {title} {\enquote {\bibinfo {title} {Quantum
  information processing with superconducting circuits: a review},}\ }\href
  {http://stacks.iop.org/0034-4885/80/i=10/a=106001} {\bibfield  {journal}
  {\bibinfo  {journal} {Reports on Progress in Physics}\ }\textbf {\bibinfo
  {volume} {80}},\ \bibinfo {pages} {106001} (\bibinfo {year}
  {2017})}\BibitemShut {NoStop}%
\bibitem [{\citenamefont {Chiorescu}\ \emph {et~al.}(2003)\citenamefont
  {Chiorescu}, \citenamefont {Nakamura}, \citenamefont {Harmans},\ and\
  \citenamefont {Mooij}}]{Chiorescu2003}%
  \BibitemOpen
  \bibfield  {author} {\bibinfo {author} {\bibfnamefont {I.}~\bibnamefont
  {Chiorescu}}, \bibinfo {author} {\bibfnamefont {Y.}~\bibnamefont {Nakamura}},
  \bibinfo {author} {\bibfnamefont {C.~J. P.~M.}\ \bibnamefont {Harmans}}, \
  and\ \bibinfo {author} {\bibfnamefont {J.~E.}\ \bibnamefont {Mooij}},\
  }\bibfield  {title} {\enquote {\bibinfo {title} {Coherent quantum dynamics of
  a superconducting flux qubit},}\ }\href {\doibase 10.1126/science.1081045}
  {\bibfield  {journal} {\bibinfo  {journal} {Science}\ }\textbf {\bibinfo
  {volume} {299}},\ \bibinfo {pages} {1869--1871} (\bibinfo {year} {2003})},\
  \Eprint
  {http://arxiv.org/abs/http://science.sciencemag.org/content/299/5614/1869.full.pdf}
  {http://science.sciencemag.org/content/299/5614/1869.full.pdf} \BibitemShut
  {NoStop}%
\bibitem [{\citenamefont {Yamamoto}\ \emph {et~al.}(2003)\citenamefont
  {Yamamoto}, \citenamefont {Pashkin}, \citenamefont {Astafiev}, \citenamefont
  {Nakamura},\ and\ \citenamefont {Tsai}}]{Yamamoto2003}%
  \BibitemOpen
  \bibfield  {author} {\bibinfo {author} {\bibfnamefont {T.}~\bibnamefont
  {Yamamoto}}, \bibinfo {author} {\bibfnamefont {Yu.~A.}\ \bibnamefont
  {Pashkin}}, \bibinfo {author} {\bibfnamefont {O.}~\bibnamefont {Astafiev}},
  \bibinfo {author} {\bibfnamefont {Y.}~\bibnamefont {Nakamura}}, \ and\
  \bibinfo {author} {\bibfnamefont {J.~S.}\ \bibnamefont {Tsai}},\ }\bibfield
  {title} {\enquote {\bibinfo {title} {Demonstration of conditional gate
  operation using superconducting charge qubits},}\ }\href
  {http://dx.doi.org/10.1038/nature02015} {\bibfield  {journal} {\bibinfo
  {journal} {Nature}\ }\textbf {\bibinfo {volume} {425}},\ \bibinfo {pages}
  {941 EP --} (\bibinfo {year} {2003})}\BibitemShut {NoStop}%
\bibitem [{\citenamefont {Wallraff}\ \emph {et~al.}(2004)\citenamefont
  {Wallraff}, \citenamefont {Schuster}, \citenamefont {Blais}, \citenamefont
  {Frunzio}, \citenamefont {Huang}, \citenamefont {Majer}, \citenamefont
  {Kumar}, \citenamefont {Girvin},\ and\ \citenamefont
  {Schoelkopf}}]{Wallraff2004}%
  \BibitemOpen
  \bibfield  {author} {\bibinfo {author} {\bibfnamefont {A.}~\bibnamefont
  {Wallraff}}, \bibinfo {author} {\bibfnamefont {D.~I.}\ \bibnamefont
  {Schuster}}, \bibinfo {author} {\bibfnamefont {A.}~\bibnamefont {Blais}},
  \bibinfo {author} {\bibfnamefont {L.}~\bibnamefont {Frunzio}}, \bibinfo
  {author} {\bibfnamefont {R.~S.}\ \bibnamefont {Huang}}, \bibinfo {author}
  {\bibfnamefont {J.}~\bibnamefont {Majer}}, \bibinfo {author} {\bibfnamefont
  {S.}~\bibnamefont {Kumar}}, \bibinfo {author} {\bibfnamefont {S.~M.}\
  \bibnamefont {Girvin}}, \ and\ \bibinfo {author} {\bibfnamefont {R.~J.}\
  \bibnamefont {Schoelkopf}},\ }\bibfield  {title} {\enquote {\bibinfo {title}
  {Strong coupling of a single photon to a superconducting qubit using circuit
  quantum electrodynamics},}\ }\href {http://dx.doi.org/10.1038/nature02851}
  {\bibfield  {journal} {\bibinfo  {journal} {Nature}\ }\textbf {\bibinfo
  {volume} {431}},\ \bibinfo {pages} {162 EP --} (\bibinfo {year}
  {2004})}\BibitemShut {NoStop}%
\bibitem [{\citenamefont {Lucero}\ \emph {et~al.}(2012)\citenamefont {Lucero},
  \citenamefont {Barends}, \citenamefont {Chen}, \citenamefont {Kelly},
  \citenamefont {Mariantoni}, \citenamefont {Megrant}, \citenamefont
  {O'Malley}, \citenamefont {Sank}, \citenamefont {Vainsencher}, \citenamefont
  {Wenner}, \citenamefont {White}, \citenamefont {Yin}, \citenamefont
  {Cleland},\ and\ \citenamefont {Martinis}}]{Lucero2012}%
  \BibitemOpen
  \bibfield  {author} {\bibinfo {author} {\bibfnamefont {Erik}\ \bibnamefont
  {Lucero}}, \bibinfo {author} {\bibfnamefont {R.}~\bibnamefont {Barends}},
  \bibinfo {author} {\bibfnamefont {Y.}~\bibnamefont {Chen}}, \bibinfo {author}
  {\bibfnamefont {J.}~\bibnamefont {Kelly}}, \bibinfo {author} {\bibfnamefont
  {M.}~\bibnamefont {Mariantoni}}, \bibinfo {author} {\bibfnamefont
  {A.}~\bibnamefont {Megrant}}, \bibinfo {author} {\bibfnamefont
  {P.}~\bibnamefont {O'Malley}}, \bibinfo {author} {\bibfnamefont
  {D.}~\bibnamefont {Sank}}, \bibinfo {author} {\bibfnamefont {A.}~\bibnamefont
  {Vainsencher}}, \bibinfo {author} {\bibfnamefont {J.}~\bibnamefont {Wenner}},
  \bibinfo {author} {\bibfnamefont {T.}~\bibnamefont {White}}, \bibinfo
  {author} {\bibfnamefont {Y.}~\bibnamefont {Yin}}, \bibinfo {author}
  {\bibfnamefont {A.~N.}\ \bibnamefont {Cleland}}, \ and\ \bibinfo {author}
  {\bibfnamefont {John~M.}\ \bibnamefont {Martinis}},\ }\bibfield  {title}
  {\enquote {\bibinfo {title} {Computing prime factors with a josephson phase
  qubit quantum processor},}\ }\href {http://dx.doi.org/10.1038/nphys2385}
  {\bibfield  {journal} {\bibinfo  {journal} {Nature Physics}\ }\textbf
  {\bibinfo {volume} {8}},\ \bibinfo {pages} {719 EP --} (\bibinfo {year}
  {2012})}\BibitemShut {NoStop}%
\bibitem [{\citenamefont {Rist{\`e}}\ \emph {et~al.}(2013)\citenamefont
  {Rist{\`e}}, \citenamefont {Dukalski}, \citenamefont {Watson}, \citenamefont
  {de~Lange}, \citenamefont {Tiggelman}, \citenamefont {Blanter}, \citenamefont
  {Lehnert}, \citenamefont {Schouten},\ and\ \citenamefont
  {DiCarlo}}]{Riste2013}%
  \BibitemOpen
  \bibfield  {author} {\bibinfo {author} {\bibfnamefont {D.}~\bibnamefont
  {Rist{\`e}}}, \bibinfo {author} {\bibfnamefont {M.}~\bibnamefont {Dukalski}},
  \bibinfo {author} {\bibfnamefont {C.~A.}\ \bibnamefont {Watson}}, \bibinfo
  {author} {\bibfnamefont {G.}~\bibnamefont {de~Lange}}, \bibinfo {author}
  {\bibfnamefont {M.~J.}\ \bibnamefont {Tiggelman}}, \bibinfo {author}
  {\bibfnamefont {Ya.~M.}\ \bibnamefont {Blanter}}, \bibinfo {author}
  {\bibfnamefont {K.~W.}\ \bibnamefont {Lehnert}}, \bibinfo {author}
  {\bibfnamefont {R.~N.}\ \bibnamefont {Schouten}}, \ and\ \bibinfo {author}
  {\bibfnamefont {L.}~\bibnamefont {DiCarlo}},\ }\bibfield  {title} {\enquote
  {\bibinfo {title} {Deterministic entanglement of superconducting qubits by
  parity measurement and feedback},}\ }\href
  {http://dx.doi.org/10.1038/nature12513} {\bibfield  {journal} {\bibinfo
  {journal} {Nature}\ }\textbf {\bibinfo {volume} {502}},\ \bibinfo {pages}
  {350 EP --} (\bibinfo {year} {2013})}\BibitemShut {NoStop}%
\bibitem [{\citenamefont {Barends}\ \emph {et~al.}(2016)\citenamefont
  {Barends}, \citenamefont {Shabani}, \citenamefont {Lamata}, \citenamefont
  {Kelly}, \citenamefont {Mezzacapo}, \citenamefont {Heras}, \citenamefont
  {Babbush}, \citenamefont {Fowler}, \citenamefont {Campbell}, \citenamefont
  {Chen}, \citenamefont {Chen}, \citenamefont {Chiaro}, \citenamefont
  {Dunsworth}, \citenamefont {Jeffrey}, \citenamefont {Lucero}, \citenamefont
  {Megrant}, \citenamefont {Mutus}, \citenamefont {Neeley}, \citenamefont
  {Neill}, \citenamefont {O'Malley}, \citenamefont {Quintana}, \citenamefont
  {Roushan}, \citenamefont {Sank}, \citenamefont {Vainsencher}, \citenamefont
  {Wenner}, \citenamefont {White}, \citenamefont {Solano}, \citenamefont
  {Neven},\ and\ \citenamefont {Martinis}}]{Barends2016}%
  \BibitemOpen
  \bibfield  {author} {\bibinfo {author} {\bibfnamefont {R.}~\bibnamefont
  {Barends}}, \bibinfo {author} {\bibfnamefont {A.}~\bibnamefont {Shabani}},
  \bibinfo {author} {\bibfnamefont {L.}~\bibnamefont {Lamata}}, \bibinfo
  {author} {\bibfnamefont {J.}~\bibnamefont {Kelly}}, \bibinfo {author}
  {\bibfnamefont {A.}~\bibnamefont {Mezzacapo}}, \bibinfo {author}
  {\bibfnamefont {U.~Las}\ \bibnamefont {Heras}}, \bibinfo {author}
  {\bibfnamefont {R.}~\bibnamefont {Babbush}}, \bibinfo {author} {\bibfnamefont
  {A.~G.}\ \bibnamefont {Fowler}}, \bibinfo {author} {\bibfnamefont
  {B.}~\bibnamefont {Campbell}}, \bibinfo {author} {\bibfnamefont
  {Yu}~\bibnamefont {Chen}}, \bibinfo {author} {\bibfnamefont {Z.}~\bibnamefont
  {Chen}}, \bibinfo {author} {\bibfnamefont {B.}~\bibnamefont {Chiaro}},
  \bibinfo {author} {\bibfnamefont {A.}~\bibnamefont {Dunsworth}}, \bibinfo
  {author} {\bibfnamefont {E.}~\bibnamefont {Jeffrey}}, \bibinfo {author}
  {\bibfnamefont {E.}~\bibnamefont {Lucero}}, \bibinfo {author} {\bibfnamefont
  {A.}~\bibnamefont {Megrant}}, \bibinfo {author} {\bibfnamefont {J.~Y.}\
  \bibnamefont {Mutus}}, \bibinfo {author} {\bibfnamefont {M.}~\bibnamefont
  {Neeley}}, \bibinfo {author} {\bibfnamefont {C.}~\bibnamefont {Neill}},
  \bibinfo {author} {\bibfnamefont {P.~J.~J.}\ \bibnamefont {O'Malley}},
  \bibinfo {author} {\bibfnamefont {C.}~\bibnamefont {Quintana}}, \bibinfo
  {author} {\bibfnamefont {P.}~\bibnamefont {Roushan}}, \bibinfo {author}
  {\bibfnamefont {D.}~\bibnamefont {Sank}}, \bibinfo {author} {\bibfnamefont
  {A.}~\bibnamefont {Vainsencher}}, \bibinfo {author} {\bibfnamefont
  {J.}~\bibnamefont {Wenner}}, \bibinfo {author} {\bibfnamefont {T.~C.}\
  \bibnamefont {White}}, \bibinfo {author} {\bibfnamefont {E.}~\bibnamefont
  {Solano}}, \bibinfo {author} {\bibfnamefont {H.}~\bibnamefont {Neven}}, \
  and\ \bibinfo {author} {\bibfnamefont {John~M.}\ \bibnamefont {Martinis}},\
  }\bibfield  {title} {\enquote {\bibinfo {title} {Digitized adiabatic quantum
  computing with a superconducting circuit},}\ }\href
  {http://dx.doi.org/10.1038/nature17658} {\bibfield  {journal} {\bibinfo
  {journal} {Nature}\ }\textbf {\bibinfo {volume} {534}},\ \bibinfo {pages}
  {222 EP --} (\bibinfo {year} {2016})}\BibitemShut {NoStop}%
\bibitem [{\citenamefont {Casparis}\ \emph {et~al.}(2016)\citenamefont
  {Casparis}, \citenamefont {Larsen}, \citenamefont {Olsen}, \citenamefont
  {Kuemmeth}, \citenamefont {Krogstrup}, \citenamefont {Nyg\aa{}rd},
  \citenamefont {Petersson},\ and\ \citenamefont {Marcus}}]{Casparis2016}%
  \BibitemOpen
  \bibfield  {author} {\bibinfo {author} {\bibfnamefont {L.}~\bibnamefont
  {Casparis}}, \bibinfo {author} {\bibfnamefont {T.~W.}\ \bibnamefont
  {Larsen}}, \bibinfo {author} {\bibfnamefont {M.~S.}\ \bibnamefont {Olsen}},
  \bibinfo {author} {\bibfnamefont {F.}~\bibnamefont {Kuemmeth}}, \bibinfo
  {author} {\bibfnamefont {P.}~\bibnamefont {Krogstrup}}, \bibinfo {author}
  {\bibfnamefont {J.}~\bibnamefont {Nyg\aa{}rd}}, \bibinfo {author}
  {\bibfnamefont {K.~D.}\ \bibnamefont {Petersson}}, \ and\ \bibinfo {author}
  {\bibfnamefont {C.~M.}\ \bibnamefont {Marcus}},\ }\bibfield  {title}
  {\enquote {\bibinfo {title} {Gatemon benchmarking and two-qubit
  operations},}\ }\href {\doibase 10.1103/PhysRevLett.116.150505} {\bibfield
  {journal} {\bibinfo  {journal} {Phys. Rev. Lett.}\ }\textbf {\bibinfo
  {volume} {116}},\ \bibinfo {pages} {150505} (\bibinfo {year}
  {2016})}\BibitemShut {NoStop}%
\bibitem [{\citenamefont {{Otterbach}}\ \emph {et~al.}(2017)\citenamefont
  {{Otterbach}}, \citenamefont {{Manenti}}, \citenamefont {{Alidoust}},
  \citenamefont {{Bestwick}}, \citenamefont {{Block}}, \citenamefont {{Bloom}},
  \citenamefont {{Caldwell}}, \citenamefont {{Didier}}, \citenamefont
  {{Schuyler Fried}}, \citenamefont {{Hong}}, \citenamefont {{Karalekas}},
  \citenamefont {{Osborn}}, \citenamefont {{Papageorge}}, \citenamefont
  {{Peterson}}, \citenamefont {{Prawiroatmodjo}}, \citenamefont {{Rubin}},
  \citenamefont {{Ryan}}, \citenamefont {{Scarabelli}}, \citenamefont
  {{Scheer}}, \citenamefont {{Sete}}, \citenamefont {{Sivarajah}},
  \citenamefont {{Smith}}, \citenamefont {{Staley}}, \citenamefont {{Tezak}},
  \citenamefont {{Zeng}}, \citenamefont {{Hudson}}, \citenamefont {{Johnson}},
  \citenamefont {{Reagor}}, \citenamefont {{da Silva}},\ and\ \citenamefont
  {{Rigetti}}}]{Otterbach2017}%
  \BibitemOpen
  \bibfield  {author} {\bibinfo {author} {\bibfnamefont {J.~S.}\ \bibnamefont
  {{Otterbach}}}, \bibinfo {author} {\bibfnamefont {R.}~\bibnamefont
  {{Manenti}}}, \bibinfo {author} {\bibfnamefont {N.}~\bibnamefont
  {{Alidoust}}}, \bibinfo {author} {\bibfnamefont {A.}~\bibnamefont
  {{Bestwick}}}, \bibinfo {author} {\bibfnamefont {M.}~\bibnamefont {{Block}}},
  \bibinfo {author} {\bibfnamefont {B.}~\bibnamefont {{Bloom}}}, \bibinfo
  {author} {\bibfnamefont {S.}~\bibnamefont {{Caldwell}}}, \bibinfo {author}
  {\bibfnamefont {N.}~\bibnamefont {{Didier}}}, \bibinfo {author}
  {\bibfnamefont {E.}~\bibnamefont {{Schuyler Fried}}}, \bibinfo {author}
  {\bibfnamefont {S.}~\bibnamefont {{Hong}}}, \bibinfo {author} {\bibfnamefont
  {P.}~\bibnamefont {{Karalekas}}}, \bibinfo {author} {\bibfnamefont {C.~B.}\
  \bibnamefont {{Osborn}}}, \bibinfo {author} {\bibfnamefont {A.}~\bibnamefont
  {{Papageorge}}}, \bibinfo {author} {\bibfnamefont {E.~C.}\ \bibnamefont
  {{Peterson}}}, \bibinfo {author} {\bibfnamefont {G.}~\bibnamefont
  {{Prawiroatmodjo}}}, \bibinfo {author} {\bibfnamefont {N.}~\bibnamefont
  {{Rubin}}}, \bibinfo {author} {\bibfnamefont {C.~A.}\ \bibnamefont {{Ryan}}},
  \bibinfo {author} {\bibfnamefont {D.}~\bibnamefont {{Scarabelli}}}, \bibinfo
  {author} {\bibfnamefont {M.}~\bibnamefont {{Scheer}}}, \bibinfo {author}
  {\bibfnamefont {E.~A.}\ \bibnamefont {{Sete}}}, \bibinfo {author}
  {\bibfnamefont {P.}~\bibnamefont {{Sivarajah}}}, \bibinfo {author}
  {\bibfnamefont {R.~S.}\ \bibnamefont {{Smith}}}, \bibinfo {author}
  {\bibfnamefont {A.}~\bibnamefont {{Staley}}}, \bibinfo {author}
  {\bibfnamefont {N.}~\bibnamefont {{Tezak}}}, \bibinfo {author} {\bibfnamefont
  {W.~J.}\ \bibnamefont {{Zeng}}}, \bibinfo {author} {\bibfnamefont
  {A.}~\bibnamefont {{Hudson}}}, \bibinfo {author} {\bibfnamefont {B.~R.}\
  \bibnamefont {{Johnson}}}, \bibinfo {author} {\bibfnamefont {M.}~\bibnamefont
  {{Reagor}}}, \bibinfo {author} {\bibfnamefont {M.~P.}\ \bibnamefont {{da
  Silva}}}, \ and\ \bibinfo {author} {\bibfnamefont {C.}~\bibnamefont
  {{Rigetti}}},\ }\bibfield  {title} {\enquote {\bibinfo {title} {{Unsupervised
  Machine Learning on a Hybrid Quantum Computer}},}\ }\href@noop {} {\bibfield
  {journal} {\bibinfo  {journal} {ArXiv e-prints}\ } (\bibinfo {year}
  {2017})},\ \Eprint {http://arxiv.org/abs/1712.05771} {arXiv:1712.05771
  [quant-ph]} \BibitemShut {NoStop}%
\bibitem [{\citenamefont {{Wang}}\ \emph {et~al.}(2018)\citenamefont {{Wang}},
  \citenamefont {{Li}}, \citenamefont {{Yin}},\ and\ \citenamefont
  {{Zeng}}}]{Wang2018}%
  \BibitemOpen
  \bibfield  {author} {\bibinfo {author} {\bibfnamefont {Y.}~\bibnamefont
  {{Wang}}}, \bibinfo {author} {\bibfnamefont {Y.}~\bibnamefont {{Li}}},
  \bibinfo {author} {\bibfnamefont {Z.-q.}\ \bibnamefont {{Yin}}}, \ and\
  \bibinfo {author} {\bibfnamefont {B.}~\bibnamefont {{Zeng}}},\ }\bibfield
  {title} {\enquote {\bibinfo {title} {{16-qubit IBM universal quantum computer
  can be fully entangled}},}\ }\href@noop {} {\bibfield  {journal} {\bibinfo
  {journal} {ArXiv e-prints}\ } (\bibinfo {year} {2018})},\ \Eprint
  {http://arxiv.org/abs/1801.03782} {arXiv:1801.03782 [quant-ph]} \BibitemShut
  {NoStop}%
\bibitem [{\citenamefont {Petta}\ \emph {et~al.}(2005)\citenamefont {Petta},
  \citenamefont {Johnson}, \citenamefont {Taylor}, \citenamefont {Laird},
  \citenamefont {Yacoby}, \citenamefont {Lukin}, \citenamefont {Marcus},
  \citenamefont {Hanson},\ and\ \citenamefont {Gossard}}]{Petta2005}%
  \BibitemOpen
  \bibfield  {author} {\bibinfo {author} {\bibfnamefont {J.~R.}\ \bibnamefont
  {Petta}}, \bibinfo {author} {\bibfnamefont {A.~C.}\ \bibnamefont {Johnson}},
  \bibinfo {author} {\bibfnamefont {J.~M.}\ \bibnamefont {Taylor}}, \bibinfo
  {author} {\bibfnamefont {E.~A.}\ \bibnamefont {Laird}}, \bibinfo {author}
  {\bibfnamefont {A.}~\bibnamefont {Yacoby}}, \bibinfo {author} {\bibfnamefont
  {M.~D.}\ \bibnamefont {Lukin}}, \bibinfo {author} {\bibfnamefont {C.~M.}\
  \bibnamefont {Marcus}}, \bibinfo {author} {\bibfnamefont {M.~P.}\
  \bibnamefont {Hanson}}, \ and\ \bibinfo {author} {\bibfnamefont {A.~C.}\
  \bibnamefont {Gossard}},\ }\bibfield  {title} {\enquote {\bibinfo {title}
  {Coherent manipulation of coupled electron spins in semiconductor quantum
  dots},}\ }\href {\doibase 10.1126/science.1116955} {\bibfield  {journal}
  {\bibinfo  {journal} {Science}\ }\textbf {\bibinfo {volume} {309}},\ \bibinfo
  {pages} {2180--2184} (\bibinfo {year} {2005})},\ \Eprint
  {http://arxiv.org/abs/http://science.sciencemag.org/content/309/5744/2180.full.pdf}
  {http://science.sciencemag.org/content/309/5744/2180.full.pdf} \BibitemShut
  {NoStop}%
\bibitem [{\citenamefont {Koppens}\ \emph {et~al.}(2006)\citenamefont
  {Koppens}, \citenamefont {Buizert}, \citenamefont {Tielrooij}, \citenamefont
  {Vink}, \citenamefont {Nowack}, \citenamefont {Meunier}, \citenamefont
  {Kouwenhoven},\ and\ \citenamefont {Vandersypen}}]{Koppens2006}%
  \BibitemOpen
  \bibfield  {author} {\bibinfo {author} {\bibfnamefont {F.~H.~L.}\
  \bibnamefont {Koppens}}, \bibinfo {author} {\bibfnamefont {C.}~\bibnamefont
  {Buizert}}, \bibinfo {author} {\bibfnamefont {K.~J.}\ \bibnamefont
  {Tielrooij}}, \bibinfo {author} {\bibfnamefont {I.~T.}\ \bibnamefont {Vink}},
  \bibinfo {author} {\bibfnamefont {K.~C.}\ \bibnamefont {Nowack}}, \bibinfo
  {author} {\bibfnamefont {T.}~\bibnamefont {Meunier}}, \bibinfo {author}
  {\bibfnamefont {L.~P.}\ \bibnamefont {Kouwenhoven}}, \ and\ \bibinfo {author}
  {\bibfnamefont {L.~M.~K.}\ \bibnamefont {Vandersypen}},\ }\bibfield  {title}
  {\enquote {\bibinfo {title} {Driven coherent oscillations of a single
  electron spin in a quantum dot},}\ }\href
  {http://dx.doi.org/10.1038/nature05065} {\bibfield  {journal} {\bibinfo
  {journal} {Nature}\ }\textbf {\bibinfo {volume} {442}},\ \bibinfo {pages}
  {766 EP --} (\bibinfo {year} {2006})}\BibitemShut {NoStop}%
\bibitem [{\citenamefont {Shulman}\ \emph {et~al.}(2012)\citenamefont
  {Shulman}, \citenamefont {Dial}, \citenamefont {Harvey}, \citenamefont
  {Bluhm}, \citenamefont {Umansky},\ and\ \citenamefont
  {Yacoby}}]{Shulman2012}%
  \BibitemOpen
  \bibfield  {author} {\bibinfo {author} {\bibfnamefont {M.~D.}\ \bibnamefont
  {Shulman}}, \bibinfo {author} {\bibfnamefont {O.~E.}\ \bibnamefont {Dial}},
  \bibinfo {author} {\bibfnamefont {S.~P.}\ \bibnamefont {Harvey}}, \bibinfo
  {author} {\bibfnamefont {H.}~\bibnamefont {Bluhm}}, \bibinfo {author}
  {\bibfnamefont {V.}~\bibnamefont {Umansky}}, \ and\ \bibinfo {author}
  {\bibfnamefont {A.}~\bibnamefont {Yacoby}},\ }\bibfield  {title} {\enquote
  {\bibinfo {title} {Demonstration of entanglement of electrostatically coupled
  singlet-triplet qubits},}\ }\href {\doibase 10.1126/science.1217692}
  {\bibfield  {journal} {\bibinfo  {journal} {Science}\ }\textbf {\bibinfo
  {volume} {336}},\ \bibinfo {pages} {202--205} (\bibinfo {year} {2012})},\
  \Eprint
  {http://arxiv.org/abs/http://science.sciencemag.org/content/336/6078/202.full.pdf}
  {http://science.sciencemag.org/content/336/6078/202.full.pdf} \BibitemShut
  {NoStop}%
\bibitem [{\citenamefont {Nichol}\ \emph {et~al.}(2017)\citenamefont {Nichol},
  \citenamefont {Orona}, \citenamefont {Harvey}, \citenamefont {Fallahi},
  \citenamefont {Gardner}, \citenamefont {Manfra},\ and\ \citenamefont
  {Yacoby}}]{Nichol2017}%
  \BibitemOpen
  \bibfield  {author} {\bibinfo {author} {\bibfnamefont {John~M.}\ \bibnamefont
  {Nichol}}, \bibinfo {author} {\bibfnamefont {Lucas~A.}\ \bibnamefont
  {Orona}}, \bibinfo {author} {\bibfnamefont {Shannon~P.}\ \bibnamefont
  {Harvey}}, \bibinfo {author} {\bibfnamefont {Saeed}\ \bibnamefont {Fallahi}},
  \bibinfo {author} {\bibfnamefont {Geoffrey~C.}\ \bibnamefont {Gardner}},
  \bibinfo {author} {\bibfnamefont {Michael~J.}\ \bibnamefont {Manfra}}, \ and\
  \bibinfo {author} {\bibfnamefont {Amir}\ \bibnamefont {Yacoby}},\ }\bibfield
  {title} {\enquote {\bibinfo {title} {High-fidelity entangling gate for
  double-quantum-dot spin qubits},}\ }\href {\doibase
  10.1038/s41534-016-0003-1} {\bibfield  {journal} {\bibinfo  {journal} {npj
  Quantum Information}\ }\textbf {\bibinfo {volume} {3}},\ \bibinfo {pages} {3}
  (\bibinfo {year} {2017})}\BibitemShut {NoStop}%
\bibitem [{\citenamefont {Stockklauser}\ \emph {et~al.}(2017)\citenamefont
  {Stockklauser}, \citenamefont {Scarlino}, \citenamefont {Koski},
  \citenamefont {Gasparinetti}, \citenamefont {Andersen}, \citenamefont
  {Reichl}, \citenamefont {Wegscheider}, \citenamefont {Ihn}, \citenamefont
  {Ensslin},\ and\ \citenamefont {Wallraff}}]{Stockklauser2017}%
  \BibitemOpen
  \bibfield  {author} {\bibinfo {author} {\bibfnamefont {A.}~\bibnamefont
  {Stockklauser}}, \bibinfo {author} {\bibfnamefont {P.}~\bibnamefont
  {Scarlino}}, \bibinfo {author} {\bibfnamefont {J.~V.}\ \bibnamefont {Koski}},
  \bibinfo {author} {\bibfnamefont {S.}~\bibnamefont {Gasparinetti}}, \bibinfo
  {author} {\bibfnamefont {C.~K.}\ \bibnamefont {Andersen}}, \bibinfo {author}
  {\bibfnamefont {C.}~\bibnamefont {Reichl}}, \bibinfo {author} {\bibfnamefont
  {W.}~\bibnamefont {Wegscheider}}, \bibinfo {author} {\bibfnamefont
  {T.}~\bibnamefont {Ihn}}, \bibinfo {author} {\bibfnamefont {K.}~\bibnamefont
  {Ensslin}}, \ and\ \bibinfo {author} {\bibfnamefont {A.}~\bibnamefont
  {Wallraff}},\ }\bibfield  {title} {\enquote {\bibinfo {title} {Strong
  coupling cavity qed with gate-defined double quantum dots enabled by a high
  impedance resonator},}\ }\href {\doibase 10.1103/PhysRevX.7.011030}
  {\bibfield  {journal} {\bibinfo  {journal} {Phys. Rev. X}\ }\textbf {\bibinfo
  {volume} {7}},\ \bibinfo {pages} {011030} (\bibinfo {year}
  {2017})}\BibitemShut {NoStop}%
\bibitem [{\citenamefont {{Watson}}\ \emph {et~al.}(2017)\citenamefont
  {{Watson}}, \citenamefont {{Philips}}, \citenamefont {{Kawakami}},
  \citenamefont {{Ward}}, \citenamefont {{Scarlino}}, \citenamefont
  {{Veldhorst}}, \citenamefont {{Savage}}, \citenamefont {{Lagally}},
  \citenamefont {{Friesen}}, \citenamefont {{Coppersmith}}, \citenamefont
  {{Eriksson}},\ and\ \citenamefont {{Vandersypen}}}]{Watson2017}%
  \BibitemOpen
  \bibfield  {author} {\bibinfo {author} {\bibfnamefont {T.~F.}\ \bibnamefont
  {{Watson}}}, \bibinfo {author} {\bibfnamefont {S.~G.~J.}\ \bibnamefont
  {{Philips}}}, \bibinfo {author} {\bibfnamefont {E.}~\bibnamefont
  {{Kawakami}}}, \bibinfo {author} {\bibfnamefont {D.~R.}\ \bibnamefont
  {{Ward}}}, \bibinfo {author} {\bibfnamefont {P.}~\bibnamefont {{Scarlino}}},
  \bibinfo {author} {\bibfnamefont {M.}~\bibnamefont {{Veldhorst}}}, \bibinfo
  {author} {\bibfnamefont {D.~E.}\ \bibnamefont {{Savage}}}, \bibinfo {author}
  {\bibfnamefont {M.~G.}\ \bibnamefont {{Lagally}}}, \bibinfo {author}
  {\bibfnamefont {M.}~\bibnamefont {{Friesen}}}, \bibinfo {author}
  {\bibfnamefont {S.~N.}\ \bibnamefont {{Coppersmith}}}, \bibinfo {author}
  {\bibfnamefont {M.~A.}\ \bibnamefont {{Eriksson}}}, \ and\ \bibinfo {author}
  {\bibfnamefont {L.~M.~K.}\ \bibnamefont {{Vandersypen}}},\ }\bibfield
  {title} {\enquote {\bibinfo {title} {{A programmable two-qubit quantum
  processor in silicon}},}\ }\href@noop {} {\bibfield  {journal} {\bibinfo
  {journal} {ArXiv e-prints}\ } (\bibinfo {year} {2017})},\ \Eprint
  {http://arxiv.org/abs/1708.04214} {arXiv:1708.04214 [cond-mat.mes-hall]}
  \BibitemShut {NoStop}%
\bibitem [{\citenamefont {Zajac}\ \emph {et~al.}(2018)\citenamefont {Zajac},
  \citenamefont {Sigillito}, \citenamefont {Russ}, \citenamefont {Borjans},
  \citenamefont {Taylor}, \citenamefont {Burkard},\ and\ \citenamefont
  {Petta}}]{Zajac2018}%
  \BibitemOpen
  \bibfield  {author} {\bibinfo {author} {\bibfnamefont {D.~M.}\ \bibnamefont
  {Zajac}}, \bibinfo {author} {\bibfnamefont {A.~J.}\ \bibnamefont
  {Sigillito}}, \bibinfo {author} {\bibfnamefont {M.}~\bibnamefont {Russ}},
  \bibinfo {author} {\bibfnamefont {F.}~\bibnamefont {Borjans}}, \bibinfo
  {author} {\bibfnamefont {J.~M.}\ \bibnamefont {Taylor}}, \bibinfo {author}
  {\bibfnamefont {G.}~\bibnamefont {Burkard}}, \ and\ \bibinfo {author}
  {\bibfnamefont {J.~R.}\ \bibnamefont {Petta}},\ }\bibfield  {title} {\enquote
  {\bibinfo {title} {Resonantly driven cnot gate for electron spins},}\ }\href
  {\doibase 10.1126/science.aao5965} {\bibfield  {journal} {\bibinfo  {journal}
  {Science}\ }\textbf {\bibinfo {volume} {359}},\ \bibinfo {pages} {439--442}
  (\bibinfo {year} {2018})},\ \Eprint
  {http://arxiv.org/abs/http://science.sciencemag.org/content/359/6374/439.full.pdf}
  {http://science.sciencemag.org/content/359/6374/439.full.pdf} \BibitemShut
  {NoStop}%
\bibitem [{\citenamefont {Wesenberg}\ \emph {et~al.}(2009)\citenamefont
  {Wesenberg}, \citenamefont {Ardavan}, \citenamefont {Briggs}, \citenamefont
  {Morton}, \citenamefont {Schoelkopf}, \citenamefont {Schuster},\ and\
  \citenamefont {M\o{}lmer}}]{Wesenberg2009}%
  \BibitemOpen
  \bibfield  {author} {\bibinfo {author} {\bibfnamefont {J.~H.}\ \bibnamefont
  {Wesenberg}}, \bibinfo {author} {\bibfnamefont {A.}~\bibnamefont {Ardavan}},
  \bibinfo {author} {\bibfnamefont {G.~A.~D.}\ \bibnamefont {Briggs}}, \bibinfo
  {author} {\bibfnamefont {J.~J.~L.}\ \bibnamefont {Morton}}, \bibinfo {author}
  {\bibfnamefont {R.~J.}\ \bibnamefont {Schoelkopf}}, \bibinfo {author}
  {\bibfnamefont {D.~I.}\ \bibnamefont {Schuster}}, \ and\ \bibinfo {author}
  {\bibfnamefont {K.}~\bibnamefont {M\o{}lmer}},\ }\bibfield  {title} {\enquote
  {\bibinfo {title} {Quantum computing with an electron spin ensemble},}\
  }\href {\doibase 10.1103/PhysRevLett.103.070502} {\bibfield  {journal}
  {\bibinfo  {journal} {Phys. Rev. Lett.}\ }\textbf {\bibinfo {volume} {103}},\
  \bibinfo {pages} {070502} (\bibinfo {year} {2009})}\BibitemShut {NoStop}%
\bibitem [{\citenamefont {Andr{\'e}}\ \emph {et~al.}(2006)\citenamefont
  {Andr{\'e}}, \citenamefont {DeMille}, \citenamefont {Doyle}, \citenamefont
  {Lukin}, \citenamefont {Maxwell}, \citenamefont {Rabl}, \citenamefont
  {Schoelkopf},\ and\ \citenamefont {Zoller}}]{Andre2006}%
  \BibitemOpen
  \bibfield  {author} {\bibinfo {author} {\bibfnamefont {A.}~\bibnamefont
  {Andr{\'e}}}, \bibinfo {author} {\bibfnamefont {D.}~\bibnamefont {DeMille}},
  \bibinfo {author} {\bibfnamefont {J.~M.}\ \bibnamefont {Doyle}}, \bibinfo
  {author} {\bibfnamefont {M.~D.}\ \bibnamefont {Lukin}}, \bibinfo {author}
  {\bibfnamefont {S.~E.}\ \bibnamefont {Maxwell}}, \bibinfo {author}
  {\bibfnamefont {P.}~\bibnamefont {Rabl}}, \bibinfo {author} {\bibfnamefont
  {R.~J.}\ \bibnamefont {Schoelkopf}}, \ and\ \bibinfo {author} {\bibfnamefont
  {P.}~\bibnamefont {Zoller}},\ }\bibfield  {title} {\enquote {\bibinfo {title}
  {A coherent all-electrical interface between polar molecules and mesoscopic
  superconducting resonators},}\ }\href {http://dx.doi.org/10.1038/nphys386}
  {\bibfield  {journal} {\bibinfo  {journal} {Nature Physics}\ }\textbf
  {\bibinfo {volume} {2}},\ \bibinfo {pages} {636 EP --} (\bibinfo {year}
  {2006})}\BibitemShut {NoStop}%
\bibitem [{\citenamefont {Dehollain}\ \emph {et~al.}(2015)\citenamefont
  {Dehollain}, \citenamefont {Simmons}, \citenamefont {Muhonen}, \citenamefont
  {Kalra}, \citenamefont {Laucht}, \citenamefont {Hudson}, \citenamefont
  {Itoh}, \citenamefont {Jamieson}, \citenamefont {McCallum}, \citenamefont
  {Dzurak},\ and\ \citenamefont {Morello}}]{Dehollain2015}%
  \BibitemOpen
  \bibfield  {author} {\bibinfo {author} {\bibfnamefont {Juan~P.}\ \bibnamefont
  {Dehollain}}, \bibinfo {author} {\bibfnamefont {Stephanie}\ \bibnamefont
  {Simmons}}, \bibinfo {author} {\bibfnamefont {Juha~T.}\ \bibnamefont
  {Muhonen}}, \bibinfo {author} {\bibfnamefont {Rachpon}\ \bibnamefont
  {Kalra}}, \bibinfo {author} {\bibfnamefont {Arne}\ \bibnamefont {Laucht}},
  \bibinfo {author} {\bibfnamefont {Fay}\ \bibnamefont {Hudson}}, \bibinfo
  {author} {\bibfnamefont {Kohei~M.}\ \bibnamefont {Itoh}}, \bibinfo {author}
  {\bibfnamefont {David~N.}\ \bibnamefont {Jamieson}}, \bibinfo {author}
  {\bibfnamefont {Jeffrey~C.}\ \bibnamefont {McCallum}}, \bibinfo {author}
  {\bibfnamefont {Andrew~S.}\ \bibnamefont {Dzurak}}, \ and\ \bibinfo {author}
  {\bibfnamefont {Andrea}\ \bibnamefont {Morello}},\ }\bibfield  {title}
  {\enquote {\bibinfo {title} {Bell's inequality violation with spins in
  silicon},}\ }\href {http://dx.doi.org/10.1038/nnano.2015.262} {\bibfield
  {journal} {\bibinfo  {journal} {Nature Nanotechnology}\ }\textbf {\bibinfo
  {volume} {11}},\ \bibinfo {pages} {242 EP --} (\bibinfo {year}
  {2015})}\BibitemShut {NoStop}%
\bibitem [{\citenamefont {Tettamanzi}\ \emph {et~al.}(2017)\citenamefont
  {Tettamanzi}, \citenamefont {Hile}, \citenamefont {House}, \citenamefont
  {Fuechsle}, \citenamefont {Rogge},\ and\ \citenamefont
  {Simmons}}]{Tettamanzi2017}%
  \BibitemOpen
  \bibfield  {author} {\bibinfo {author} {\bibfnamefont {Giuseppe~Carlo}\
  \bibnamefont {Tettamanzi}}, \bibinfo {author} {\bibfnamefont {Samuel~James}\
  \bibnamefont {Hile}}, \bibinfo {author} {\bibfnamefont {Matthew~Gregory}\
  \bibnamefont {House}}, \bibinfo {author} {\bibfnamefont {Martin}\
  \bibnamefont {Fuechsle}}, \bibinfo {author} {\bibfnamefont {Sven}\
  \bibnamefont {Rogge}}, \ and\ \bibinfo {author} {\bibfnamefont {Michelle~Y.}\
  \bibnamefont {Simmons}},\ }\bibfield  {title} {\enquote {\bibinfo {title}
  {Probing the quantum states of a single atom transistor at microwave
  frequencies},}\ }\bibfield  {booktitle} {\emph {\bibinfo {booktitle} {ACS
  Nano}},\ }\href {\doibase 10.1021/acsnano.6b06362} {\bibfield  {journal}
  {\bibinfo  {journal} {ACS Nano}\ }\textbf {\bibinfo {volume} {11}},\ \bibinfo
  {pages} {2444--2451} (\bibinfo {year} {2017})}\BibitemShut {NoStop}%
\bibitem [{\citenamefont {Aspelmeyer}\ \emph
  {et~al.}(2014{\natexlab{a}})\citenamefont {Aspelmeyer}, \citenamefont
  {Kippenberg},\ and\ \citenamefont {Marquardt}}]{Aspelmeyer2013}%
  \BibitemOpen
  \bibfield  {author} {\bibinfo {author} {\bibfnamefont {Markus}\ \bibnamefont
  {Aspelmeyer}}, \bibinfo {author} {\bibfnamefont {Tobias~J.}\ \bibnamefont
  {Kippenberg}}, \ and\ \bibinfo {author} {\bibfnamefont {Florian}\
  \bibnamefont {Marquardt}},\ }\bibfield  {title} {\enquote {\bibinfo {title}
  {Cavity optomechanics},}\ }\href {\doibase 10.1103/RevModPhys.86.1391}
  {\bibfield  {journal} {\bibinfo  {journal} {Rev. Mod. Phys.}\ }\textbf
  {\bibinfo {volume} {86}},\ \bibinfo {pages} {1391--1452} (\bibinfo {year}
  {2014}{\natexlab{a}})}\BibitemShut {NoStop}%
\bibitem [{\citenamefont {Aspelmeyer}\ \emph
  {et~al.}(2014{\natexlab{b}})\citenamefont {Aspelmeyer}, \citenamefont
  {Kippenberg},\ and\ \citenamefont {Marquardt}}]{CavityOptomechSpringer}%
  \BibitemOpen
  \bibinfo {editor} {\bibfnamefont {Markus}\ \bibnamefont {Aspelmeyer}},
  \bibinfo {editor} {\bibfnamefont {Tobias~J.}\ \bibnamefont {Kippenberg}}, \
  and\ \bibinfo {editor} {\bibfnamefont {Florian}\ \bibnamefont {Marquardt}},\
  eds.,\ \href@noop {} {\emph {\bibinfo {title} {Cavity Optomechanics: Nano-
  and Micromechanical Resonators Interacting with Light}}},\ \bibinfo {edition}
  {1st}\ ed.,\ Quantum Science and Technology\ (\bibinfo  {publisher}
  {Springer-Verlag Berlin Heidelberg},\ \bibinfo {year} {2014})\BibitemShut
  {NoStop}%
\bibitem [{\citenamefont {Danilishin}\ and\ \citenamefont
  {Khalili}(2012)}]{Danilishin2012}%
  \BibitemOpen
  \bibfield  {author} {\bibinfo {author} {\bibfnamefont {Stefan~L.}\
  \bibnamefont {Danilishin}}\ and\ \bibinfo {author} {\bibfnamefont
  {Farid~Ya.}\ \bibnamefont {Khalili}},\ }\bibfield  {title} {\enquote
  {\bibinfo {title} {Quantum measurement theory in gravitational-wave
  detectors},}\ }\href {\doibase 10.12942/lrr-2012-5} {\bibfield  {journal}
  {\bibinfo  {journal} {Living Reviews in Relativity}\ }\textbf {\bibinfo
  {volume} {15}} (\bibinfo {year} {2012}),\ 10.12942/lrr-2012-5}\BibitemShut
  {NoStop}%
\bibitem [{\citenamefont {Arcizet}\ \emph {et~al.}(2006)\citenamefont
  {Arcizet}, \citenamefont {Cohadon}, \citenamefont {Briant}, \citenamefont
  {Pinard},\ and\ \citenamefont {Heidmann}}]{Arcizet2006}%
  \BibitemOpen
  \bibfield  {author} {\bibinfo {author} {\bibfnamefont {O.}~\bibnamefont
  {Arcizet}}, \bibinfo {author} {\bibfnamefont {P.~F.}\ \bibnamefont
  {Cohadon}}, \bibinfo {author} {\bibfnamefont {T.}~\bibnamefont {Briant}},
  \bibinfo {author} {\bibfnamefont {M.}~\bibnamefont {Pinard}}, \ and\ \bibinfo
  {author} {\bibfnamefont {A.}~\bibnamefont {Heidmann}},\ }\bibfield  {title}
  {\enquote {\bibinfo {title} {Radiation-pressure cooling and optomechanical
  instability of a micromirror},}\ }\href
  {http://dx.doi.org/10.1038/nature05244} {\bibfield  {journal} {\bibinfo
  {journal} {Nature}\ }\textbf {\bibinfo {volume} {444}},\ \bibinfo {pages}
  {71--74} (\bibinfo {year} {2006})}\BibitemShut {NoStop}%
\bibitem [{\citenamefont {Schliesser}\ \emph {et~al.}(2006)\citenamefont
  {Schliesser}, \citenamefont {Del'Haye}, \citenamefont {Nooshi}, \citenamefont
  {Vahala},\ and\ \citenamefont {Kippenberg}}]{Schliesser2006}%
  \BibitemOpen
  \bibfield  {author} {\bibinfo {author} {\bibfnamefont {A.}~\bibnamefont
  {Schliesser}}, \bibinfo {author} {\bibfnamefont {P.}~\bibnamefont
  {Del'Haye}}, \bibinfo {author} {\bibfnamefont {N.}~\bibnamefont {Nooshi}},
  \bibinfo {author} {\bibfnamefont {K.~J.}\ \bibnamefont {Vahala}}, \ and\
  \bibinfo {author} {\bibfnamefont {T.~J.}\ \bibnamefont {Kippenberg}},\
  }\bibfield  {title} {\enquote {\bibinfo {title} {Radiation pressure cooling
  of a micromechanical oscillator using dynamical backaction},}\ }\href
  {\doibase 10.1103/PhysRevLett.97.243905} {\bibfield  {journal} {\bibinfo
  {journal} {Phys. Rev. Lett.}\ }\textbf {\bibinfo {volume} {97}},\ \bibinfo
  {pages} {243905} (\bibinfo {year} {2006})}\BibitemShut {NoStop}%
\bibitem [{\citenamefont {Gigan}\ \emph {et~al.}(2006)\citenamefont {Gigan},
  \citenamefont {Bohm}, \citenamefont {Paternostro}, \citenamefont {Blaser},
  \citenamefont {Langer}, \citenamefont {Hertzberg}, \citenamefont {Schwab},
  \citenamefont {Bauerle}, \citenamefont {Aspelmeyer},\ and\ \citenamefont
  {Zeilinger}}]{Gigan2006}%
  \BibitemOpen
  \bibfield  {author} {\bibinfo {author} {\bibfnamefont {S.}~\bibnamefont
  {Gigan}}, \bibinfo {author} {\bibfnamefont {H.~R.}\ \bibnamefont {Bohm}},
  \bibinfo {author} {\bibfnamefont {M.}~\bibnamefont {Paternostro}}, \bibinfo
  {author} {\bibfnamefont {F.}~\bibnamefont {Blaser}}, \bibinfo {author}
  {\bibfnamefont {G.}~\bibnamefont {Langer}}, \bibinfo {author} {\bibfnamefont
  {J.~B.}\ \bibnamefont {Hertzberg}}, \bibinfo {author} {\bibfnamefont {K.~C.}\
  \bibnamefont {Schwab}}, \bibinfo {author} {\bibfnamefont {D.}~\bibnamefont
  {Bauerle}}, \bibinfo {author} {\bibfnamefont {M.}~\bibnamefont {Aspelmeyer}},
  \ and\ \bibinfo {author} {\bibfnamefont {A.}~\bibnamefont {Zeilinger}},\
  }\bibfield  {title} {\enquote {\bibinfo {title} {Self-cooling of a
  micromirror by radiation pressure},}\ }\href
  {http://dx.doi.org/10.1038/nature05273} {\bibfield  {journal} {\bibinfo
  {journal} {Nature}\ }\textbf {\bibinfo {volume} {444}},\ \bibinfo {pages}
  {67--70} (\bibinfo {year} {2006})}\BibitemShut {NoStop}%
\bibitem [{\citenamefont {Thompson}\ \emph {et~al.}(2008)\citenamefont
  {Thompson}, \citenamefont {Zwickl}, \citenamefont {Jayich}, \citenamefont
  {Marquardt}, \citenamefont {Girvin},\ and\ \citenamefont
  {Harris}}]{Thompson2008}%
  \BibitemOpen
  \bibfield  {author} {\bibinfo {author} {\bibfnamefont {J.~D.}\ \bibnamefont
  {Thompson}}, \bibinfo {author} {\bibfnamefont {B.~M.}\ \bibnamefont
  {Zwickl}}, \bibinfo {author} {\bibfnamefont {A.~M.}\ \bibnamefont {Jayich}},
  \bibinfo {author} {\bibfnamefont {Florian}\ \bibnamefont {Marquardt}},
  \bibinfo {author} {\bibfnamefont {S.~M.}\ \bibnamefont {Girvin}}, \ and\
  \bibinfo {author} {\bibfnamefont {J.~G.~E.}\ \bibnamefont {Harris}},\
  }\bibfield  {title} {\enquote {\bibinfo {title} {Strong dispersive coupling
  of a high-finesse cavity to a micromechanical membrane},}\ }\href
  {http://dx.doi.org/10.1038/nature06715} {\bibfield  {journal} {\bibinfo
  {journal} {Nature}\ }\textbf {\bibinfo {volume} {452}},\ \bibinfo {pages}
  {72--75} (\bibinfo {year} {2008})}\BibitemShut {NoStop}%
\bibitem [{\citenamefont {Wilson}\ \emph {et~al.}(2009)\citenamefont {Wilson},
  \citenamefont {Regal}, \citenamefont {Papp},\ and\ \citenamefont
  {Kimble}}]{Wilson2009}%
  \BibitemOpen
  \bibfield  {author} {\bibinfo {author} {\bibfnamefont {D.~J.}\ \bibnamefont
  {Wilson}}, \bibinfo {author} {\bibfnamefont {C.~A.}\ \bibnamefont {Regal}},
  \bibinfo {author} {\bibfnamefont {S.~B.}\ \bibnamefont {Papp}}, \ and\
  \bibinfo {author} {\bibfnamefont {H.~J.}\ \bibnamefont {Kimble}},\ }\bibfield
   {title} {\enquote {\bibinfo {title} {Cavity optomechanics with
  stoichiometric sin films},}\ }\href {\doibase 10.1103/PhysRevLett.103.207204}
  {\bibfield  {journal} {\bibinfo  {journal} {Phys. Rev. Lett.}\ }\textbf
  {\bibinfo {volume} {103}},\ \bibinfo {pages} {207204} (\bibinfo {year}
  {2009})}\BibitemShut {NoStop}%
\bibitem [{\citenamefont {Chan}\ \emph {et~al.}(2011)\citenamefont {Chan},
  \citenamefont {Alegre}, \citenamefont {Safavi-Naeini}, \citenamefont {Hill},
  \citenamefont {Krause}, \citenamefont {Groblacher}, \citenamefont
  {Aspelmeyer},\ and\ \citenamefont {Painter}}]{Chan2011}%
  \BibitemOpen
  \bibfield  {author} {\bibinfo {author} {\bibfnamefont {Jasper}\ \bibnamefont
  {Chan}}, \bibinfo {author} {\bibfnamefont {T.~P.~Mayer}\ \bibnamefont
  {Alegre}}, \bibinfo {author} {\bibfnamefont {Amir~H.}\ \bibnamefont
  {Safavi-Naeini}}, \bibinfo {author} {\bibfnamefont {Jeff~T.}\ \bibnamefont
  {Hill}}, \bibinfo {author} {\bibfnamefont {Alex}\ \bibnamefont {Krause}},
  \bibinfo {author} {\bibfnamefont {Simon}\ \bibnamefont {Groblacher}},
  \bibinfo {author} {\bibfnamefont {Markus}\ \bibnamefont {Aspelmeyer}}, \ and\
  \bibinfo {author} {\bibfnamefont {Oskar}\ \bibnamefont {Painter}},\
  }\bibfield  {title} {\enquote {\bibinfo {title} {Laser cooling of a
  nanomechanical oscillator into its quantum ground state},}\ }\href
  {http://dx.doi.org/10.1038/nature10461} {\bibfield  {journal} {\bibinfo
  {journal} {Nature}\ }\textbf {\bibinfo {volume} {478}},\ \bibinfo {pages}
  {89--92} (\bibinfo {year} {2011})}\BibitemShut {NoStop}%
\bibitem [{\citenamefont {Verhagen}\ \emph {et~al.}(2012)\citenamefont
  {Verhagen}, \citenamefont {Deleglise}, \citenamefont {Weis}, \citenamefont
  {Schliesser},\ and\ \citenamefont {Kippenberg}}]{Verhagen2012}%
  \BibitemOpen
  \bibfield  {author} {\bibinfo {author} {\bibfnamefont {E.}~\bibnamefont
  {Verhagen}}, \bibinfo {author} {\bibfnamefont {S.}~\bibnamefont {Deleglise}},
  \bibinfo {author} {\bibfnamefont {S.}~\bibnamefont {Weis}}, \bibinfo {author}
  {\bibfnamefont {A.}~\bibnamefont {Schliesser}}, \ and\ \bibinfo {author}
  {\bibfnamefont {T.~J.}\ \bibnamefont {Kippenberg}},\ }\bibfield  {title}
  {\enquote {\bibinfo {title} {Quantum-coherent coupling of a mechanical
  oscillator to an optical cavity mode},}\ }\href
  {http://dx.doi.org/10.1038/nature10787} {\bibfield  {journal} {\bibinfo
  {journal} {Nature}\ }\textbf {\bibinfo {volume} {482}},\ \bibinfo {pages}
  {63--67} (\bibinfo {year} {2012})}\BibitemShut {NoStop}%
\bibitem [{\citenamefont {Underwood}\ \emph {et~al.}(2015)\citenamefont
  {Underwood}, \citenamefont {Mason}, \citenamefont {Lee}, \citenamefont {Xu},
  \citenamefont {Jiang}, \citenamefont {Shkarin}, \citenamefont {B\o{}rkje},
  \citenamefont {Girvin},\ and\ \citenamefont {Harris}}]{Underwood2015}%
  \BibitemOpen
  \bibfield  {author} {\bibinfo {author} {\bibfnamefont {M.}~\bibnamefont
  {Underwood}}, \bibinfo {author} {\bibfnamefont {D.}~\bibnamefont {Mason}},
  \bibinfo {author} {\bibfnamefont {D.}~\bibnamefont {Lee}}, \bibinfo {author}
  {\bibfnamefont {H.}~\bibnamefont {Xu}}, \bibinfo {author} {\bibfnamefont
  {L.}~\bibnamefont {Jiang}}, \bibinfo {author} {\bibfnamefont {A.~B.}\
  \bibnamefont {Shkarin}}, \bibinfo {author} {\bibfnamefont {K.}~\bibnamefont
  {B\o{}rkje}}, \bibinfo {author} {\bibfnamefont {S.~M.}\ \bibnamefont
  {Girvin}}, \ and\ \bibinfo {author} {\bibfnamefont {J.~G.~E.}\ \bibnamefont
  {Harris}},\ }\bibfield  {title} {\enquote {\bibinfo {title} {Measurement of
  the motional sidebands of a nanogram-scale oscillator in the quantum
  regime},}\ }\href {\doibase 10.1103/PhysRevA.92.061801} {\bibfield  {journal}
  {\bibinfo  {journal} {Phys. Rev. A}\ }\textbf {\bibinfo {volume} {92}},\
  \bibinfo {pages} {061801} (\bibinfo {year} {2015})}\BibitemShut {NoStop}%
\bibitem [{\citenamefont {Peterson}\ \emph {et~al.}(2016)\citenamefont
  {Peterson}, \citenamefont {Purdy}, \citenamefont {Kampel}, \citenamefont
  {Andrews}, \citenamefont {Yu}, \citenamefont {Lehnert},\ and\ \citenamefont
  {Regal}}]{Peterson2016}%
  \BibitemOpen
  \bibfield  {author} {\bibinfo {author} {\bibfnamefont {R.~W.}\ \bibnamefont
  {Peterson}}, \bibinfo {author} {\bibfnamefont {T.~P.}\ \bibnamefont {Purdy}},
  \bibinfo {author} {\bibfnamefont {N.~S.}\ \bibnamefont {Kampel}}, \bibinfo
  {author} {\bibfnamefont {R.~W.}\ \bibnamefont {Andrews}}, \bibinfo {author}
  {\bibfnamefont {P.-L.}\ \bibnamefont {Yu}}, \bibinfo {author} {\bibfnamefont
  {K.~W.}\ \bibnamefont {Lehnert}}, \ and\ \bibinfo {author} {\bibfnamefont
  {C.~A.}\ \bibnamefont {Regal}},\ }\bibfield  {title} {\enquote {\bibinfo
  {title} {Laser cooling of a micromechanical membrane to the quantum
  backaction limit},}\ }\href {\doibase 10.1103/PhysRevLett.116.063601}
  {\bibfield  {journal} {\bibinfo  {journal} {Phys. Rev. Lett.}\ }\textbf
  {\bibinfo {volume} {116}},\ \bibinfo {pages} {063601} (\bibinfo {year}
  {2016})}\BibitemShut {NoStop}%
\bibitem [{\citenamefont {Nielsen}\ \emph {et~al.}(2017)\citenamefont
  {Nielsen}, \citenamefont {Tsaturyan}, \citenamefont {M{\o}ller},
  \citenamefont {Polzik},\ and\ \citenamefont {Schliesser}}]{Nielsen2017}%
  \BibitemOpen
  \bibfield  {author} {\bibinfo {author} {\bibfnamefont {William
  Hvidtfelt~Padk{\ae}r}\ \bibnamefont {Nielsen}}, \bibinfo {author}
  {\bibfnamefont {Yeghishe}\ \bibnamefont {Tsaturyan}}, \bibinfo {author}
  {\bibfnamefont {Christoffer~Bo}\ \bibnamefont {M{\o}ller}}, \bibinfo {author}
  {\bibfnamefont {Eugene~S.}\ \bibnamefont {Polzik}}, \ and\ \bibinfo {author}
  {\bibfnamefont {Albert}\ \bibnamefont {Schliesser}},\ }\bibfield  {title}
  {\enquote {\bibinfo {title} {Multimode optomechanical system in the quantum
  regime},}\ }\href {\doibase 10.1073/pnas.1608412114} {\bibfield  {journal}
  {\bibinfo  {journal} {Proceedings of the National Academy of Sciences}\
  }\textbf {\bibinfo {volume} {114}},\ \bibinfo {pages} {62--66} (\bibinfo
  {year} {2017})},\ \Eprint
  {http://arxiv.org/abs/http://www.pnas.org/content/114/1/62.full.pdf}
  {http://www.pnas.org/content/114/1/62.full.pdf} \BibitemShut {NoStop}%
\bibitem [{\citenamefont {O'Connell}\ \emph {et~al.}(2010)\citenamefont
  {O'Connell}, \citenamefont {Hofheinz}, \citenamefont {Ansmann}, \citenamefont
  {Bialczak}, \citenamefont {Lenander}, \citenamefont {Lucero}, \citenamefont
  {Neeley}, \citenamefont {Sank}, \citenamefont {Wang}, \citenamefont {Weides},
  \citenamefont {Wenner}, \citenamefont {Martinis},\ and\ \citenamefont
  {Cleland}}]{OConnell2010}%
  \BibitemOpen
  \bibfield  {author} {\bibinfo {author} {\bibfnamefont {A.~D.}\ \bibnamefont
  {O'Connell}}, \bibinfo {author} {\bibfnamefont {M.}~\bibnamefont {Hofheinz}},
  \bibinfo {author} {\bibfnamefont {M.}~\bibnamefont {Ansmann}}, \bibinfo
  {author} {\bibfnamefont {Radoslaw~C.}\ \bibnamefont {Bialczak}}, \bibinfo
  {author} {\bibfnamefont {M.}~\bibnamefont {Lenander}}, \bibinfo {author}
  {\bibfnamefont {Erik}\ \bibnamefont {Lucero}}, \bibinfo {author}
  {\bibfnamefont {M.}~\bibnamefont {Neeley}}, \bibinfo {author} {\bibfnamefont
  {D.}~\bibnamefont {Sank}}, \bibinfo {author} {\bibfnamefont {H.}~\bibnamefont
  {Wang}}, \bibinfo {author} {\bibfnamefont {M.}~\bibnamefont {Weides}},
  \bibinfo {author} {\bibfnamefont {J.}~\bibnamefont {Wenner}}, \bibinfo
  {author} {\bibfnamefont {John~M.}\ \bibnamefont {Martinis}}, \ and\ \bibinfo
  {author} {\bibfnamefont {A.~N.}\ \bibnamefont {Cleland}},\ }\bibfield
  {title} {\enquote {\bibinfo {title} {Quantum ground state and single-phonon
  control of a mechanical resonator},}\ }\href
  {http://dx.doi.org/10.1038/nature08967} {\bibfield  {journal} {\bibinfo
  {journal} {Nature}\ }\textbf {\bibinfo {volume} {464}},\ \bibinfo {pages}
  {697--703} (\bibinfo {year} {2010})}\BibitemShut {NoStop}%
\bibitem [{\citenamefont {Rocheleau}\ \emph {et~al.}(2010)\citenamefont
  {Rocheleau}, \citenamefont {Ndukum}, \citenamefont {Macklin}, \citenamefont
  {Hertzberg}, \citenamefont {Clerk},\ and\ \citenamefont
  {Schwab}}]{Rocheleau2010}%
  \BibitemOpen
  \bibfield  {author} {\bibinfo {author} {\bibfnamefont {T.}~\bibnamefont
  {Rocheleau}}, \bibinfo {author} {\bibfnamefont {T.}~\bibnamefont {Ndukum}},
  \bibinfo {author} {\bibfnamefont {C.}~\bibnamefont {Macklin}}, \bibinfo
  {author} {\bibfnamefont {J.~B.}\ \bibnamefont {Hertzberg}}, \bibinfo {author}
  {\bibfnamefont {A.~A.}\ \bibnamefont {Clerk}}, \ and\ \bibinfo {author}
  {\bibfnamefont {K.~C.}\ \bibnamefont {Schwab}},\ }\bibfield  {title}
  {\enquote {\bibinfo {title} {Preparation and detection of a mechanical
  resonator near the ground state of motion},}\ }\href
  {http://dx.doi.org/10.1038/nature08681} {\bibfield  {journal} {\bibinfo
  {journal} {Nature}\ }\textbf {\bibinfo {volume} {463}},\ \bibinfo {pages}
  {72--75} (\bibinfo {year} {2010})}\BibitemShut {NoStop}%
\bibitem [{\citenamefont {Teufel}\ \emph {et~al.}(2011)\citenamefont {Teufel},
  \citenamefont {Donner}, \citenamefont {Li}, \citenamefont {Harlow},
  \citenamefont {Allman}, \citenamefont {Cicak}, \citenamefont {Sirois},
  \citenamefont {Whittaker}, \citenamefont {Lehnert},\ and\ \citenamefont
  {Simmonds}}]{Teufel2011}%
  \BibitemOpen
  \bibfield  {author} {\bibinfo {author} {\bibfnamefont {J.~D.}\ \bibnamefont
  {Teufel}}, \bibinfo {author} {\bibfnamefont {T.}~\bibnamefont {Donner}},
  \bibinfo {author} {\bibfnamefont {Dale}\ \bibnamefont {Li}}, \bibinfo
  {author} {\bibfnamefont {J.~W.}\ \bibnamefont {Harlow}}, \bibinfo {author}
  {\bibfnamefont {M.~S.}\ \bibnamefont {Allman}}, \bibinfo {author}
  {\bibfnamefont {K.}~\bibnamefont {Cicak}}, \bibinfo {author} {\bibfnamefont
  {A.~J.}\ \bibnamefont {Sirois}}, \bibinfo {author} {\bibfnamefont {J.~D.}\
  \bibnamefont {Whittaker}}, \bibinfo {author} {\bibfnamefont {K.~W.}\
  \bibnamefont {Lehnert}}, \ and\ \bibinfo {author} {\bibfnamefont {R.~W.}\
  \bibnamefont {Simmonds}},\ }\bibfield  {title} {\enquote {\bibinfo {title}
  {Sideband cooling of micromechanical motion to the quantum ground state},}\
  }\href {http://dx.doi.org/10.1038/nature10261} {\bibfield  {journal}
  {\bibinfo  {journal} {Nature}\ }\textbf {\bibinfo {volume} {475}},\ \bibinfo
  {pages} {359--363} (\bibinfo {year} {2011})}\BibitemShut {NoStop}%
\bibitem [{\citenamefont {Palomaki}\ \emph
  {et~al.}(2013{\natexlab{a}})\citenamefont {Palomaki}, \citenamefont {Harlow},
  \citenamefont {Teufel}, \citenamefont {Simmonds},\ and\ \citenamefont
  {Lehnert}}]{Palomaki2013}%
  \BibitemOpen
  \bibfield  {author} {\bibinfo {author} {\bibfnamefont {T.~A.}\ \bibnamefont
  {Palomaki}}, \bibinfo {author} {\bibfnamefont {J.~W.}\ \bibnamefont
  {Harlow}}, \bibinfo {author} {\bibfnamefont {J.~D.}\ \bibnamefont {Teufel}},
  \bibinfo {author} {\bibfnamefont {R.~W.}\ \bibnamefont {Simmonds}}, \ and\
  \bibinfo {author} {\bibfnamefont {K.~W.}\ \bibnamefont {Lehnert}},\
  }\bibfield  {title} {\enquote {\bibinfo {title} {Coherent state transfer
  between itinerant microwave fields and a mechanical oscillator},}\ }\href
  {http://dx.doi.org/10.1038/nature11915} {\bibfield  {journal} {\bibinfo
  {journal} {Nature}\ }\textbf {\bibinfo {volume} {495}},\ \bibinfo {pages}
  {210--214} (\bibinfo {year} {2013}{\natexlab{a}})}\BibitemShut {NoStop}%
\bibitem [{\citenamefont {Reed}\ \emph {et~al.}(2017)\citenamefont {Reed},
  \citenamefont {Mayer}, \citenamefont {Teufel}, \citenamefont {Burkhart},
  \citenamefont {Pfaff}, \citenamefont {Reagor}, \citenamefont {Sletten},
  \citenamefont {Ma}, \citenamefont {Schoelkopf}, \citenamefont {Knill},\ and\
  \citenamefont {Lehnert}}]{Reed2017}%
  \BibitemOpen
  \bibfield  {author} {\bibinfo {author} {\bibfnamefont {A.~P.}\ \bibnamefont
  {Reed}}, \bibinfo {author} {\bibfnamefont {K.~H.}\ \bibnamefont {Mayer}},
  \bibinfo {author} {\bibfnamefont {J.~D.}\ \bibnamefont {Teufel}}, \bibinfo
  {author} {\bibfnamefont {L.~D.}\ \bibnamefont {Burkhart}}, \bibinfo {author}
  {\bibfnamefont {W.}~\bibnamefont {Pfaff}}, \bibinfo {author} {\bibfnamefont
  {M.}~\bibnamefont {Reagor}}, \bibinfo {author} {\bibfnamefont
  {L.}~\bibnamefont {Sletten}}, \bibinfo {author} {\bibfnamefont
  {X.}~\bibnamefont {Ma}}, \bibinfo {author} {\bibfnamefont {R.~J.}\
  \bibnamefont {Schoelkopf}}, \bibinfo {author} {\bibfnamefont
  {E.}~\bibnamefont {Knill}}, \ and\ \bibinfo {author} {\bibfnamefont {K.~W.}\
  \bibnamefont {Lehnert}},\ }\bibfield  {title} {\enquote {\bibinfo {title}
  {Faithful conversion of propagating quantum information to mechanical
  motion},}\ }\href {http://dx.doi.org/10.1038/nphys4251} {\bibfield  {journal}
  {\bibinfo  {journal} {Nature Physics}\ ,\ \bibinfo {pages} {EP --}} (\bibinfo
  {year} {2017})}\BibitemShut {NoStop}%
\bibitem [{\citenamefont {Palomaki}\ \emph
  {et~al.}(2013{\natexlab{b}})\citenamefont {Palomaki}, \citenamefont {Teufel},
  \citenamefont {Simmonds},\ and\ \citenamefont {Lehnert}}]{Palomaki2013b}%
  \BibitemOpen
  \bibfield  {author} {\bibinfo {author} {\bibfnamefont {T.~A.}\ \bibnamefont
  {Palomaki}}, \bibinfo {author} {\bibfnamefont {J.~D.}\ \bibnamefont
  {Teufel}}, \bibinfo {author} {\bibfnamefont {R.~W.}\ \bibnamefont
  {Simmonds}}, \ and\ \bibinfo {author} {\bibfnamefont {K.~W.}\ \bibnamefont
  {Lehnert}},\ }\bibfield  {title} {\enquote {\bibinfo {title} {Entangling
  mechanical motion with microwave fields},}\ }\href {\doibase
  10.1126/science.1244563} {\bibfield  {journal} {\bibinfo  {journal}
  {Science}\ }\textbf {\bibinfo {volume} {342}},\ \bibinfo {pages} {710--713}
  (\bibinfo {year} {2013}{\natexlab{b}})},\ \Eprint
  {http://arxiv.org/abs/http://science.sciencemag.org/content/342/6159/710.full.pdf}
  {http://science.sciencemag.org/content/342/6159/710.full.pdf} \BibitemShut
  {NoStop}%
\bibitem [{\citenamefont {Massel}\ \emph {et~al.}(2011)\citenamefont {Massel},
  \citenamefont {Heikkila}, \citenamefont {Pirkkalainen}, \citenamefont {Cho},
  \citenamefont {Saloniemi}, \citenamefont {Hakonen},\ and\ \citenamefont
  {Sillanpaa}}]{Massel2011}%
  \BibitemOpen
  \bibfield  {author} {\bibinfo {author} {\bibfnamefont {F.}~\bibnamefont
  {Massel}}, \bibinfo {author} {\bibfnamefont {T.~T.}\ \bibnamefont
  {Heikkila}}, \bibinfo {author} {\bibfnamefont {J.~M.}\ \bibnamefont
  {Pirkkalainen}}, \bibinfo {author} {\bibfnamefont {S.~U.}\ \bibnamefont
  {Cho}}, \bibinfo {author} {\bibfnamefont {H.}~\bibnamefont {Saloniemi}},
  \bibinfo {author} {\bibfnamefont {P.~J.}\ \bibnamefont {Hakonen}}, \ and\
  \bibinfo {author} {\bibfnamefont {M.~A.}\ \bibnamefont {Sillanpaa}},\
  }\bibfield  {title} {\enquote {\bibinfo {title} {Microwave amplification with
  nanomechanical resonators},}\ }\href {http://dx.doi.org/10.1038/nature10628}
  {\bibfield  {journal} {\bibinfo  {journal} {Nature}\ }\textbf {\bibinfo
  {volume} {480}},\ \bibinfo {pages} {351--354} (\bibinfo {year}
  {2011})}\BibitemShut {NoStop}%
\bibitem [{\citenamefont {Ockeloen-Korppi}\ \emph {et~al.}(2016)\citenamefont
  {Ockeloen-Korppi}, \citenamefont {Damsk\"agg}, \citenamefont {Pirkkalainen},
  \citenamefont {Heikkil\"a}, \citenamefont {Massel},\ and\ \citenamefont
  {Sillanp\"a\"a}}]{Ockeloen2016}%
  \BibitemOpen
  \bibfield  {author} {\bibinfo {author} {\bibfnamefont {C.~F.}\ \bibnamefont
  {Ockeloen-Korppi}}, \bibinfo {author} {\bibfnamefont {E.}~\bibnamefont
  {Damsk\"agg}}, \bibinfo {author} {\bibfnamefont {J.-M.}\ \bibnamefont
  {Pirkkalainen}}, \bibinfo {author} {\bibfnamefont {T.~T.}\ \bibnamefont
  {Heikkil\"a}}, \bibinfo {author} {\bibfnamefont {F.}~\bibnamefont {Massel}},
  \ and\ \bibinfo {author} {\bibfnamefont {M.~A.}\ \bibnamefont
  {Sillanp\"a\"a}},\ }\bibfield  {title} {\enquote {\bibinfo {title} {Low-noise
  amplification and frequency conversion with a multiport microwave
  optomechanical device},}\ }\href {\doibase 10.1103/PhysRevX.6.041024}
  {\bibfield  {journal} {\bibinfo  {journal} {Phys. Rev. X}\ }\textbf {\bibinfo
  {volume} {6}},\ \bibinfo {pages} {041024} (\bibinfo {year}
  {2016})}\BibitemShut {NoStop}%
\bibitem [{\citenamefont {T{\'o}th}\ \emph {et~al.}(2017)\citenamefont
  {T{\'o}th}, \citenamefont {Bernier}, \citenamefont {Nunnenkamp},
  \citenamefont {Feofanov},\ and\ \citenamefont {Kippenberg}}]{Toth2017}%
  \BibitemOpen
  \bibfield  {author} {\bibinfo {author} {\bibfnamefont {L.~D.}\ \bibnamefont
  {T{\'o}th}}, \bibinfo {author} {\bibfnamefont {N.~R.}\ \bibnamefont
  {Bernier}}, \bibinfo {author} {\bibfnamefont {A.}~\bibnamefont {Nunnenkamp}},
  \bibinfo {author} {\bibfnamefont {A.~K.}\ \bibnamefont {Feofanov}}, \ and\
  \bibinfo {author} {\bibfnamefont {T.~J.}\ \bibnamefont {Kippenberg}},\
  }\bibfield  {title} {\enquote {\bibinfo {title} {A dissipative quantum
  reservoir for microwave light using a mechanical oscillator},}\ }\href
  {http://dx.doi.org/10.1038/nphys4121} {\bibfield  {journal} {\bibinfo
  {journal} {Nature Physics}\ }\textbf {\bibinfo {volume} {13}},\ \bibinfo
  {pages} {787 EP --} (\bibinfo {year} {2017})}\BibitemShut {NoStop}%
\bibitem [{\citenamefont {LaHaye}\ \emph {et~al.}(2009)\citenamefont {LaHaye},
  \citenamefont {Suh}, \citenamefont {Echternach}, \citenamefont {Schwab},\
  and\ \citenamefont {Roukes}}]{LaHaye2009}%
  \BibitemOpen
  \bibfield  {author} {\bibinfo {author} {\bibfnamefont {M.~D.}\ \bibnamefont
  {LaHaye}}, \bibinfo {author} {\bibfnamefont {J.}~\bibnamefont {Suh}},
  \bibinfo {author} {\bibfnamefont {P.~M.}\ \bibnamefont {Echternach}},
  \bibinfo {author} {\bibfnamefont {K.~C.}\ \bibnamefont {Schwab}}, \ and\
  \bibinfo {author} {\bibfnamefont {M.~L.}\ \bibnamefont {Roukes}},\ }\bibfield
   {title} {\enquote {\bibinfo {title} {Nanomechanical measurements of a
  superconducting qubit},}\ }\href {http://dx.doi.org/10.1038/nature08093}
  {\bibfield  {journal} {\bibinfo  {journal} {Nature}\ }\textbf {\bibinfo
  {volume} {459}},\ \bibinfo {pages} {960 EP --} (\bibinfo {year}
  {2009})}\BibitemShut {NoStop}%
\bibitem [{\citenamefont {Rouxinol}\ \emph {et~al.}(2016)\citenamefont
  {Rouxinol}, \citenamefont {Hao}, \citenamefont {Brito}, \citenamefont
  {Caldeira}, \citenamefont {Irish},\ and\ \citenamefont
  {LaHaye}}]{Rouxinal2016}%
  \BibitemOpen
  \bibfield  {author} {\bibinfo {author} {\bibfnamefont {F}~\bibnamefont
  {Rouxinol}}, \bibinfo {author} {\bibfnamefont {Y}~\bibnamefont {Hao}},
  \bibinfo {author} {\bibfnamefont {F}~\bibnamefont {Brito}}, \bibinfo {author}
  {\bibfnamefont {A~O}\ \bibnamefont {Caldeira}}, \bibinfo {author}
  {\bibfnamefont {E~K}\ \bibnamefont {Irish}}, \ and\ \bibinfo {author}
  {\bibfnamefont {M~D}\ \bibnamefont {LaHaye}},\ }\bibfield  {title} {\enquote
  {\bibinfo {title} {Measurements of nanoresonator-qubit interactions in a
  hybrid quantum electromechanical system},}\ }\href
  {http://stacks.iop.org/0957-4484/27/i=36/a=364003} {\bibfield  {journal}
  {\bibinfo  {journal} {Nanotechnology}\ }\textbf {\bibinfo {volume} {27}},\
  \bibinfo {pages} {364003} (\bibinfo {year} {2016})}\BibitemShut {NoStop}%
\bibitem [{\citenamefont {Chu}\ \emph {et~al.}(2017)\citenamefont {Chu},
  \citenamefont {Kharel}, \citenamefont {Renninger}, \citenamefont {Burkhart},
  \citenamefont {Frunzio}, \citenamefont {Rakich},\ and\ \citenamefont
  {Schoelkopf}}]{Chu2017}%
  \BibitemOpen
  \bibfield  {author} {\bibinfo {author} {\bibfnamefont {Yiwen}\ \bibnamefont
  {Chu}}, \bibinfo {author} {\bibfnamefont {Prashanta}\ \bibnamefont {Kharel}},
  \bibinfo {author} {\bibfnamefont {William~H.}\ \bibnamefont {Renninger}},
  \bibinfo {author} {\bibfnamefont {Luke~D.}\ \bibnamefont {Burkhart}},
  \bibinfo {author} {\bibfnamefont {Luigi}\ \bibnamefont {Frunzio}}, \bibinfo
  {author} {\bibfnamefont {Peter~T.}\ \bibnamefont {Rakich}}, \ and\ \bibinfo
  {author} {\bibfnamefont {Robert~J.}\ \bibnamefont {Schoelkopf}},\ }\bibfield
  {title} {\enquote {\bibinfo {title} {Quantum acoustics with superconducting
  qubits},}\ }\href {\doibase 10.1126/science.aao1511} {\bibfield  {journal}
  {\bibinfo  {journal} {Science}\ }\textbf {\bibinfo {volume} {358}},\ \bibinfo
  {pages} {199--202} (\bibinfo {year} {2017})},\ \Eprint
  {http://arxiv.org/abs/http://science.sciencemag.org/content/358/6360/199.full.pdf}
  {http://science.sciencemag.org/content/358/6360/199.full.pdf} \BibitemShut
  {NoStop}%
\bibitem [{\citenamefont {Safavi-Naeini}\ and\ \citenamefont
  {Painter}(2011)}]{Safavi-Naeini2011}%
  \BibitemOpen
  \bibfield  {author} {\bibinfo {author} {\bibfnamefont {Amir~H}\ \bibnamefont
  {Safavi-Naeini}}\ and\ \bibinfo {author} {\bibfnamefont {Oskar}\ \bibnamefont
  {Painter}},\ }\bibfield  {title} {\enquote {\bibinfo {title} {Proposal for an
  optomechanical traveling wave phonon--photon translator},}\ }\href
  {http://stacks.iop.org/1367-2630/13/i=1/a=013017} {\bibfield  {journal}
  {\bibinfo  {journal} {New Journal of Physics}\ }\textbf {\bibinfo {volume}
  {13}},\ \bibinfo {pages} {013017} (\bibinfo {year} {2011})}\BibitemShut
  {NoStop}%
\bibitem [{\citenamefont {Regal}\ and\ \citenamefont
  {Lehnert}(2011)}]{Regal2011}%
  \BibitemOpen
  \bibfield  {author} {\bibinfo {author} {\bibfnamefont {C.~A.}\ \bibnamefont
  {Regal}}\ and\ \bibinfo {author} {\bibfnamefont {K.~W.}\ \bibnamefont
  {Lehnert}},\ }\bibfield  {title} {\enquote {\bibinfo {title} {From cavity
  electromechanics to cavity optomechanics},}\ }\href@noop {} {\bibfield
  {journal} {\bibinfo  {journal} {Journal of Physics: Conference Series}\
  }\textbf {\bibinfo {volume} {264}},\ \bibinfo {pages} {012025} (\bibinfo
  {year} {2011})}\BibitemShut {NoStop}%
\bibitem [{\citenamefont {Taylor}\ \emph {et~al.}(2011)\citenamefont {Taylor},
  \citenamefont {S{\o}rensen}, \citenamefont {Marcus},\ and\ \citenamefont
  {Polzik}}]{Taylor2011}%
  \BibitemOpen
  \bibfield  {author} {\bibinfo {author} {\bibfnamefont {J.~M.}\ \bibnamefont
  {Taylor}}, \bibinfo {author} {\bibfnamefont {A.~S.}\ \bibnamefont
  {S{\o}rensen}}, \bibinfo {author} {\bibfnamefont {C.~M.}\ \bibnamefont
  {Marcus}}, \ and\ \bibinfo {author} {\bibfnamefont {E.~S.}\ \bibnamefont
  {Polzik}},\ }\bibfield  {title} {\enquote {\bibinfo {title} {Laser cooling
  and optical detection of excitations in a lc electrical circuit},}\
  }\href@noop {} {\bibfield  {journal} {\bibinfo  {journal} {Phys. Rev. Lett.}\
  }\textbf {\bibinfo {volume} {107}},\ \bibinfo {pages} {273601} (\bibinfo
  {year} {2011})}\BibitemShut {NoStop}%
\bibitem [{\citenamefont {Tian}(2015)}]{Tian2015}%
  \BibitemOpen
  \bibfield  {author} {\bibinfo {author} {\bibfnamefont {Lin}\ \bibnamefont
  {Tian}},\ }\bibfield  {title} {\enquote {\bibinfo {title}
  {Optoelectromechanical transducer: Reversible conversion between microwave
  and optical photons},}\ }\href {\doibase 10.1002/andp.201400116} {\bibfield
  {journal} {\bibinfo  {journal} {Annalen der Physik}\ }\textbf {\bibinfo
  {volume} {527}},\ \bibinfo {pages} {1--14} (\bibinfo {year}
  {2015})}\BibitemShut {NoStop}%
\bibitem [{\citenamefont {{{\v C}ernot{\'{\i}}k}}\ \emph
  {et~al.}(2017)\citenamefont {{{\v C}ernot{\'{\i}}k}}, \citenamefont
  {{Mahmoodian}},\ and\ \citenamefont {{Hammerer}}}]{Ondrej2017}%
  \BibitemOpen
  \bibfield  {author} {\bibinfo {author} {\bibfnamefont {O.}~\bibnamefont {{{\v
  C}ernot{\'{\i}}k}}}, \bibinfo {author} {\bibfnamefont {S.}~\bibnamefont
  {{Mahmoodian}}}, \ and\ \bibinfo {author} {\bibfnamefont {K.}~\bibnamefont
  {{Hammerer}}},\ }\bibfield  {title} {\enquote {\bibinfo {title} {{Spatially
  Adiabatic Frequency Conversion in Optoelectromechanical Arrays}},}\
  }\href@noop {} {\bibfield  {journal} {\bibinfo  {journal} {ArXiv e-prints}\ }
  (\bibinfo {year} {2017})},\ \Eprint {http://arxiv.org/abs/1707.03339}
  {arXiv:1707.03339 [quant-ph]} \BibitemShut {NoStop}%
\bibitem [{\citenamefont {Barzanjeh}\ \emph {et~al.}(2012)\citenamefont
  {Barzanjeh}, \citenamefont {Abdi}, \citenamefont {Milburn}, \citenamefont
  {Tombesi},\ and\ \citenamefont {Vitali}}]{Barzanjeh2012}%
  \BibitemOpen
  \bibfield  {author} {\bibinfo {author} {\bibfnamefont {Sh.}\ \bibnamefont
  {Barzanjeh}}, \bibinfo {author} {\bibfnamefont {M.}~\bibnamefont {Abdi}},
  \bibinfo {author} {\bibfnamefont {G.~J.}\ \bibnamefont {Milburn}}, \bibinfo
  {author} {\bibfnamefont {P.}~\bibnamefont {Tombesi}}, \ and\ \bibinfo
  {author} {\bibfnamefont {D.}~\bibnamefont {Vitali}},\ }\bibfield  {title}
  {\enquote {\bibinfo {title} {Reversible optical-to-microwave quantum
  interface},}\ }\href {\doibase 10.1103/PhysRevLett.109.130503} {\bibfield
  {journal} {\bibinfo  {journal} {Phys. Rev. Lett.}\ }\textbf {\bibinfo
  {volume} {109}},\ \bibinfo {pages} {130503} (\bibinfo {year}
  {2012})}\BibitemShut {NoStop}%
\bibitem [{\citenamefont {{\ifmmode \check{C}\else \v{C}\fi{}ernot\'{\i}k,
  Ond\ifmmode \check{r}\else \v{r}\fi{}ej and Hammerer,
  Klemens}}(2016)}]{Ondrej2016}%
  \BibitemOpen
  \bibfield  {author} {\bibinfo {author} {\bibnamefont {{\ifmmode
  \check{C}\else \v{C}\fi{}ernot\'{\i}k, Ond\ifmmode \check{r}\else
  \v{r}\fi{}ej and Hammerer, Klemens}}},\ }\bibfield  {title} {\enquote
  {\bibinfo {title} {Measurement-induced long-distance entanglement of
  superconducting qubits using optomechanical transducers},}\ }\href {\doibase
  10.1103/PhysRevA.94.012340} {\bibfield  {journal} {\bibinfo  {journal} {Phys.
  Rev. A}\ }\textbf {\bibinfo {volume} {94}},\ \bibinfo {pages} {012340}
  (\bibinfo {year} {2016})}\BibitemShut {NoStop}%
\bibitem [{\citenamefont {Jayich}\ \emph {et~al.}(2008)\citenamefont {Jayich},
  \citenamefont {Sankey}, \citenamefont {Zwickl}, \citenamefont {Yang},
  \citenamefont {Thompson}, \citenamefont {Girvin}, \citenamefont {Clerk},
  \citenamefont {Marquardt},\ and\ \citenamefont {Harris}}]{Jayich2008}%
  \BibitemOpen
  \bibfield  {author} {\bibinfo {author} {\bibfnamefont {A~M}\ \bibnamefont
  {Jayich}}, \bibinfo {author} {\bibfnamefont {J~C}\ \bibnamefont {Sankey}},
  \bibinfo {author} {\bibfnamefont {B~M}\ \bibnamefont {Zwickl}}, \bibinfo
  {author} {\bibfnamefont {C}~\bibnamefont {Yang}}, \bibinfo {author}
  {\bibfnamefont {J~D}\ \bibnamefont {Thompson}}, \bibinfo {author}
  {\bibfnamefont {S~M}\ \bibnamefont {Girvin}}, \bibinfo {author}
  {\bibfnamefont {A~A}\ \bibnamefont {Clerk}}, \bibinfo {author} {\bibfnamefont
  {F}~\bibnamefont {Marquardt}}, \ and\ \bibinfo {author} {\bibfnamefont
  {J~G~E}\ \bibnamefont {Harris}},\ }\bibfield  {title} {\enquote {\bibinfo
  {title} {Dispersive optomechanics: a membrane inside a cavity},}\ }\href@noop
  {} {\bibfield  {journal} {\bibinfo  {journal} {New Journal of Physics}\
  }\textbf {\bibinfo {volume} {10}},\ \bibinfo {pages} {095008} (\bibinfo
  {year} {2008})}\BibitemShut {NoStop}%
\bibitem [{\citenamefont {Unterreithmeier}\ \emph {et~al.}(2009)\citenamefont
  {Unterreithmeier}, \citenamefont {Weig},\ and\ \citenamefont
  {Kotthaus}}]{Unterreithmeier2009}%
  \BibitemOpen
  \bibfield  {author} {\bibinfo {author} {\bibfnamefont {Quirin~P.}\
  \bibnamefont {Unterreithmeier}}, \bibinfo {author} {\bibfnamefont {Eva~M.}\
  \bibnamefont {Weig}}, \ and\ \bibinfo {author} {\bibfnamefont {Jorg~P.}\
  \bibnamefont {Kotthaus}},\ }\bibfield  {title} {\enquote {\bibinfo {title}
  {Universal transduction scheme for nanomechanical systems based on dielectric
  forces},}\ }\href {http://dx.doi.org/10.1038/nature07932} {\bibfield
  {journal} {\bibinfo  {journal} {Nature}\ }\textbf {\bibinfo {volume} {458}},\
  \bibinfo {pages} {1001--1004} (\bibinfo {year} {2009})}\BibitemShut {NoStop}%
\bibitem [{\citenamefont {Schmid}\ \emph {et~al.}(2014)\citenamefont {Schmid},
  \citenamefont {Bagci}, \citenamefont {Zeuthen}, \citenamefont {Taylor},
  \citenamefont {Herring}, \citenamefont {Cassidy}, \citenamefont {Marcus},
  \citenamefont {Guillermo~Villanueva}, \citenamefont {Amato}, \citenamefont
  {Boisen}, \citenamefont {Cheol~Shin}, \citenamefont {Kong}, \citenamefont
  {S{\o}rensen}, \citenamefont {Usami},\ and\ \citenamefont
  {Polzik}}]{Schmid2014}%
  \BibitemOpen
  \bibfield  {author} {\bibinfo {author} {\bibfnamefont {Silvan}\ \bibnamefont
  {Schmid}}, \bibinfo {author} {\bibfnamefont {Tolga}\ \bibnamefont {Bagci}},
  \bibinfo {author} {\bibfnamefont {Emil}\ \bibnamefont {Zeuthen}}, \bibinfo
  {author} {\bibfnamefont {Jacob~M.}\ \bibnamefont {Taylor}}, \bibinfo {author}
  {\bibfnamefont {Patrick~K.}\ \bibnamefont {Herring}}, \bibinfo {author}
  {\bibfnamefont {Maja~C.}\ \bibnamefont {Cassidy}}, \bibinfo {author}
  {\bibfnamefont {Charles~M.}\ \bibnamefont {Marcus}}, \bibinfo {author}
  {\bibfnamefont {Luis}\ \bibnamefont {Guillermo~Villanueva}}, \bibinfo
  {author} {\bibfnamefont {Bartolo}\ \bibnamefont {Amato}}, \bibinfo {author}
  {\bibfnamefont {Anja}\ \bibnamefont {Boisen}}, \bibinfo {author}
  {\bibfnamefont {Yong}\ \bibnamefont {Cheol~Shin}}, \bibinfo {author}
  {\bibfnamefont {Jing}\ \bibnamefont {Kong}}, \bibinfo {author} {\bibfnamefont
  {Anders~S.}\ \bibnamefont {S{\o}rensen}}, \bibinfo {author} {\bibfnamefont
  {Koji}\ \bibnamefont {Usami}}, \ and\ \bibinfo {author} {\bibfnamefont
  {Eugene~S.}\ \bibnamefont {Polzik}},\ }\bibfield  {title} {\enquote {\bibinfo
  {title} {Single-layer graphene on silicon nitride micromembrane
  resonators},}\ }\href {\doibase http://dx.doi.org/10.1063/1.4862296}
  {\bibfield  {journal} {\bibinfo  {journal} {Journal of Applied Physics}\
  }\textbf {\bibinfo {volume} {115}},\ \bibinfo {eid} {054513} (\bibinfo {year}
  {2014})}\BibitemShut {NoStop}%
\bibitem [{\citenamefont {Bagci}\ \emph {et~al.}(2014)\citenamefont {Bagci},
  \citenamefont {Simonsen}, \citenamefont {Schmid}, \citenamefont {Villanueva},
  \citenamefont {Zeuthen}, \citenamefont {Appel}, \citenamefont {Taylor},
  \citenamefont {S\o{}rensen}, \citenamefont {Usami}, \citenamefont
  {Schliesser},\ and\ \citenamefont {Polzik}}]{Bagci2014}%
  \BibitemOpen
  \bibfield  {author} {\bibinfo {author} {\bibfnamefont {T.}~\bibnamefont
  {Bagci}}, \bibinfo {author} {\bibfnamefont {A.}~\bibnamefont {Simonsen}},
  \bibinfo {author} {\bibfnamefont {S.}~\bibnamefont {Schmid}}, \bibinfo
  {author} {\bibfnamefont {L.~G.}\ \bibnamefont {Villanueva}}, \bibinfo
  {author} {\bibfnamefont {E.}~\bibnamefont {Zeuthen}}, \bibinfo {author}
  {\bibfnamefont {J.}~\bibnamefont {Appel}}, \bibinfo {author} {\bibfnamefont
  {J.~M.}\ \bibnamefont {Taylor}}, \bibinfo {author} {\bibfnamefont
  {A.}~\bibnamefont {S\o{}rensen}}, \bibinfo {author} {\bibfnamefont
  {K.}~\bibnamefont {Usami}}, \bibinfo {author} {\bibfnamefont
  {A.}~\bibnamefont {Schliesser}}, \ and\ \bibinfo {author} {\bibfnamefont
  {E.~S.}\ \bibnamefont {Polzik}},\ }\bibfield  {title} {\enquote {\bibinfo
  {title} {Optical detection of radio waves through a nanomechanical
  transducer},}\ }\href {http://dx.doi.org/10.1038/nature13029} {\bibfield
  {journal} {\bibinfo  {journal} {Nature}\ }\textbf {\bibinfo {volume} {507}},\
  \bibinfo {pages} {81--85} (\bibinfo {year} {2014})}\BibitemShut {NoStop}%
\bibitem [{\citenamefont {Andrews}\ \emph {et~al.}(2014)\citenamefont
  {Andrews}, \citenamefont {Peterson}, \citenamefont {Purdy}, \citenamefont
  {Cicak}, \citenamefont {Simmonds}, \citenamefont {Regal},\ and\ \citenamefont
  {Lehnert}}]{Andrews2014}%
  \BibitemOpen
  \bibfield  {author} {\bibinfo {author} {\bibfnamefont {R.~W.}\ \bibnamefont
  {Andrews}}, \bibinfo {author} {\bibfnamefont {R.~W.}\ \bibnamefont
  {Peterson}}, \bibinfo {author} {\bibfnamefont {T.~P.}\ \bibnamefont {Purdy}},
  \bibinfo {author} {\bibfnamefont {K.}~\bibnamefont {Cicak}}, \bibinfo
  {author} {\bibfnamefont {R.~W.}\ \bibnamefont {Simmonds}}, \bibinfo {author}
  {\bibfnamefont {C.~A.}\ \bibnamefont {Regal}}, \ and\ \bibinfo {author}
  {\bibfnamefont {K.~W.}\ \bibnamefont {Lehnert}},\ }\bibfield  {title}
  {\enquote {\bibinfo {title} {Bidirectional and efficient conversion between
  microwave and optical light},}\ }\href {http://dx.doi.org/10.1038/nphys2911}
  {\bibfield  {journal} {\bibinfo  {journal} {Nat Phys}\ }\textbf {\bibinfo
  {volume} {10}},\ \bibinfo {pages} {321--326} (\bibinfo {year}
  {2014})}\BibitemShut {NoStop}%
\bibitem [{\citenamefont {Pitanti}\ \emph {et~al.}(2015)\citenamefont
  {Pitanti}, \citenamefont {Fink}, \citenamefont {Safavi-Naeini}, \citenamefont
  {Hill}, \citenamefont {Lei}, \citenamefont {Tredicucci},\ and\ \citenamefont
  {Painter}}]{Pitanti2015}%
  \BibitemOpen
  \bibfield  {author} {\bibinfo {author} {\bibfnamefont {Alessandro}\
  \bibnamefont {Pitanti}}, \bibinfo {author} {\bibfnamefont {Johannes~M.}\
  \bibnamefont {Fink}}, \bibinfo {author} {\bibfnamefont {Amir~H.}\
  \bibnamefont {Safavi-Naeini}}, \bibinfo {author} {\bibfnamefont {Jeff~T.}\
  \bibnamefont {Hill}}, \bibinfo {author} {\bibfnamefont {Chan~U.}\
  \bibnamefont {Lei}}, \bibinfo {author} {\bibfnamefont {Alessandro}\
  \bibnamefont {Tredicucci}}, \ and\ \bibinfo {author} {\bibfnamefont {Oskar}\
  \bibnamefont {Painter}},\ }\bibfield  {title} {\enquote {\bibinfo {title}
  {Strong opto-electro-mechanical coupling in a silicon photonic crystal
  cavity},}\ }\href {\doibase 10.1364/OE.23.003196} {\bibfield  {journal}
  {\bibinfo  {journal} {Opt. Express}\ }\textbf {\bibinfo {volume} {23}},\
  \bibinfo {pages} {3196--3208} (\bibinfo {year} {2015})}\BibitemShut {NoStop}%
\bibitem [{\citenamefont {Fink}\ \emph {et~al.}(2016)\citenamefont {Fink},
  \citenamefont {Kalaee}, \citenamefont {Pitanti}, \citenamefont {Norte},
  \citenamefont {Heinzle}, \citenamefont {Davan{\c c}o}, \citenamefont
  {Srinivasan},\ and\ \citenamefont {Painter}}]{Fink2016}%
  \BibitemOpen
  \bibfield  {author} {\bibinfo {author} {\bibfnamefont {J.~M.}\ \bibnamefont
  {Fink}}, \bibinfo {author} {\bibfnamefont {M.}~\bibnamefont {Kalaee}},
  \bibinfo {author} {\bibfnamefont {A.}~\bibnamefont {Pitanti}}, \bibinfo
  {author} {\bibfnamefont {R.}~\bibnamefont {Norte}}, \bibinfo {author}
  {\bibfnamefont {L.}~\bibnamefont {Heinzle}}, \bibinfo {author} {\bibfnamefont
  {M.}~\bibnamefont {Davan{\c c}o}}, \bibinfo {author} {\bibfnamefont
  {K.}~\bibnamefont {Srinivasan}}, \ and\ \bibinfo {author} {\bibfnamefont
  {O.}~\bibnamefont {Painter}},\ }\bibfield  {title} {\enquote {\bibinfo
  {title} {Quantum electromechanics on silicon nitride nanomembranes},}\ }\href
  {http://dx.doi.org/10.1038/ncomms12396} {\bibfield  {journal} {\bibinfo
  {journal} {Nature Communications}\ }\textbf {\bibinfo {volume} {7}},\
  \bibinfo {pages} {12396 EP --} (\bibinfo {year} {2016})}\BibitemShut
  {NoStop}%
\bibitem [{\citenamefont {Menke}\ \emph {et~al.}(2017)\citenamefont {Menke},
  \citenamefont {Burns}, \citenamefont {Higginbotham}, \citenamefont {Kampel},
  \citenamefont {Peterson}, \citenamefont {Cicak}, \citenamefont {Simmonds},
  \citenamefont {Regal},\ and\ \citenamefont {Lehnert}}]{Menke2017}%
  \BibitemOpen
  \bibfield  {author} {\bibinfo {author} {\bibfnamefont {T.}~\bibnamefont
  {Menke}}, \bibinfo {author} {\bibfnamefont {P.~S.}\ \bibnamefont {Burns}},
  \bibinfo {author} {\bibfnamefont {A.~P.}\ \bibnamefont {Higginbotham}},
  \bibinfo {author} {\bibfnamefont {N.~S.}\ \bibnamefont {Kampel}}, \bibinfo
  {author} {\bibfnamefont {R.~W.}\ \bibnamefont {Peterson}}, \bibinfo {author}
  {\bibfnamefont {K.}~\bibnamefont {Cicak}}, \bibinfo {author} {\bibfnamefont
  {R.~W.}\ \bibnamefont {Simmonds}}, \bibinfo {author} {\bibfnamefont {C.~A.}\
  \bibnamefont {Regal}}, \ and\ \bibinfo {author} {\bibfnamefont {K.~W.}\
  \bibnamefont {Lehnert}},\ }\bibfield  {title} {\enquote {\bibinfo {title}
  {Reconfigurable re-entrant cavity for wireless coupling to an
  electro-optomechanical device},}\ }\href {\doibase 10.1063/1.5000973}
  {\bibfield  {journal} {\bibinfo  {journal} {Review of Scientific
  Instruments}\ }\textbf {\bibinfo {volume} {88}},\ \bibinfo {pages} {094701}
  (\bibinfo {year} {2017})},\ \Eprint
  {http://arxiv.org/abs/https://doi.org/10.1063/1.5000973}
  {https://doi.org/10.1063/1.5000973} \BibitemShut {NoStop}%
\bibitem [{\citenamefont {Bochmann}\ \emph {et~al.}(2013)\citenamefont
  {Bochmann}, \citenamefont {Vainsencher}, \citenamefont {Awschalom},\ and\
  \citenamefont {Cleland}}]{Bochmann2013}%
  \BibitemOpen
  \bibfield  {author} {\bibinfo {author} {\bibfnamefont {Joerg}\ \bibnamefont
  {Bochmann}}, \bibinfo {author} {\bibfnamefont {Amit}\ \bibnamefont
  {Vainsencher}}, \bibinfo {author} {\bibfnamefont {David~D.}\ \bibnamefont
  {Awschalom}}, \ and\ \bibinfo {author} {\bibfnamefont {Andrew~N.}\
  \bibnamefont {Cleland}},\ }\bibfield  {title} {\enquote {\bibinfo {title}
  {Nanomechanical coupling between microwave and optical photons},}\ }\href
  {http://dx.doi.org/10.1038/nphys2748} {\bibfield  {journal} {\bibinfo
  {journal} {Nat Phys}\ }\textbf {\bibinfo {volume} {9}},\ \bibinfo {pages}
  {712--716} (\bibinfo {year} {2013})}\BibitemShut {NoStop}%
\bibitem [{\citenamefont {Vainsencher}\ \emph {et~al.}(2016)\citenamefont
  {Vainsencher}, \citenamefont {Satzinger}, \citenamefont {Peairs},\ and\
  \citenamefont {Cleland}}]{Vainsencher2016}%
  \BibitemOpen
  \bibfield  {author} {\bibinfo {author} {\bibfnamefont {Amit}\ \bibnamefont
  {Vainsencher}}, \bibinfo {author} {\bibfnamefont {K.~J.}\ \bibnamefont
  {Satzinger}}, \bibinfo {author} {\bibfnamefont {G.~A.}\ \bibnamefont
  {Peairs}}, \ and\ \bibinfo {author} {\bibfnamefont {A.~N.}\ \bibnamefont
  {Cleland}},\ }\bibfield  {title} {\enquote {\bibinfo {title} {Bi-directional
  conversion between microwave and optical frequencies in a piezoelectric
  optomechanical device},}\ }\href {\doibase 10.1063/1.4955408} {\bibfield
  {journal} {\bibinfo  {journal} {Applied Physics Letters}\ }\textbf {\bibinfo
  {volume} {109}},\ \bibinfo {pages} {033107} (\bibinfo {year} {2016})},\
  \Eprint {http://arxiv.org/abs/http://dx.doi.org/10.1063/1.4955408}
  {http://dx.doi.org/10.1063/1.4955408} \BibitemShut {NoStop}%
\bibitem [{\citenamefont {Balram}\ \emph {et~al.}(2016)\citenamefont {Balram},
  \citenamefont {Davan{\c c}o}, \citenamefont {Song},\ and\ \citenamefont
  {Srinivasan}}]{Balram2016}%
  \BibitemOpen
  \bibfield  {author} {\bibinfo {author} {\bibfnamefont {Krishna~C.}\
  \bibnamefont {Balram}}, \bibinfo {author} {\bibfnamefont {Marcelo~I.}\
  \bibnamefont {Davan{\c c}o}}, \bibinfo {author} {\bibfnamefont {Jin~Dong}\
  \bibnamefont {Song}}, \ and\ \bibinfo {author} {\bibfnamefont {Kartik}\
  \bibnamefont {Srinivasan}},\ }\bibfield  {title} {\enquote {\bibinfo {title}
  {Coherent coupling between radiofrequency, optical and acoustic waves in
  piezo-optomechanical circuits},}\ }\href
  {http://dx.doi.org/10.1038/nphoton.2016.46} {\bibfield  {journal} {\bibinfo
  {journal} {Nat Photon}\ }\textbf {\bibinfo {volume} {10}},\ \bibinfo {pages}
  {346--352} (\bibinfo {year} {2016})}\BibitemShut {NoStop}%
\bibitem [{\citenamefont {Zou}\ \emph {et~al.}(2016)\citenamefont {Zou},
  \citenamefont {Han}, \citenamefont {Jiang},\ and\ \citenamefont
  {Tang}}]{Zou2016}%
  \BibitemOpen
  \bibfield  {author} {\bibinfo {author} {\bibfnamefont {Chang-Ling}\
  \bibnamefont {Zou}}, \bibinfo {author} {\bibfnamefont {Xu}~\bibnamefont
  {Han}}, \bibinfo {author} {\bibfnamefont {Liang}\ \bibnamefont {Jiang}}, \
  and\ \bibinfo {author} {\bibfnamefont {Hong~X.}\ \bibnamefont {Tang}},\
  }\bibfield  {title} {\enquote {\bibinfo {title} {Cavity piezomechanical
  strong coupling and frequency conversion on an aluminum nitride chip},}\
  }\href {\doibase 10.1103/PhysRevA.94.013812} {\bibfield  {journal} {\bibinfo
  {journal} {Phys. Rev. A}\ }\textbf {\bibinfo {volume} {94}},\ \bibinfo
  {pages} {013812} (\bibinfo {year} {2016})}\BibitemShut {NoStop}%
\bibitem [{\citenamefont {{Takeda}}\ \emph {et~al.}(2017)\citenamefont
  {{Takeda}}, \citenamefont {{Nagasaka}}, \citenamefont {{Noguchi}},
  \citenamefont {{Yamazaki}}, \citenamefont {{Nakamura}}, \citenamefont
  {{Iwase}}, \citenamefont {{Taylor}},\ and\ \citenamefont
  {{Usami}}}]{Takeda2017}%
  \BibitemOpen
  \bibfield  {author} {\bibinfo {author} {\bibfnamefont {K.}~\bibnamefont
  {{Takeda}}}, \bibinfo {author} {\bibfnamefont {K.}~\bibnamefont
  {{Nagasaka}}}, \bibinfo {author} {\bibfnamefont {A.}~\bibnamefont
  {{Noguchi}}}, \bibinfo {author} {\bibfnamefont {R.}~\bibnamefont
  {{Yamazaki}}}, \bibinfo {author} {\bibfnamefont {Y.}~\bibnamefont
  {{Nakamura}}}, \bibinfo {author} {\bibfnamefont {E.}~\bibnamefont {{Iwase}}},
  \bibinfo {author} {\bibfnamefont {J.~M.}\ \bibnamefont {{Taylor}}}, \ and\
  \bibinfo {author} {\bibfnamefont {K.}~\bibnamefont {{Usami}}},\ }\bibfield
  {title} {\enquote {\bibinfo {title} {{Electro-mechano-optical NMR
  detection}},}\ }\href@noop {} {\bibfield  {journal} {\bibinfo  {journal}
  {ArXiv e-prints}\ } (\bibinfo {year} {2017})},\ \Eprint
  {http://arxiv.org/abs/1706.00532} {arXiv:1706.00532 [quant-ph]} \BibitemShut
  {NoStop}%
\bibitem [{\citenamefont {Hudson}\ and\ \citenamefont
  {Parthasarathy}(1984)}]{Hudson1984}%
  \BibitemOpen
  \bibfield  {author} {\bibinfo {author} {\bibfnamefont {R.~L.}\ \bibnamefont
  {Hudson}}\ and\ \bibinfo {author} {\bibfnamefont {K.~R.}\ \bibnamefont
  {Parthasarathy}},\ }\bibfield  {title} {\enquote {\bibinfo {title} {Quantum
  ito's formula and stochastic evolutions},}\ }\href
  {http://projecteuclid.org/euclid.cmp/1103941122} {\bibfield  {journal}
  {\bibinfo  {journal} {Comm. Math. Phys.}\ }\textbf {\bibinfo {volume} {93}},\
  \bibinfo {pages} {301--323} (\bibinfo {year} {1984})}\BibitemShut {NoStop}%
\bibitem [{\citenamefont {Gardiner}\ and\ \citenamefont
  {Collett}(1985)}]{Collett1985}%
  \BibitemOpen
  \bibfield  {author} {\bibinfo {author} {\bibfnamefont {C.~W.}\ \bibnamefont
  {Gardiner}}\ and\ \bibinfo {author} {\bibfnamefont {M.~J.}\ \bibnamefont
  {Collett}},\ }\bibfield  {title} {\enquote {\bibinfo {title} {Input and
  output in damped quantum systems: Quantum stochastic differential equations
  and the master equation},}\ }\href {\doibase 10.1103/PhysRevA.31.3761}
  {\bibfield  {journal} {\bibinfo  {journal} {Phys. Rev. A}\ }\textbf {\bibinfo
  {volume} {31}},\ \bibinfo {pages} {3761--3774} (\bibinfo {year}
  {1985})}\BibitemShut {NoStop}%
\bibitem [{\citenamefont {Yurke}\ and\ \citenamefont
  {Denker}(1984)}]{Yurke1984}%
  \BibitemOpen
  \bibfield  {author} {\bibinfo {author} {\bibfnamefont {Bernard}\ \bibnamefont
  {Yurke}}\ and\ \bibinfo {author} {\bibfnamefont {John~S.}\ \bibnamefont
  {Denker}},\ }\bibfield  {title} {\enquote {\bibinfo {title} {Quantum network
  theory},}\ }\href {\doibase 10.1103/PhysRevA.29.1419} {\bibfield  {journal}
  {\bibinfo  {journal} {Phys. Rev. A}\ }\textbf {\bibinfo {volume} {29}},\
  \bibinfo {pages} {1419--1437} (\bibinfo {year} {1984})}\BibitemShut {NoStop}%
\bibitem [{\citenamefont {Wang}\ and\ \citenamefont {Clerk}(2012)}]{Wang2012}%
  \BibitemOpen
  \bibfield  {author} {\bibinfo {author} {\bibfnamefont {Ying-Dan}\
  \bibnamefont {Wang}}\ and\ \bibinfo {author} {\bibfnamefont {Aashish~A.}\
  \bibnamefont {Clerk}},\ }\bibfield  {title} {\enquote {\bibinfo {title}
  {Using interference for high fidelity quantum state transfer in
  optomechanics},}\ }\href {\doibase 10.1103/PhysRevLett.108.153603} {\bibfield
   {journal} {\bibinfo  {journal} {Phys. Rev. Lett.}\ }\textbf {\bibinfo
  {volume} {108}},\ \bibinfo {pages} {153603} (\bibinfo {year}
  {2012})}\BibitemShut {NoStop}%
\bibitem [{\citenamefont {Tian}(2012)}]{Tian2012}%
  \BibitemOpen
  \bibfield  {author} {\bibinfo {author} {\bibfnamefont {Lin}\ \bibnamefont
  {Tian}},\ }\bibfield  {title} {\enquote {\bibinfo {title} {Adiabatic state
  conversion and pulse transmission in optomechanical systems},}\ }\href
  {\doibase 10.1103/PhysRevLett.108.153604} {\bibfield  {journal} {\bibinfo
  {journal} {Phys. Rev. Lett.}\ }\textbf {\bibinfo {volume} {108}},\ \bibinfo
  {pages} {153604} (\bibinfo {year} {2012})}\BibitemShut {NoStop}%
\bibitem [{\citenamefont {Zeuthen}(2016)}]{EZPhD}%
  \BibitemOpen
  \bibfield  {author} {\bibinfo {author} {\bibfnamefont {Emil}\ \bibnamefont
  {Zeuthen}},\ }\href@noop {} {Ph.D. thesis},\ \bibinfo  {school} {Niels Bohr
  Institute, University of Copenhagen} (\bibinfo {year} {2016})\BibitemShut
  {NoStop}%
\bibitem [{\citenamefont {Tilmans}(1996)}]{Tilmans1996}%
  \BibitemOpen
  \bibfield  {author} {\bibinfo {author} {\bibfnamefont {Harrie A~C}\
  \bibnamefont {Tilmans}},\ }\bibfield  {title} {\enquote {\bibinfo {title}
  {Equivalent circuit representation of electromechanical transducers: I.
  lumped-parameter systems},}\ }\href
  {http://stacks.iop.org/0960-1317/6/i=1/a=036} {\bibfield  {journal} {\bibinfo
   {journal} {Journal of Micromechanics and Microengineering}\ }\textbf
  {\bibinfo {volume} {6}},\ \bibinfo {pages} {157} (\bibinfo {year}
  {1996})}\BibitemShut {NoStop}%
\bibitem [{\citenamefont {Lin}\ \emph {et~al.}(1998)\citenamefont {Lin},
  \citenamefont {Howe},\ and\ \citenamefont {Pisano}}]{Liwei1998}%
  \BibitemOpen
  \bibfield  {author} {\bibinfo {author} {\bibfnamefont {Liwei}\ \bibnamefont
  {Lin}}, \bibinfo {author} {\bibfnamefont {R.T.}\ \bibnamefont {Howe}}, \ and\
  \bibinfo {author} {\bibfnamefont {A.P.}\ \bibnamefont {Pisano}},\ }\bibfield
  {title} {\enquote {\bibinfo {title} {Microelectromechanical filters for
  signal processing},}\ }\href {\doibase 10.1109/84.709645} {\bibfield
  {journal} {\bibinfo  {journal} {Microelectromechanical Systems, Journal of}\
  }\textbf {\bibinfo {volume} {7}},\ \bibinfo {pages} {286--294} (\bibinfo
  {year} {1998})}\BibitemShut {NoStop}%
\bibitem [{\citenamefont {O'Connell}\ and\ \citenamefont
  {Cleland}(2014)}]{OConnell2014}%
  \BibitemOpen
  \bibfield  {author} {\bibinfo {author} {\bibfnamefont {Aaron}\ \bibnamefont
  {O'Connell}}\ and\ \bibinfo {author} {\bibfnamefont {Andrew~N.}\ \bibnamefont
  {Cleland}},\ }\enquote {\bibinfo {title} {Microwave-frequency mechanical
  resonators operated in the quantum limit},}\ in\ \href
  {http://dx.doi.org/10.1007/978-3-642-55312-7_12} {\emph {\bibinfo {booktitle}
  {Cavity Optomechanics}}},\ \bibinfo {editor} {edited by\ \bibinfo {editor}
  {\bibfnamefont {Markus}\ \bibnamefont {Aspelmeyer}}, \bibinfo {editor}
  {\bibfnamefont {Tobias~J.}\ \bibnamefont {Kippenberg}}, \ and\ \bibinfo
  {editor} {\bibfnamefont {Florian}\ \bibnamefont {Marquardt}}}\ (\bibinfo
  {publisher} {Springer Berlin Heidelberg},\ \bibinfo {year} {2014})\ pp.\
  \bibinfo {pages} {253--281}\BibitemShut {NoStop}%
\bibitem [{\citenamefont {Brown}\ \emph {et~al.}(2007)\citenamefont {Brown},
  \citenamefont {Britton}, \citenamefont {Epstein}, \citenamefont {Chiaverini},
  \citenamefont {Leibfried},\ and\ \citenamefont {Wineland}}]{Brown2007}%
  \BibitemOpen
  \bibfield  {author} {\bibinfo {author} {\bibfnamefont {K.~R.}\ \bibnamefont
  {Brown}}, \bibinfo {author} {\bibfnamefont {J.}~\bibnamefont {Britton}},
  \bibinfo {author} {\bibfnamefont {R.~J.}\ \bibnamefont {Epstein}}, \bibinfo
  {author} {\bibfnamefont {J.}~\bibnamefont {Chiaverini}}, \bibinfo {author}
  {\bibfnamefont {D.}~\bibnamefont {Leibfried}}, \ and\ \bibinfo {author}
  {\bibfnamefont {D.~J.}\ \bibnamefont {Wineland}},\ }\bibfield  {title}
  {\enquote {\bibinfo {title} {Passive cooling of a micromechanical oscillator
  with a resonant electric circuit},}\ }\href {\doibase
  10.1103/PhysRevLett.99.137205} {\bibfield  {journal} {\bibinfo  {journal}
  {Phys. Rev. Lett.}\ }\textbf {\bibinfo {volume} {99}},\ \bibinfo {pages}
  {137205} (\bibinfo {year} {2007})}\BibitemShut {NoStop}%
\bibitem [{\citenamefont {Gardiner}\ and\ \citenamefont
  {Zoller}(2000)}]{Gardiner}%
  \BibitemOpen
  \bibfield  {author} {\bibinfo {author} {\bibfnamefont {Crispin~W.}\
  \bibnamefont {Gardiner}}\ and\ \bibinfo {author} {\bibfnamefont {Peter}\
  \bibnamefont {Zoller}},\ }\href@noop {} {\emph {\bibinfo {title} {Quantum
  Noise}}},\ \bibinfo {edition} {second enlarged edition}\ ed.,\ \bibinfo
  {series} {Springer series in synergetics}, Vol.~\bibinfo {volume} {56}\
  (\bibinfo  {publisher} {Springer},\ \bibinfo {year} {2000})\BibitemShut
  {NoStop}%
\bibitem [{\citenamefont {Giovannetti}\ and\ \citenamefont
  {Vitali}(2001)}]{Giovannetti2001}%
  \BibitemOpen
  \bibfield  {author} {\bibinfo {author} {\bibfnamefont {Vittorio}\
  \bibnamefont {Giovannetti}}\ and\ \bibinfo {author} {\bibfnamefont {David}\
  \bibnamefont {Vitali}},\ }\bibfield  {title} {\enquote {\bibinfo {title}
  {Phase-noise measurement in a cavity with a movable mirror undergoing quantum
  brownian motion},}\ }\href {\doibase 10.1103/PhysRevA.63.023812} {\bibfield
  {journal} {\bibinfo  {journal} {Phys. Rev. A}\ }\textbf {\bibinfo {volume}
  {63}},\ \bibinfo {pages} {023812} (\bibinfo {year} {2001})}\BibitemShut
  {NoStop}%
\bibitem [{\citenamefont {Massel}\ \emph {et~al.}(2012)\citenamefont {Massel},
  \citenamefont {Cho}, \citenamefont {Pirkkalainen}, \citenamefont {Hakonen},
  \citenamefont {Heikkil{\"a}},\ and\ \citenamefont
  {Sillanp{\"a}{\"a}}}]{Massel2012}%
  \BibitemOpen
  \bibfield  {author} {\bibinfo {author} {\bibfnamefont {Francesco}\
  \bibnamefont {Massel}}, \bibinfo {author} {\bibfnamefont {Sung~Un}\
  \bibnamefont {Cho}}, \bibinfo {author} {\bibfnamefont {Juha-Matti}\
  \bibnamefont {Pirkkalainen}}, \bibinfo {author} {\bibfnamefont {Pertti~J.}\
  \bibnamefont {Hakonen}}, \bibinfo {author} {\bibfnamefont {Tero~T.}\
  \bibnamefont {Heikkil{\"a}}}, \ and\ \bibinfo {author} {\bibfnamefont
  {Mika~A.}\ \bibnamefont {Sillanp{\"a}{\"a}}},\ }\bibfield  {title} {\enquote
  {\bibinfo {title} {Multimode circuit optomechanics near the quantum limit},}\
  }\href {http://dx.doi.org/10.1038/ncomms1993} {\bibfield  {journal} {\bibinfo
   {journal} {Nat Commun}\ }\textbf {\bibinfo {volume} {3}},\ \bibinfo {pages}
  {987} (\bibinfo {year} {2012})}\BibitemShut {NoStop}%
\bibitem [{\citenamefont {Schmid}\ \emph {et~al.}(2010)\citenamefont {Schmid},
  \citenamefont {Hierold},\ and\ \citenamefont {Boisen}}]{Schmid2010}%
  \BibitemOpen
  \bibfield  {author} {\bibinfo {author} {\bibfnamefont {Silvan}\ \bibnamefont
  {Schmid}}, \bibinfo {author} {\bibfnamefont {Christofer}\ \bibnamefont
  {Hierold}}, \ and\ \bibinfo {author} {\bibfnamefont {Anja}\ \bibnamefont
  {Boisen}},\ }\bibfield  {title} {\enquote {\bibinfo {title} {Modeling the
  kelvin polarization force actuation of micro- and nanomechanical systems},}\
  }\href {\doibase 10.1063/1.3309027} {\bibfield  {journal} {\bibinfo
  {journal} {Journal of Applied Physics}\ }\textbf {\bibinfo {volume} {107}},\
  \bibinfo {eid} {054510} (\bibinfo {year} {2010})}\BibitemShut {NoStop}%
\bibitem [{\citenamefont {Zhou}\ \emph {et~al.}(2013)\citenamefont {Zhou},
  \citenamefont {Hocke}, \citenamefont {Schliesser}, \citenamefont {Marx},
  \citenamefont {Huebl}, \citenamefont {Gross},\ and\ \citenamefont
  {Kippenberg}}]{Zhou2013}%
  \BibitemOpen
  \bibfield  {author} {\bibinfo {author} {\bibfnamefont {X.}~\bibnamefont
  {Zhou}}, \bibinfo {author} {\bibfnamefont {F.}~\bibnamefont {Hocke}},
  \bibinfo {author} {\bibfnamefont {A.}~\bibnamefont {Schliesser}}, \bibinfo
  {author} {\bibfnamefont {A.}~\bibnamefont {Marx}}, \bibinfo {author}
  {\bibfnamefont {H.}~\bibnamefont {Huebl}}, \bibinfo {author} {\bibfnamefont
  {R.}~\bibnamefont {Gross}}, \ and\ \bibinfo {author} {\bibfnamefont {T.~J.}\
  \bibnamefont {Kippenberg}},\ }\bibfield  {title} {\enquote {\bibinfo {title}
  {Slowing, advancing and switching of microwave signals using circuit
  nanoelectromechanics},}\ }\href {http://dx.doi.org/10.1038/nphys2527}
  {\bibfield  {journal} {\bibinfo  {journal} {Nat Phys}\ }\textbf {\bibinfo
  {volume} {9}},\ \bibinfo {pages} {179--184} (\bibinfo {year}
  {2013})}\BibitemShut {NoStop}%
\bibitem [{\citenamefont {Crandall}\ \emph {et~al.}(1968)\citenamefont
  {Crandall}, \citenamefont {Karnopp}, \citenamefont {Edward F.~Kurtz},\ and\
  \citenamefont {Pridmore-Brown}}]{Crandall1968}%
  \BibitemOpen
  \bibfield  {author} {\bibinfo {author} {\bibfnamefont {Stephen~H.}\
  \bibnamefont {Crandall}}, \bibinfo {author} {\bibfnamefont {Dean~C.}\
  \bibnamefont {Karnopp}}, \bibinfo {author} {\bibfnamefont {Jr.}\ \bibnamefont
  {Edward F.~Kurtz}}, \ and\ \bibinfo {author} {\bibfnamefont {David~C.}\
  \bibnamefont {Pridmore-Brown}},\ }\href@noop {} {\emph {\bibinfo {title}
  {Dynamics of mechanical and electromechanical systems}}},\ edited by\
  \bibinfo {editor} {\bibfnamefont {Stephen~H.}\ \bibnamefont {Crandall}}\
  (\bibinfo  {publisher} {McGraw-Hill},\ \bibinfo {year} {1968})\BibitemShut
  {NoStop}%
\bibitem [{\citenamefont {Elste}\ \emph {et~al.}(2009)\citenamefont {Elste},
  \citenamefont {Girvin},\ and\ \citenamefont {Clerk}}]{Elste2009}%
  \BibitemOpen
  \bibfield  {author} {\bibinfo {author} {\bibfnamefont {Florian}\ \bibnamefont
  {Elste}}, \bibinfo {author} {\bibfnamefont {S.~M.}\ \bibnamefont {Girvin}}, \
  and\ \bibinfo {author} {\bibfnamefont {A.~A.}\ \bibnamefont {Clerk}},\
  }\bibfield  {title} {\enquote {\bibinfo {title} {Quantum noise interference
  and backaction cooling in cavity nanomechanics},}\ }\href {\doibase
  10.1103/PhysRevLett.102.207209} {\bibfield  {journal} {\bibinfo  {journal}
  {Phys. Rev. Lett.}\ }\textbf {\bibinfo {volume} {102}},\ \bibinfo {pages}
  {207209} (\bibinfo {year} {2009})}\BibitemShut {NoStop}%
\bibitem [{\citenamefont {Xuereb}\ \emph {et~al.}(2012)\citenamefont {Xuereb},
  \citenamefont {Genes},\ and\ \citenamefont {Dantan}}]{Xuereb2012}%
  \BibitemOpen
  \bibfield  {author} {\bibinfo {author} {\bibfnamefont {Andr\'e}\ \bibnamefont
  {Xuereb}}, \bibinfo {author} {\bibfnamefont {Claudiu}\ \bibnamefont {Genes}},
  \ and\ \bibinfo {author} {\bibfnamefont {Aur\'elien}\ \bibnamefont
  {Dantan}},\ }\bibfield  {title} {\enquote {\bibinfo {title} {Strong coupling
  and long-range collective interactions in optomechanical arrays},}\ }\href
  {\doibase 10.1103/PhysRevLett.109.223601} {\bibfield  {journal} {\bibinfo
  {journal} {Phys. Rev. Lett.}\ }\textbf {\bibinfo {volume} {109}},\ \bibinfo
  {pages} {223601} (\bibinfo {year} {2012})}\BibitemShut {NoStop}%
\bibitem [{\citenamefont {Shkarin}\ \emph {et~al.}(2014)\citenamefont
  {Shkarin}, \citenamefont {Flowers-Jacobs}, \citenamefont {Hoch},
  \citenamefont {Kashkanova}, \citenamefont {Deutsch}, \citenamefont
  {Reichel},\ and\ \citenamefont {Harris}}]{Shkarin2014}%
  \BibitemOpen
  \bibfield  {author} {\bibinfo {author} {\bibfnamefont {A.~B.}\ \bibnamefont
  {Shkarin}}, \bibinfo {author} {\bibfnamefont {N.~E.}\ \bibnamefont
  {Flowers-Jacobs}}, \bibinfo {author} {\bibfnamefont {S.~W.}\ \bibnamefont
  {Hoch}}, \bibinfo {author} {\bibfnamefont {A.~D.}\ \bibnamefont
  {Kashkanova}}, \bibinfo {author} {\bibfnamefont {C.}~\bibnamefont {Deutsch}},
  \bibinfo {author} {\bibfnamefont {J.}~\bibnamefont {Reichel}}, \ and\
  \bibinfo {author} {\bibfnamefont {J.~G.~E.}\ \bibnamefont {Harris}},\
  }\bibfield  {title} {\enquote {\bibinfo {title} {Optically mediated
  hybridization between two mechanical modes},}\ }\href {\doibase
  10.1103/PhysRevLett.112.013602} {\bibfield  {journal} {\bibinfo  {journal}
  {Phys. Rev. Lett.}\ }\textbf {\bibinfo {volume} {112}},\ \bibinfo {pages}
  {013602} (\bibinfo {year} {2014})}\BibitemShut {NoStop}%
\bibitem [{\citenamefont {Dorf}\ and\ \citenamefont
  {Svoboda}(2010)}]{dorf2010}%
  \BibitemOpen
  \bibfield  {author} {\bibinfo {author} {\bibfnamefont {R.C.}\ \bibnamefont
  {Dorf}}\ and\ \bibinfo {author} {\bibfnamefont {J.A.}\ \bibnamefont
  {Svoboda}},\ }\enquote {\bibinfo {title} {Introduction to electric
  circuits},}\ \ (\bibinfo  {publisher} {John Wiley \& Sons},\ \bibinfo {year}
  {2010})\ Chap.\ \bibinfo {chapter} {5 Circuit Theorems},\ \bibinfo {edition}
  {8th}\ ed.\BibitemShut {Stop}%
\bibitem [{\citenamefont {Pozar}(2012)}]{Pozar}%
  \BibitemOpen
  \bibfield  {author} {\bibinfo {author} {\bibfnamefont {David~M.}\
  \bibnamefont {Pozar}},\ }\href@noop {} {\emph {\bibinfo {title} {Microwave
  Engineering}}},\ \bibinfo {edition} {4th}\ ed.\ (\bibinfo  {publisher} {John
  Wiley \& Sons, Inc.},\ \bibinfo {year} {2012})\BibitemShut {NoStop}%
\bibitem [{\citenamefont {Nyquist}(1928)}]{Nyquist1928}%
  \BibitemOpen
  \bibfield  {author} {\bibinfo {author} {\bibfnamefont {H.}~\bibnamefont
  {Nyquist}},\ }\bibfield  {title} {\enquote {\bibinfo {title} {Thermal
  agitation of electric charge in conductors},}\ }\href {\doibase
  10.1103/PhysRev.32.110} {\bibfield  {journal} {\bibinfo  {journal} {Phys.
  Rev.}\ }\textbf {\bibinfo {volume} {32}},\ \bibinfo {pages} {110--113}
  (\bibinfo {year} {1928})}\BibitemShut {NoStop}%
\bibitem [{\citenamefont {{Zeuthen}}\ \emph {et~al.}(2016)\citenamefont
  {{Zeuthen}}, \citenamefont {{Schliesser}}, \citenamefont {{S{\o}rensen}},\
  and\ \citenamefont {{Taylor}}}]{alpha}%
  \BibitemOpen
  \bibfield  {author} {\bibinfo {author} {\bibfnamefont {E.}~\bibnamefont
  {{Zeuthen}}}, \bibinfo {author} {\bibfnamefont {A.}~\bibnamefont
  {{Schliesser}}}, \bibinfo {author} {\bibfnamefont {A.~S.}\ \bibnamefont
  {{S{\o}rensen}}}, \ and\ \bibinfo {author} {\bibfnamefont {J.~M.}\
  \bibnamefont {{Taylor}}},\ }\bibfield  {title} {\enquote {\bibinfo {title}
  {{Figures of merit for quantum transducers}},}\ }\href@noop {} {\bibfield
  {journal} {\bibinfo  {journal} {ArXiv e-prints}\ } (\bibinfo {year}
  {2016})},\ \Eprint {http://arxiv.org/abs/1610.01099} {arXiv:1610.01099
  [quant-ph]} \BibitemShut {NoStop}%
\bibitem [{\citenamefont {Weis}\ \emph {et~al.}(2010)\citenamefont {Weis},
  \citenamefont {Rivi{\`e}re}, \citenamefont {Del{\'e}glise}, \citenamefont
  {Gavartin}, \citenamefont {Arcizet}, \citenamefont {Schliesser},\ and\
  \citenamefont {Kippenberg}}]{Weis2010}%
  \BibitemOpen
  \bibfield  {author} {\bibinfo {author} {\bibfnamefont {Stefan}\ \bibnamefont
  {Weis}}, \bibinfo {author} {\bibfnamefont {R{\'e}mi}\ \bibnamefont
  {Rivi{\`e}re}}, \bibinfo {author} {\bibfnamefont {Samuel}\ \bibnamefont
  {Del{\'e}glise}}, \bibinfo {author} {\bibfnamefont {Emanuel}\ \bibnamefont
  {Gavartin}}, \bibinfo {author} {\bibfnamefont {Olivier}\ \bibnamefont
  {Arcizet}}, \bibinfo {author} {\bibfnamefont {Albert}\ \bibnamefont
  {Schliesser}}, \ and\ \bibinfo {author} {\bibfnamefont {Tobias~J.}\
  \bibnamefont {Kippenberg}},\ }\bibfield  {title} {\enquote {\bibinfo {title}
  {Optomechanically induced transparency},}\ }\href {\doibase
  10.1126/science.1195596} {\bibfield  {journal} {\bibinfo  {journal}
  {Science}\ }\textbf {\bibinfo {volume} {330}},\ \bibinfo {pages} {1520--1523}
  (\bibinfo {year} {2010})},\ \Eprint
  {http://arxiv.org/abs/http://www.sciencemag.org/content/330/6010/1520.full.pdf}
  {http://www.sciencemag.org/content/330/6010/1520.full.pdf} \BibitemShut
  {NoStop}%
\bibitem [{\citenamefont {Safavi-Naeini}\ \emph
  {et~al.}(2013{\natexlab{a}})\citenamefont {Safavi-Naeini}, \citenamefont
  {Chan}, \citenamefont {Hill}, \citenamefont {Gr{\"o}blacher}, \citenamefont
  {Miao}, \citenamefont {Chen}, \citenamefont {Aspelmeyer},\ and\ \citenamefont
  {Painter}}]{Safavi-Naeini2013b}%
  \BibitemOpen
  \bibfield  {author} {\bibinfo {author} {\bibfnamefont {Amir~H}\ \bibnamefont
  {Safavi-Naeini}}, \bibinfo {author} {\bibfnamefont {Jasper}\ \bibnamefont
  {Chan}}, \bibinfo {author} {\bibfnamefont {Jeff~T}\ \bibnamefont {Hill}},
  \bibinfo {author} {\bibfnamefont {Simon}\ \bibnamefont {Gr{\"o}blacher}},
  \bibinfo {author} {\bibfnamefont {Haixing}\ \bibnamefont {Miao}}, \bibinfo
  {author} {\bibfnamefont {Yanbei}\ \bibnamefont {Chen}}, \bibinfo {author}
  {\bibfnamefont {Markus}\ \bibnamefont {Aspelmeyer}}, \ and\ \bibinfo {author}
  {\bibfnamefont {Oskar}\ \bibnamefont {Painter}},\ }\bibfield  {title}
  {\enquote {\bibinfo {title} {Laser noise in cavity-optomechanical cooling and
  thermometry},}\ }\href {http://stacks.iop.org/1367-2630/15/i=3/a=035007}
  {\bibfield  {journal} {\bibinfo  {journal} {New Journal of Physics}\ }\textbf
  {\bibinfo {volume} {15}},\ \bibinfo {pages} {035007} (\bibinfo {year}
  {2013}{\natexlab{a}})}\BibitemShut {NoStop}%
\bibitem [{\citenamefont {Clerk}\ \emph {et~al.}(2010)\citenamefont {Clerk},
  \citenamefont {Devoret}, \citenamefont {Girvin}, \citenamefont {Marquardt},\
  and\ \citenamefont {Schoelkopf}}]{QNreview}%
  \BibitemOpen
  \bibfield  {author} {\bibinfo {author} {\bibfnamefont {A.~A.}\ \bibnamefont
  {Clerk}}, \bibinfo {author} {\bibfnamefont {M.~H.}\ \bibnamefont {Devoret}},
  \bibinfo {author} {\bibfnamefont {S.~M.}\ \bibnamefont {Girvin}}, \bibinfo
  {author} {\bibfnamefont {Florian}\ \bibnamefont {Marquardt}}, \ and\ \bibinfo
  {author} {\bibfnamefont {R.~J.}\ \bibnamefont {Schoelkopf}},\ }\bibfield
  {title} {\enquote {\bibinfo {title} {Introduction to quantum noise,
  measurement, and amplification},}\ }\href {\doibase
  10.1103/RevModPhys.82.1155} {\bibfield  {journal} {\bibinfo  {journal} {Rev.
  Mod. Phys.}\ }\textbf {\bibinfo {volume} {82}},\ \bibinfo {pages}
  {1155--1208} (\bibinfo {year} {2010})}\BibitemShut {NoStop}%
\bibitem [{\citenamefont {Safavi-Naeini}\ \emph
  {et~al.}(2013{\natexlab{b}})\citenamefont {Safavi-Naeini}, \citenamefont
  {Groblacher}, \citenamefont {Hill}, \citenamefont {Chan}, \citenamefont
  {Aspelmeyer},\ and\ \citenamefont {Painter}}]{Safavi-Naeini2013}%
  \BibitemOpen
  \bibfield  {author} {\bibinfo {author} {\bibfnamefont {Amir~H.}\ \bibnamefont
  {Safavi-Naeini}}, \bibinfo {author} {\bibfnamefont {Simon}\ \bibnamefont
  {Groblacher}}, \bibinfo {author} {\bibfnamefont {Jeff~T.}\ \bibnamefont
  {Hill}}, \bibinfo {author} {\bibfnamefont {Jasper}\ \bibnamefont {Chan}},
  \bibinfo {author} {\bibfnamefont {Markus}\ \bibnamefont {Aspelmeyer}}, \ and\
  \bibinfo {author} {\bibfnamefont {Oskar}\ \bibnamefont {Painter}},\
  }\bibfield  {title} {\enquote {\bibinfo {title} {Squeezed light from a
  silicon micromechanical resonator},}\ }\href
  {http://dx.doi.org/10.1038/nature12307} {\bibfield  {journal} {\bibinfo
  {journal} {Nature}\ }\textbf {\bibinfo {volume} {500}},\ \bibinfo {pages}
  {185--189} (\bibinfo {year} {2013}{\natexlab{b}})}\BibitemShut {NoStop}%
\bibitem [{\citenamefont {Purdy}\ \emph {et~al.}(2013)\citenamefont {Purdy},
  \citenamefont {Yu}, \citenamefont {Peterson}, \citenamefont {Kampel},\ and\
  \citenamefont {Regal}}]{Purdy2013}%
  \BibitemOpen
  \bibfield  {author} {\bibinfo {author} {\bibfnamefont {T.~P.}\ \bibnamefont
  {Purdy}}, \bibinfo {author} {\bibfnamefont {P.-L.}\ \bibnamefont {Yu}},
  \bibinfo {author} {\bibfnamefont {R.~W.}\ \bibnamefont {Peterson}}, \bibinfo
  {author} {\bibfnamefont {N.~S.}\ \bibnamefont {Kampel}}, \ and\ \bibinfo
  {author} {\bibfnamefont {C.~A.}\ \bibnamefont {Regal}},\ }\bibfield  {title}
  {\enquote {\bibinfo {title} {Strong optomechanical squeezing of light},}\
  }\href {\doibase 10.1103/PhysRevX.3.031012} {\bibfield  {journal} {\bibinfo
  {journal} {Phys. Rev. X}\ }\textbf {\bibinfo {volume} {3}},\ \bibinfo {pages}
  {031012} (\bibinfo {year} {2013})}\BibitemShut {NoStop}%
\bibitem [{\citenamefont {{Lvovsky}}(2014)}]{Lvovsky2014}%
  \BibitemOpen
  \bibfield  {author} {\bibinfo {author} {\bibfnamefont {A.~I.}\ \bibnamefont
  {{Lvovsky}}},\ }\bibfield  {title} {\enquote {\bibinfo {title} {{Squeezed
  light}},}\ }\href@noop {} {\bibfield  {journal} {\bibinfo  {journal} {ArXiv
  e-prints}\ } (\bibinfo {year} {2014})},\ \Eprint
  {http://arxiv.org/abs/1401.4118} {arXiv:1401.4118 [quant-ph]} \BibitemShut
  {NoStop}%
\bibitem [{\citenamefont {Mancini}\ and\ \citenamefont
  {Tombesi}(1994)}]{Mancini1994}%
  \BibitemOpen
  \bibfield  {author} {\bibinfo {author} {\bibfnamefont {S.}~\bibnamefont
  {Mancini}}\ and\ \bibinfo {author} {\bibfnamefont {P.}~\bibnamefont
  {Tombesi}},\ }\bibfield  {title} {\enquote {\bibinfo {title} {Quantum noise
  reduction by radiation pressure},}\ }\href {\doibase
  10.1103/PhysRevA.49.4055} {\bibfield  {journal} {\bibinfo  {journal} {Phys.
  Rev. A}\ }\textbf {\bibinfo {volume} {49}},\ \bibinfo {pages} {4055--4065}
  (\bibinfo {year} {1994})}\BibitemShut {NoStop}%
\bibitem [{\citenamefont {Fabre}\ \emph {et~al.}(1994)\citenamefont {Fabre},
  \citenamefont {Pinard}, \citenamefont {Bourzeix}, \citenamefont {Heidmann},
  \citenamefont {Giacobino},\ and\ \citenamefont {Reynaud}}]{Fabre1994}%
  \BibitemOpen
  \bibfield  {author} {\bibinfo {author} {\bibfnamefont {C.}~\bibnamefont
  {Fabre}}, \bibinfo {author} {\bibfnamefont {M.}~\bibnamefont {Pinard}},
  \bibinfo {author} {\bibfnamefont {S.}~\bibnamefont {Bourzeix}}, \bibinfo
  {author} {\bibfnamefont {A.}~\bibnamefont {Heidmann}}, \bibinfo {author}
  {\bibfnamefont {E.}~\bibnamefont {Giacobino}}, \ and\ \bibinfo {author}
  {\bibfnamefont {S.}~\bibnamefont {Reynaud}},\ }\bibfield  {title} {\enquote
  {\bibinfo {title} {Quantum-noise reduction using a cavity with a movable
  mirror},}\ }\href {\doibase 10.1103/PhysRevA.49.1337} {\bibfield  {journal}
  {\bibinfo  {journal} {Phys. Rev. A}\ }\textbf {\bibinfo {volume} {49}},\
  \bibinfo {pages} {1337--1343} (\bibinfo {year} {1994})}\BibitemShut {NoStop}%
\bibitem [{\citenamefont {Dobrindt}\ \emph {et~al.}(2008)\citenamefont
  {Dobrindt}, \citenamefont {Wilson-Rae},\ and\ \citenamefont
  {Kippenberg}}]{Dobrindt2008}%
  \BibitemOpen
  \bibfield  {author} {\bibinfo {author} {\bibfnamefont {J.~M.}\ \bibnamefont
  {Dobrindt}}, \bibinfo {author} {\bibfnamefont {I.}~\bibnamefont
  {Wilson-Rae}}, \ and\ \bibinfo {author} {\bibfnamefont {T.~J.}\ \bibnamefont
  {Kippenberg}},\ }\bibfield  {title} {\enquote {\bibinfo {title} {Parametric
  normal-mode splitting in cavity optomechanics},}\ }\href {\doibase
  10.1103/PhysRevLett.101.263602} {\bibfield  {journal} {\bibinfo  {journal}
  {Phys. Rev. Lett.}\ }\textbf {\bibinfo {volume} {101}},\ \bibinfo {pages}
  {263602} (\bibinfo {year} {2008})}\BibitemShut {NoStop}%
\bibitem [{\citenamefont {Stannigel}\ \emph {et~al.}(2010)\citenamefont
  {Stannigel}, \citenamefont {Rabl}, \citenamefont {S\o{}rensen}, \citenamefont
  {Zoller},\ and\ \citenamefont {Lukin}}]{Stannigel2010}%
  \BibitemOpen
  \bibfield  {author} {\bibinfo {author} {\bibfnamefont {K.}~\bibnamefont
  {Stannigel}}, \bibinfo {author} {\bibfnamefont {P.}~\bibnamefont {Rabl}},
  \bibinfo {author} {\bibfnamefont {A.~S.}\ \bibnamefont {S\o{}rensen}},
  \bibinfo {author} {\bibfnamefont {P.}~\bibnamefont {Zoller}}, \ and\ \bibinfo
  {author} {\bibfnamefont {M.~D.}\ \bibnamefont {Lukin}},\ }\bibfield  {title}
  {\enquote {\bibinfo {title} {Optomechanical transducers for long-distance
  quantum communication},}\ }\href {\doibase 10.1103/PhysRevLett.105.220501}
  {\bibfield  {journal} {\bibinfo  {journal} {Phys. Rev. Lett.}\ }\textbf
  {\bibinfo {volume} {105}},\ \bibinfo {pages} {220501} (\bibinfo {year}
  {2010})}\BibitemShut {NoStop}%
\bibitem [{\citenamefont {Burkard}\ \emph {et~al.}(2004)\citenamefont
  {Burkard}, \citenamefont {Koch},\ and\ \citenamefont
  {DiVincenzo}}]{Burkard2004}%
  \BibitemOpen
  \bibfield  {author} {\bibinfo {author} {\bibfnamefont {Guido}\ \bibnamefont
  {Burkard}}, \bibinfo {author} {\bibfnamefont {Roger~H.}\ \bibnamefont
  {Koch}}, \ and\ \bibinfo {author} {\bibfnamefont {David~P.}\ \bibnamefont
  {DiVincenzo}},\ }\bibfield  {title} {\enquote {\bibinfo {title} {Multilevel
  quantum description of decoherence in superconducting qubits},}\ }\href
  {\doibase 10.1103/PhysRevB.69.064503} {\bibfield  {journal} {\bibinfo
  {journal} {Phys. Rev. B}\ }\textbf {\bibinfo {volume} {69}},\ \bibinfo
  {pages} {064503} (\bibinfo {year} {2004})}\BibitemShut {NoStop}%
\bibitem [{\citenamefont {Burkard}(2005)}]{Burkard2005}%
  \BibitemOpen
  \bibfield  {author} {\bibinfo {author} {\bibfnamefont {Guido}\ \bibnamefont
  {Burkard}},\ }\bibfield  {title} {\enquote {\bibinfo {title} {Circuit theory
  for decoherence in superconducting charge qubits},}\ }\href {\doibase
  10.1103/PhysRevB.71.144511} {\bibfield  {journal} {\bibinfo  {journal} {Phys.
  Rev. B}\ }\textbf {\bibinfo {volume} {71}},\ \bibinfo {pages} {144511}
  (\bibinfo {year} {2005})}\BibitemShut {NoStop}%
\bibitem [{\citenamefont {Nigg}\ \emph {et~al.}(2012)\citenamefont {Nigg},
  \citenamefont {Paik}, \citenamefont {Vlastakis}, \citenamefont {Kirchmair},
  \citenamefont {Shankar}, \citenamefont {Frunzio}, \citenamefont {Devoret},
  \citenamefont {Schoelkopf},\ and\ \citenamefont {Girvin}}]{Nigg2012}%
  \BibitemOpen
  \bibfield  {author} {\bibinfo {author} {\bibfnamefont {Simon~E.}\
  \bibnamefont {Nigg}}, \bibinfo {author} {\bibfnamefont {Hanhee}\ \bibnamefont
  {Paik}}, \bibinfo {author} {\bibfnamefont {Brian}\ \bibnamefont {Vlastakis}},
  \bibinfo {author} {\bibfnamefont {Gerhard}\ \bibnamefont {Kirchmair}},
  \bibinfo {author} {\bibfnamefont {S.}~\bibnamefont {Shankar}}, \bibinfo
  {author} {\bibfnamefont {Luigi}\ \bibnamefont {Frunzio}}, \bibinfo {author}
  {\bibfnamefont {M.~H.}\ \bibnamefont {Devoret}}, \bibinfo {author}
  {\bibfnamefont {R.~J.}\ \bibnamefont {Schoelkopf}}, \ and\ \bibinfo {author}
  {\bibfnamefont {S.~M.}\ \bibnamefont {Girvin}},\ }\bibfield  {title}
  {\enquote {\bibinfo {title} {Black-box superconducting circuit
  quantization},}\ }\href {\doibase 10.1103/PhysRevLett.108.240502} {\bibfield
  {journal} {\bibinfo  {journal} {Phys. Rev. Lett.}\ }\textbf {\bibinfo
  {volume} {108}},\ \bibinfo {pages} {240502} (\bibinfo {year}
  {2012})}\BibitemShut {NoStop}%
\end{thebibliography}%

\end{document}